\documentclass[10pt,conference]{IEEEtran}
\IEEEoverridecommandlockouts
\usepackage{cite}
\usepackage{lscape}
\usepackage{booktabs}
\usepackage{makecell}
\usepackage[table,xcdraw]{xcolor}
\usepackage{float}
\usepackage{threeparttable}
\usepackage{adjustbox}
\usepackage{amsmath,amssymb,amsfonts}
\usepackage{algorithmic}
\usepackage{graphicx}
\usepackage{textcomp}
\usepackage{url}
\usepackage{fontawesome}
\usepackage{adjustbox}

\usepackage{breakurl}
\usepackage[breaklinks]{hyperref}
\usepackage{xcolor,pifont}
\def\BibTeX{{\rm B\kern-.05em{\sc i\kern-.025em b}\kern-.08em
    T\kern-.1667em\lower.7ex\hbox{E}\kern-.125emX}}
\usepackage{flushend}
\usepackage{multicol}
\usepackage{placeins}
\usepackage{indentfirst}
\usepackage{fancybox}
\usepackage[most]{tcolorbox}
\usepackage{multirow}
\usepackage[normalem]{ulem}
\useunder{\uline}{\ul}{}

\usepackage{MnSymbol,wasysym}

\usepackage{color}
\usepackage{framed}
\definecolor{Gray}{gray}{0.7}
\definecolor{lightGray}{gray}{0.9}

\begin{document}

\title{
% An Empirical Study on Software Bill of Materials: Where We Stand Now, and the Road Ahead
An Empirical Study on Software Bill of Materials: Where We Stand and the Road Ahead
% Towards a Roadmap for Operationalizing Software Bill of Materials (SBOM)
}
% \author{
% \IEEEauthorblockN{Anoymous Authors}

% % \and
% % \IEEEauthorblockN{3\textsuperscript{rd} Given Name Surname}
% % \IEEEauthorblockA{\textit{dept. name of organization (of Aff.)} \\
% % \textit{name of organization (of Aff.)}\\
% % City, Country \\
% % email address or ORCID}
% % \and
% % \IEEEauthorblockN{4\textsuperscript{th} Given Name Surname}
% % \IEEEauthorblockA{\textit{dept. name of organization (of Aff.)} \\
% % \textit{name of organization (of Aff.)}\\
% % City, Country \\
% % email address or ORCID}
% % \and
% % \IEEEauthorblockN{5\textsuperscript{th} Given Name Surname}
% % \IEEEauthorblockA{\textit{dept. name of organization (of Aff.)} \\
% % \textit{name of organization (of Aff.)}\\
% % City, Country \\
% % email address or ORCID}
% % \and
% % \IEEEauthorblockN{6\textsuperscript{th} Given Name Surname}
% % \IEEEauthorblockA{\textit{dept. name of organization (of Aff.)} \\
% % \textit{name of organization (of Aff.)}\\
% % City, Country \\
% % email address or ORCID}
% }
\author{
\IEEEauthorblockN{Boming Xia\IEEEauthorrefmark{1}\IEEEauthorrefmark{2}, Tingting Bi\IEEEauthorrefmark{1}\IEEEauthorrefmark{3}, Zhenchang Xing\IEEEauthorrefmark{1}\IEEEauthorrefmark{4}, Qinghua Lu\IEEEauthorrefmark{1}\IEEEauthorrefmark{2}, Liming Zhu\IEEEauthorrefmark{1}\IEEEauthorrefmark{2}}
\IEEEauthorblockA{\IEEEauthorrefmark{1}\textit{CSIRO's Data61, Sydney, Australia}\\
\IEEEauthorrefmark{2}\textit{University of New South Wales, Sydney, Australia}\\
\IEEEauthorrefmark{3}\textit{Monash University, Melbourne, Australia}\\
\IEEEauthorrefmark{4}\textit{Australian National University, Canberra, Australia}\\
% firstname.lastname@data61.csiro.au
}
% \and
% \and
% \and
% \and
}

\maketitle

\begin{abstract}
%The prevalence of software and spiking 
The rapid growth of software supply chain attacks has attracted considerable attention to software bill of materials (SBOM).
SBOMs are a crucial building block to ensure the transparency of software supply chains that helps improve software supply chain security.
Although there are significant efforts from academia and industry to facilitate SBOM development, 
it is still unclear how practitioners perceive SBOMs and what are the challenges of adopting SBOMs in practice.
Furthermore, existing SBOM-related studies tend to be ad-hoc and lack software engineering focuses.
To bridge this gap, we conducted the first empirical study to interview and survey SBOM practitioners. We applied a mixed qualitative and quantitative method for gathering data from 17 interviewees and 65 survey respondents from 15 countries across five continents to understand how practitioners perceive the SBOM field. We summarized 26 statements and grouped them into three topics on SBOM's states of practice.
Based on the study results, we derived a goal model and highlighted future directions where practitioners can put in their effort.
% \note{This paper is distinguished from existing work (e.g., Linux Foundation's SBOM and Cybersecurity Readiness report) from the following four aspects: a) more up-to-date view of the SBOM field; b) software engineering-oriented focus; c) investigation of the SBOM status and expectations from practitioners' perspective; and d) a systematic and mixed research methodology.}
\end{abstract}


\section{Introduction}
% \zc{Too much ``boring'' stuff before these eye-catching, high-profile incidents. Should make these high-profile incidents very up-front, what damages they caused or may cause in the context of software supply chain security.}

% \zc{Maybe start with a sentence like this statement from LF report ``industry leaders have transformed to software-defined models, enabled by cloud computing, edge computing, artificial intelligence software, and embedded systems. Along with this digital transformation opportunity comes increasing cybersecurity risk if software assets are not sourced and managed appropriately.'' This gives the big context where SBOM plays critical role, and it naturally leads to software supply chain security.}

% {\color{blue} TB: Can we put the second paragraph ahead? like state software supply chain..then what was the issues in the software supply, like vulnerabilities or attacks (the first paragraph}

% The prevalence of software systems plays an essential role in boosting daily organizational operations\cite{Furrer2019}. Despite the opportunities and benefits of the software-enabled and connected services, the underlying security and integrity issues of software are not to be neglected. Especially,
Modern software products are assembled through intricate and dynamic supply chains \cite{framing_ntia}, while recent attacks against software supply chains (SSC) have increased significantly (e.g., SolarWinds attack\cite{enwiki:1104117684}).
% A software supply chain consists of the components, libraries, tools, and processes used to develop, build, and publish a software artifact \cite{deming2020good}, including open source software/components (OSS) as the majority of the software organizations use OSS \cite{lf_2022}.
According to Sonatype's report\cite{sonatype_report}, there was a 650\% year-over-year increase in SSC attacks in 2021, and the number was 430\% in 2020.
% Especially, recent infamous vulnerabilities or even attacks (e.g., Apache Log4j \cite{}, SolarWinds attack \cite{}) 
% have made their case that it's imperative for software vendors to have systematic visibility into the software systems and their supply chains, so that timely and accurate identification of impacted components can be done.
SSC attacks mainly aim at the upstream open source software/components (OSS) \cite{ohm2020backstabber}, yet OSS is heavily relied upon in software development \cite{lf_2022}. The reliance on OSS leads to additional risks, such as the lack of reliable maintenance and support compared to proprietary software/components\cite{singh2015open}.
% For example, an identified issued may not be timely addressed since OSS can be maintained by a group of develop
% For example, in March 2022, the primary maintainer of a popular OSS package, node-ipc, intentionally injected malware into an update that overwrote file systems in specific geographical locations \cite{9810126}. 
% The pervasiveness of software and yet 
The security risks of software and its supply chain call for improved visibility into the SSC, with which timely and accurate identification of the impacted software/components could be carried out in case of a vulnerability or an SSC attack.

A software bill of materials (SBOM) is a formal machine-readable inventory of the components (and their dependency relationships) used for producing a software product \cite{sbom_mini}. SBOMs enhance the security of both the proprietary and open source components in SSCs\cite{gartner_sbom} through improved transparency. 
According to Linux Foundation's SBOM and Cybersecurity Readiness report (SBOM readiness report for short) \cite{lf_2022}, SBOMs are critical for enhancing SSC security. 90\% of the surveyed organizations have started or are planning their SBOM journey, with 54\% already addressing SBOMs. The report also estimated 66\% and 13\% growth in SBOM production or consumption in 2022 and 2023, respectively.
Nonetheless, some organizations are still concerned about how SBOM adoption and application will evolve (e.g., 40\% are uncertain about industrial SBOM commitment and 39\% seek consensus on SBOM data fields).

Despite SBOMs' essentiality for software and SSC security, there remain questions to answer and problems to solve. 
% \zc{Maybe a bit more background from LF report Executive Summary (point 1-7) to show SBOM is important and world-wide concern. Then say there are many unanswered questions ...}
% However, according to Linux Foundation's SBOM and Cybersecurity Readiness report \cite{lf_2022}, there are three major concerns on SBOMs: a) how/when/where/who to integrate SBOMs production and consumption in the DevOps process; b) how to integrate SBOMs into the governance, risk, and compliance (GRC) process; and, c) industrial consensus on SBOMs.
Motivated by the value of SBOMs and the existing gaps, with the overarching goal of investigating the SBOM status quo from practitioners' perspectives, this paper aims to answer the following research questions (RQs).

\textbf{RQ1: What is the current state of SBOM practice?} 

% This RQ investigates the following key SBOM aspects in practice: SBOM benefits, adoption, generation, distribution, validation, and SBOM for vulnerability management.
Despite the benefits of SBOMs and the SBOM readiness report showing an overall 90\% of SBOM readiness, how practitioners perceive SBOMs and how SBOMs are being addressed in practice needs further investigation. To answer this question, we analyzed the SBOM practice status from SBOM generation, distribution and sharing, validation and verification, and vulnerability and exploitability management. We summarized the current SBOM practices and what practitioners expect.

% our interview and survey results are not as optimistic, and SBOMs are not widely adopted in practice.
% SBOM generation and distribution also require further standardization and more mature mechanisms.
% Furthermore, the validation of SBOM data is also generally neglected in current practice (agreed by 49.2\% of survey respondents), which causes potential risks as a malicious party can easily tamper with SBOMs.
% In addition, for the typical SBOM use case of vulnerability management, the exploitability status should be more than binary as a seemingly unexploitable vulnerability may still be exploited.\\
\textbf{RQ2: What is the current state of SBOM tooling support?}

% This RQ focuses on SBOM tooling status including the .
Despite the proliferation of the SBOM tooling market, this RQ focuses on the SBOM tooling status from the practitioners' perspective. We investigated the practitioners' attitudes towards existing tools from the following aspects: the necessity/availability/usability/integrity of SBOM tools.
While exploring the current SBOM tooling state, we also looked into practitioners' expectations of SBOM tools.
% Although some SBOM tools take the existing data (e.g., package manager) and output the information in a more unified way as SBOMs, the necessity of SBOM tool are recognized.

% Although some interviewees think SBOMs are merely existing data (e.g., packager manager) fed into SBOM tools to output standard formats, 83.1\% of respondents acknowledge the importance and necessity of SBOM tools.
% However, even though there are many existing SBOM tools and the SBOM tools market is expected to burgeon in 2022 and 2023\cite{lf_2022}, most existing tools are open source and focus on SBOM generation, while SBOM consumption on the procurers' side is significantly deficient.
% Moreover, 64.6\% of respondents agree that existing SBOM tools can be hard to use for various reasons (e.g., complexity).
% Another issue for SBOM tooling is how to validate SBOM tools, such as their ability to generate accurate and complete SBOMs and their security against malicious manipulation.\\
\textbf{RQ3: What are the main concerns for SBOM?}

% This RQ focuses on the main concerns practitioners have for SBOMs.
This RQ investigates the most outstanding concerns SBOM practitioners have.
Although the prospect of SBOMs is promising, there are still challenges to resolve.
With this RQ, we aim to provide a reference for the most imminent issues for future research and development on SBOMs.

% During the interviews, SBOM tool developers’ common concern was that the existing SBOM standard formats could not fully support their needs.
% The second concern is that organizations are worried that the SBOMs with detailed information about the software products can serve as ``roadmaps for attackers". The NTIA has justified this\cite{NTIA_myth}, but 63.1\% of respondents remain concerned.
% The most fundamental concern lies with the lack of SBOM adoption and education, agreed by 80\% of respondents.

% To answer the research questions above, we can provide a clear picture of the SBOM field as well as point out the remaining issues that require further attention.

Our research aims to unveil the state of the SBOM field, investigating what practitioners have and how they are addressing SBOMs, versus what they expect. Our work on SBOMs has four main distinctions and contributions compared to existing work represented by the SBOM readiness report: 
\begin{enumerate}
    \item \textbf{Timeliness}: The SBOM readiness report was published in January 2022, with the survey launched in June 2021. \href{https://trends.google.com/trends/explore?date=2021-07-01 2022-11-14&q=SBOM}{Google Trends} shows a significant increase in SBOM interest since July 2021. This study provides a more updated view of SBOMs.
    \item \textbf{Software engineering (SE) angles}: The SBOM readiness report is industry-oriented and security-focused, whereas this research broadens and deepens the report and supplements SE angles, evidenced by considerations such as SBOM generation/update throughout software development lifecycle (Finding 5) and AIBOM (Section \ref{aibom}). The detailed comparison between this research and the SBOM readiness report is presented in Table \ref{compare}.
    % . We summarize the 
    % evidenced by additional considerations on SBOM lifecycle (e.g., dynamic SBOM generation (Finding 3), SBOM re-generation and versioning (Section \ref{sbomgepoint}) and AIBOM (Section \ref{aibom}).
    \item \textbf{Different objectives}: The SBOM readiness report aims to investigate whether and to what extent organizations are prepared for SBOM production and consumption (i.e., readiness). In contrast, this study focuses on current SBOM practices and expectations from practitioners' perspectives. We provide a set of implications, including a goal model, for future endeavors towards further operationalizing SBOMs.
    \item \textbf{Systematic methodology}: We conducted the first empirical study on the SBOM status from practitioners' perspectives, using a mixed methodology. Instead of using predefined questions as in SBOM readiness report, we qualitatively and quantitatively coded in-depth opinions from 17 interviewees into 26 representative statements, which were validated in a survey with 65 valid respondents from 15 countries.
\end{enumerate}

% \begin{figure}[]
% \includegraphics[width=\columnwidth]{RQ.png}
% \caption{Research questions.}
% \label{rq}
% \end{figure}
% The main contributions of this paper are:
% \begin{itemize}
%     \item %We investigated how the practitioners perceive SBOM. %the SBOM field. 
%     To the best of our knowledge, this is the first empirical study on the SBOM status quo. We interviewed 17 SBOM practitioners and conducted an online survey of 65 valid respondents from 15 countries. The study results can shed light on how practitioners perceive the SBOM field.
%     \item We examined the current issues and concerns that practitioners have for SBOMs; we provided a set of implications, including a goal model for future endeavors towards further operationalizing SBOMs.
% \end{itemize}  

% \zc{If need space, this can be cut.}
The remainder of this paper is organized as follows. Section \ref{background} describes the context of the SBOM field.
Section \ref{3method} presents the methodology of our study.
In section \ref{4results} we present the study results.
Section \ref{Dis} discusses the implications of this study.
Section \ref{6related} discusses related work, and section \ref{7conclusion} draws conclusions and outlines avenues for future work.

\section{Background: What is SBOM?}
\label{background}

\begin{figure}[htb]
\includegraphics[width=\columnwidth]{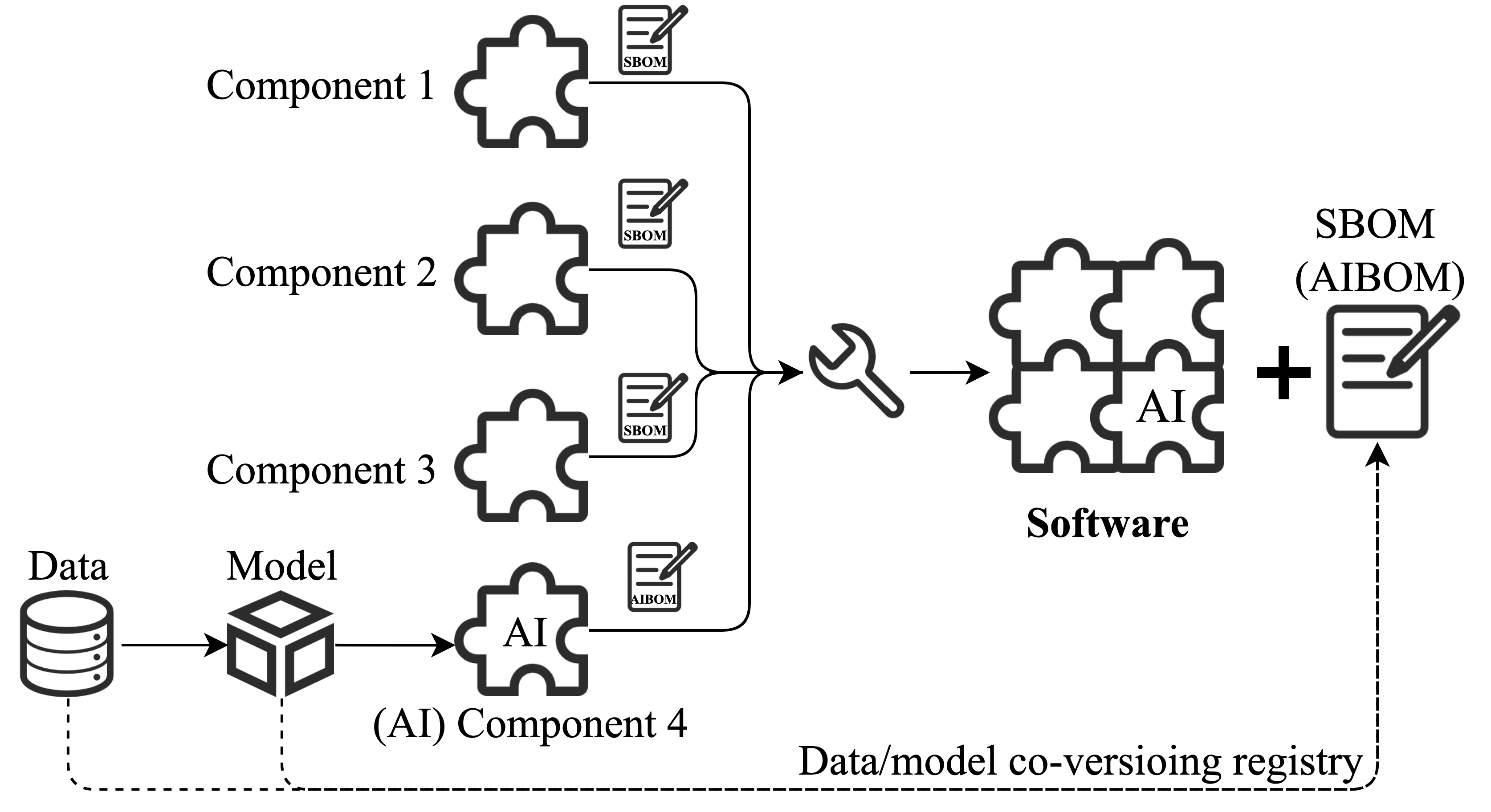}
\caption{(AI) Software supply chain and SBOM\cite{google_white_paper}.}
\label{aissc_fig}
\end{figure}

A bill of materials (BOM) was initially used in the manufacturing industry as an inventory list of all the sub-assemblies and components in a parent assembly\cite{jiao2000generic}.
Sharing the same origin, an SBOM as the building block to enhanced software supply chain security is a software ``BOM" (see Fig. \ref{aissc_fig}).

% \begin{figure}[]
% \centering
% \includegraphics[width=\columnwidth]{ssc.png}
% \caption{Software supply chain and SBOM\cite{google_white_paper}.}
% \label{ssc_fig}
% \end{figure}

There are three main SBOM standard formats: a) Software Package Data eXchange (\href{https://spdx.dev/}{SPDX}), b) \href{https://cyclonedx.org/}{CycloneDX}, and c) Software Identification (\href{https://csrc.nist.gov/projects/Software-Identification-SWID}{SWID}) Tagging, while the first two are most adopted\cite{sbom_his}.
SPDX is an open-source \href{https://www.iso.org/standard/81870.html}{international standard} hosted by the Linux Foundation, emphasizing licence compliance.
% SPDX has a lightweight version (i.e., SPDX Lite) as a trade-off between actual industrial workflow and the standard.
CycloneDX was designed by OWASP in 2017 whose primary focus is security.
SWID Tagging is also an international standard maintained by the US National Institute of Standards and Technology, focusing on providing a transparent software/components identification mechanism.

The above three formats all have the corresponding tooling to help operationalize the formats into practice that are listed on their respective websites. 
 It is worth mentioning that the SBOM formats and tooling working group under the US National Telecommunications and Information Administration (NTIA) had an effort to summarize all tools supporting different standards (i.e., \href{https://docs.google.com/document/d/1A1jFIYihB-IyT0gv7E_KoSjLbwNGmu_wOXBs6siemXA/edit}{SPDX list}, \href{https://docs.google.com/document/d/1biwYXrtoRc_LF7Pw10TO2TGIhlM6jwkDG23nc9M_RiE/edit}{CycloneDX list}, \href{https://docs.google.com/document/d/1oebYvHcOhtMG8Uhnd5he0l_vhty7MsTjp6fYCOwUmwM/edit}{SWID list}).
%  \note{Although the lists were a work in progress and have not been updated since February 2022, they provide a reference on the SBOM tooling playground, where most of the SBOM tools are open source.}

% Despite the promising results, the report also summarized three main concerns regarding SBOMs: a) How to integrate SBOMs production and consumption into DevOps process; b) How to integrate SBOMs production and consumption into GRC (governance, risk, and compliance); and c) Industry consensus on SBOMs evolvement. Inspired by the identified gap, this work aims to narrow it by providing an SBOMs-incorporated DevOps process that is summarized by interviewing and surveying software practitioners, as well as providing insights on SBOMs with GRC and SBOMs evolvement.

In terms of government-side SBOM efforts, 
with incidents such as the SolarWinds attack ringing the alarm of SSC security, the US government issued an executive order on enhancing cybersecurity\cite{house_executive_2021} in May 2021, explicitly mandating all companies trading with the US government to provide SBOMs.
\href{https://www.ntia.gov/SBOM}{NTIA} has published a series of documents and guidelines (e.g., SBOM minimum elements\cite{sbom_mini}) to facilitate SBOM development.
The Cybersecurity and Infrastructure Security Agency (\href{https://www.cisa.gov/sbom}{CISA}) is also actively working on SBOM facilitation by regularly hosting listening sessions with the SBOM industrial community. 
% \zc{??from which perspective? ??add reference}

Notably, the Linux Foundation published the SBOM readiness report\cite{lf_2022} in January 2022, which surveyed 412 organizations across the globe.
According to the report, 98\% of the surveyed organizations are concerned about software security, over 80\% are aware of the US executive order, and 90\% have started their SBOM journey. Although the evolvement of SBOM adoption and application remains a concern, the report predicts that the SBOM tool market is expected to explode in 2022 and 2023.
As of January 2022, there were about 20 SBOM tool vendors in the market, and some were from adjacent markets like Software Composition Analysis (SCA) which dates back to 2002\cite{sca}.
Meanwhile, various open source tools are available with a focus on SBOM generation.

However, while the SBOM readiness report comprehensively covers a broad category, their reports were based on multi-choice questions with limited choices, which constrained the possible answers.
The guidelines and documents published by NTIA provide a reference for understanding SBOM and SBOM practice but lack the actual perception from the SBOM practitioners.
To fill the gaps, we interviewed 17 SBOM practitioners (see Table \ref{interviewee-info}) with open-ended questions on how they perceive current SBOM practice and SBOM tooling. We then organized an online survey containing statements from the interviews, allowing more practitioners to validate the statements.
Compared with the SBOM readiness report and NTIA's documents, we focus on how actual SBOM practitioners think about SBOMs and how they are addressing SBOMs in production without limiting the possible answers.

\section{Research Methodology}
\label{3method}
\begin{figure}[htb]
\includegraphics[width=\columnwidth]{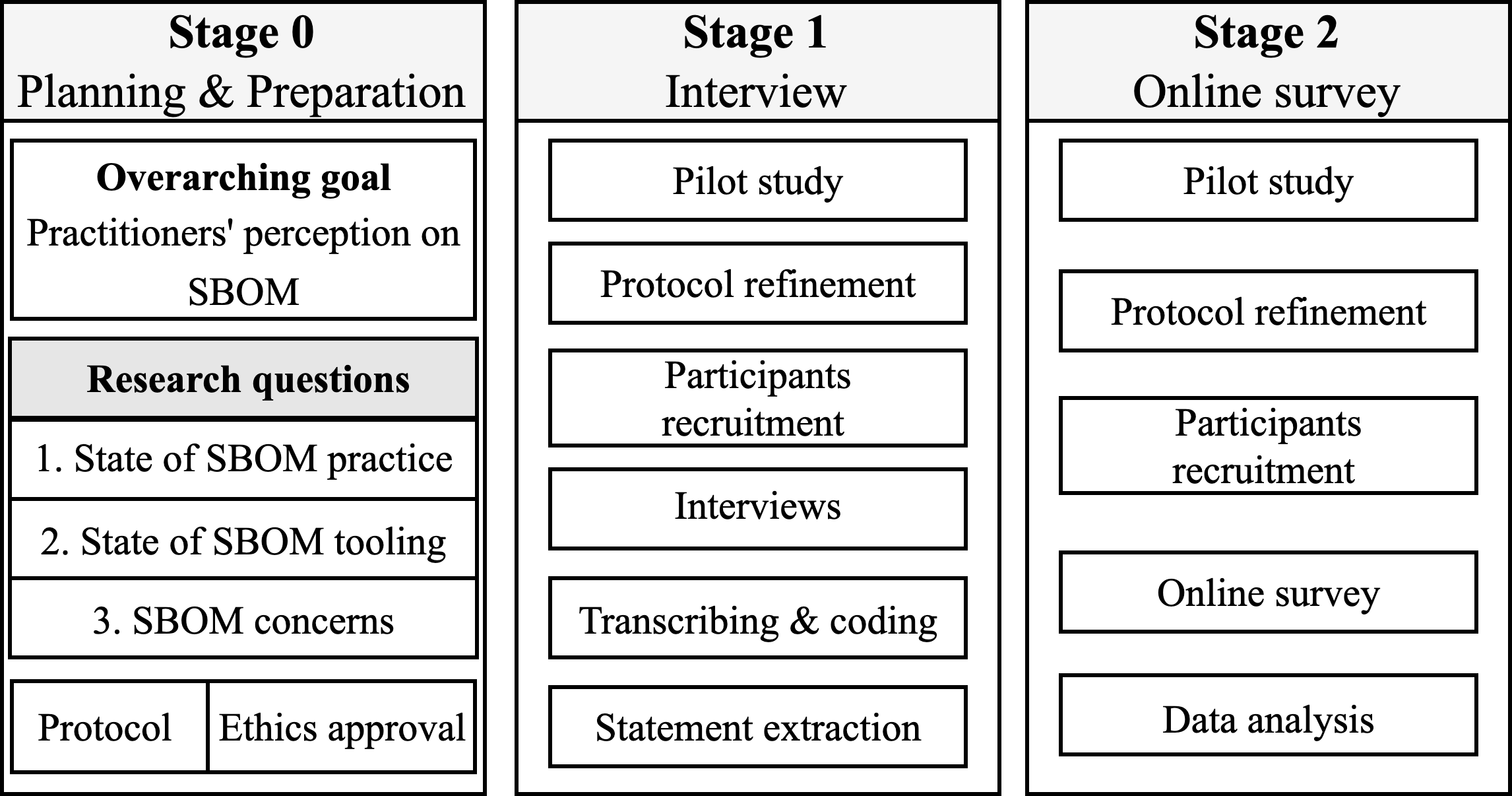}
\caption{Overall research methodology.}
\label{method_fig}
\end{figure}

This study presents an exploratory empirical study on SBOM status in practice.
Fig. \ref{method_fig} shows the overall methodology adopted in this paper that consists of three stages, following a mixed qualitative and quantitative approach \cite{easterbrook2008selecting}.
We described the planning and preparation stage in Section \ref{subsec_planning}; the data collection and analysis processes of the interviews and the online survey are presented in Sections \ref{subsec_interview} and \ref{subsec_survey}.

\subsection{Stage Zero: Planning and Preparation}

\label{subsec_planning}

% Double blind review, we should not mention any institute.

%At this stage, we brainstormed and drafted the research protocol (research questions, interview questions etc.), and 

At the planning stage, we prepared a research protocol\footnote{\url{https://drive.google.com/drive/folders/1gylBoeF09sJRg2xEpJh2tWnjp0cCMhxM?usp=sharing}} and drafted two types of interview questions: demographics and open-ended. For demographics, we ask about the participants' background information, such as job roles and experiences. For the open-ended questions, we asked how the participants perceive SBOMs.
We obtained ethics approvals for this study.
% (Reference number: HC220324 (UNSW), 095/22 (CSIRO)).
% For the ease of the participants, the first-author translated all designed questions (both interviews and online survey) to Chinese, which was double-checked by the second author.

\subsection{Stage One: Interview}
\label{subsec_interview}

\textbf{Pilot interview and protocol refinement}. Before the formal interviews, we conducted a small-scale pilot interview with 3 participants from our connections. Based on their feedback and suggestions, we adjusted some interview questions.

\textbf{Participant recruitment}. We recruited 17 SBOM practitioners from 13 organizations (e.g., CISA, Oracle) across 7 countries (see Table \ref{interviewee-info}). Interviewees were recruited by: a) emailing our contacts who helped further disseminate the invitation emails to their colleagues; b) emailing developers on GitHub working on SBOM-related projects whose email addresses are public; c) advertising on Twitter and LinkedIn and interested people can contact the first author. The 17 interviewees have worked in software-related fields for around 14 years on average (min 4 years and max 30 years), while they have been working actively in SBOM-related fields for around 1.4 years on average (min 2 months and max 5 years). We will refer to the 17 interviewees as I1 to I17.

% \usepackage{graphicx}
% \usepackage{graphicx}
\begin{table}[]
\renewcommand\arraystretch{1.1}
\centering
\caption{Interviewee Demographics*}
\label{interviewee-info}
\resizebox{\columnwidth}{!}{%
\begin{tabular}{lllcc}
\hline
\textbf{ID} &
  \multicolumn{1}{c}{\textbf{Field of work}} &
  \multicolumn{1}{c}{\textbf{Country}} &
  \textbf{Work exp.} &
  \textbf{SBOM exp.} \\ \hline
I1  & Dev.                     & China     & 10  & 0.5 \\
I2  & Dev.                     & China     & 10  & 0.2 \\
I3  & Dev.                     & China     & 10  & 0.2 \\
I4  & Dev. \& Sec.             & Australia & 20  & 3.0 \\
I5  & Dev.                     & US        & 18  & 0.5 \\
I6  & Dev.                     & US        & 20  & 0.5 \\
I7  & Sec.                     & Brazil    & 4   & 0.5 \\
I8  & Dev. \& Cnslt.           & Ireland   & 15  & 1.5 \\
I9  & Sec.                     & India     & 5.5 & 1.0 \\
I10 & Cnslt. \& Adv.           & US        & 20  & 3.0 \\
I11 & Dev. \& Sec. (SBOM tool) & Israel    & 12  & 1.5 \\
I12 & Cnslt. \& Adv.           & Israel    & 15  & 0.3 \\
I13 & Dev.\&Sec.\&Res.         & Australia & 30  & 2.0 \\
I14 & Dev. \& Sec. \& Res.     & US        & 10  & 3.0 \\
I15 & Dev.                     & India     & 8   & 1.5 \\
I16 & Dev. (SBOM tool)         & US        & 20  & 1.0 \\
I17 & Cnslt. \& Adv.           & US        & 15  & 5.0 \\ \hline
\multicolumn{5}{l}{\begin{tabular}[c]{@{}l@{}}\small\\ \small *Dev./Sec.: Software Development/Security; Cnslt.: Consultant; Adv.:\\ \small Advisor; Res.: Researcher. Experiences are listed in years, as of July 2022\end{tabular}}
\end{tabular}%
}
\end{table}

\textbf{Transcribing and coding}. 1) Transcribing. The interviews were audio-recorded. The first author transcribed the audio recordings, and the second author double-checked the transcripts. 2) Pilot coding. The first two authors (i.e., coders) conducted a pilot coding of the 3 pilot interview transcripts. They discussed the initial coding results and reached a certain level of preliminary agreement on the granularity of thematic coding. 3) Code generation. The coders then performed thematic coding to qualitatively analyze the interview transcripts \cite{smith2015qualitative,bi2022accessibility} of 17 interviewees using \href{https://www.maxqda.com/}{MAXQDA2022} tool.
The first coder generated 574 codes under 86 cards (i.e., repetitive and similar codes classified into the same category). The second coder generated 364 codes under 41 cards. After discussing the coding results with a third author, the coders further cleared the coding granularity, combined similar cards, and disposed of cards with limited value. Finally, a total of 54 unique cards were generated.
% The third author then reviewed the generated codes and provided feedback. After adjustment based on the feedback, we generated a total of 938 cards that contained the codes and ended up with 600 unique codes after merging the results from the coders and duplication removal. 
% During the interview process, when saturation was reached after interviewing xx people, where no new codes could be generated, when considered the codes list to be stable.

% (Copied from Accessibility paper) We then conducted thematic coding analysis of the transcripts [Ass 12]. We dropped sentences during the coding process that are not related to "software design and development for accessibility in practice". The first two authors read and coded the contents of transcripts. We used the MAXQDA3 tool for analyzing and coding the qualitative data. To ensure the quality of codes, we invited a Ph.D. candidate to verify the first two authors’ initial codes and provided suggestions for improvement. After incorporating these suggestions, we generated a total of 288 cards that contain the codes. After merging the codes with the same words or meaning, we had a total of 122 unique codes. We noticed that when our interview transcripts reached saturation and new codes did not appear anymore, our list of codes was then considered stable.

\textbf{Data analysis and open card sorting}. The coders separately sorted the 54 generated cards into potential themes (not predefined) given thematic similarity. After the sorting process, the coders calculated Cohen's Kappa value \cite{cohen1960coefficient} to assess their agreement level. The overall value was 0.77, indicating substantial agreement. The coders discussed their disagreements to reach a common ground. he coders reviewed and agreed on the final themes to reduce card sorting bias. Eventually, we derived 26 statements (see Table \ref{statements}) under 3 themes: State of SBOMs Practice (T1), SBOM Tooling Support (T2), and SBOM Issues and Concerns (T3). All the authors have double-checked our coding results to ensure the reported results are accurate and consistent.
% The two coders separately analyzed the codes and classified them into potential statements under different topics which were not pre-defined.  was adopted to measure the agreement between the two authors. The overall value was 0.83, indicating a general consensus between the two coders. The coders discussed the discrepancy and reached an agreement. The coders then reviewed the final results to reduce bias. This process resulted in 26 statements (see Table \ref{statements}).

\subsection{Stage Two: Online Survey}
\label{subsec_survey}

We conducted an online survey to confirm or refute the extracted statements. We designed the survey following Kitchenham and Pfleeger’s guideline \cite{kitchenham2008personal}. The survey was anonymous, and all information collected was non-identifiable. 

\textbf{Survey design and pilot study}. The survey was published via \href{https://www.qualtrics.com/}{Qualtrics}. Different types of questions were included in the survey (e.g., multiple choice and free text). The statements are scored on a 5-point Likert scale (Strongly disagree, Disagree, Neutral, Agree, Strongly agree), with an additional ``Not sure".
% in case the respondents do not understand the statements. 
We piloted the survey with six participants from Australia and Singapore and then refined the survey. The pilot study results were excluded from the final results. The formal survey consists of 7 sections:
demographics, SBOM status quo, generation, distribution, tooling,  benefits, and concerns.
% \begin{itemize}
%   \item Demographics. To better understand the background of the participants, a set of questions was designed to collect demographic information, including geographical location, fields of work, years of software- and SBOM-related experience, team size, and familiarity with SBOMs.
%   \item SBOM status quo. This section list 8 statements on SBOM's current situation.
%   \item SBOM generation. This section list 5 statements on how SBOMs are being generated and re-generated.
%   \item SBOM distribution. This section list 4 statements on how SBOMs are and should be distributed from SBOM producers to downstream procurers
%   \item SBOM tooling. This section mainly investigates the limitations of the existing SBOM tools with 7 statements.
%   \item SBOM benefit. This section list 3 significant benefits commonly brought up by the interviewees. Since there are already surveys on more detailed benefits of SBOMs (e.g., \cite{lf_2022}), here we focus on higher-level benefits.
%   \item SBOM concern. This section list 3 major concerns mentioned by the interviewees. For the same reason, we focus on higher level concerns.
% \end{itemize}

\textbf{Participants recruitment}. To increase the number of participants, we adopted the following strategy for recruitment:
\begin{itemize}
    \item We contacted industrial practitioners from several companies worldwide and asked for their help in disseminating the survey invitation emails.
    \item We sent invitation emails to over 2000 developers from GitHub whose email addresses are publicly available.
    % . Since there were only limited SBOMs-related projects on GitHub, we emailed project contributors based on the project stars even though no SBOM or similar concept was explicitly mentioned in their projects.
    \item We posted the recruitment advertisement on social media platforms (i.e., Twitter and LinkedIn).
\end{itemize}

% \usepackage{booktabs}
% \usepackage{graphicx}
\begin{table}[]
\renewcommand\arraystretch{1.1}
\centering
\caption{Survey respondents demographics*}
\label{survey_demo}
\resizebox{\columnwidth}{!}{%
\begin{tabular}{@{}llll@{}}
\toprule
\multicolumn{1}{c}{\textbf{Field of work}} &
  \multicolumn{1}{c}{\textbf{Project team}} &
  \multicolumn{1}{c}{\textbf{Work exp.}} &
  \multicolumn{1}{c}{\textbf{SBOM exp.}} \\ \midrule
Dev. (38.6\%)        & \textless 10 ppl. (16)    & \textless 1 year (1)       & \textless 6 months (17)   \\
Sec. (30.1\%)        & 10-20 ppl. (20)           & 1-3 years (11)             & 0.5-1 year (17)           \\
Cnslt./Adv. (10.8\%) & 20-50 ppl. (15)           & 3-5 years (13)             & 1-2 years (13)            \\
Mgmt (9.7\%)         & \textgreater 50 ppl. (15) & 5-10 years (9)             & \textgreater 2 years (19) \\
Res. (9.7\%)         & -                         & \textgreater 10 years (32) & -                         \\
Other (1.1\%)        & -                         & -                          & -                         \\ \midrule
\multicolumn{4}{l}{\begin{tabular}[c]{@{}l@{}}\footnotesize{*Mgmt.: Management; ppl.: people. Since multiple answers are supported}\\ for respondents’ work, field of work is listed in percentages while the\\ others are listed with response numbers.\end{tabular}}
\end{tabular}%
}
\end{table}
                                          
% \begin{figure}[]
% \includegraphics[width=\columnwidth]{surveydemo.png}
% \caption{Survey respondents demographics\protect\footnotemark.}
% \label{surveydemo_fig}
% \end{figure}
% \footnotetext{Considering we support multiple answers for respondents' work, field of work is listed in percentages while the others are listed with response numbers.} 
We received a total of 129 responses, including 27 with respondents selecting ``(Very) unfamiliar with SBOM". 
Note that there could be more people unfamiliar with SBOM who did not respond to our survey.
After removing them and the incomplete responses and responses completed within 2 minutes, we had 65 valid responses.
We acknowledge that the number of responses is not as ideal as similar empirical studies (e.g., \cite{bi2022accessibility,xiaxin}). However, we believe this is consistent with our findings on the lack of SBOM adoption and education (i.e., Findings 1 and 10).
% Compared with \zc{our previous work \cite{} ??double blind} with an average of 200 responses, we had a more challenging time trying to recruit more survey participants. This also supports one of the major concerns regarding SBOM adoption.
The 65 participants come from 15 countries across 5 continents. The top 3 countries where the participants reside are Australia, China, and the US. An overview of the survey respondents' demographics is presented in Table \ref{survey_demo}.
It is worth noting that although nearly half (47.7\%) of the respondents have worked in the software field for over 10 years, only one quarter (27.7\%) have worked on SBOMs for over 2 years, indicating that SBOM is still a relatively fresh concept to software practitioners.

\textbf{Data analysis}. Apart from the demographics, SBOM familiarity questions, and a final optional free-text question, all statements are presented as Likert-scale questions (see the bar charts in Table \ref{statements}) for the evaluation of the agreement degree.

\begin{table*}[]
\centering
\caption{interview and survey results on SBOM statements}
\label{statements}
\resizebox{\textwidth}{!}{%
\begin{tabular}{llcc}
\hline
\multicolumn{2}{c|}{} &
  \multicolumn{2}{c}{\textbf{Likert distribution}} \\ \cline{3-4} 
\multicolumn{2}{c|}{\multirow{-2}{*}{\textbf{Statement}}} &
  Graph &
  Score \\ \hline
\multicolumn{4}{l}{\textbf{T1. State of SBOM practice}} \\ \hline
\multicolumn{1}{l|}{S1. Improving transparency and visibility into the software products is the biggest benefit of SBOMs.} &
  \multicolumn{1}{l|}{} &
  \includegraphics[width = 0.85cm, height = 0.35 cm]{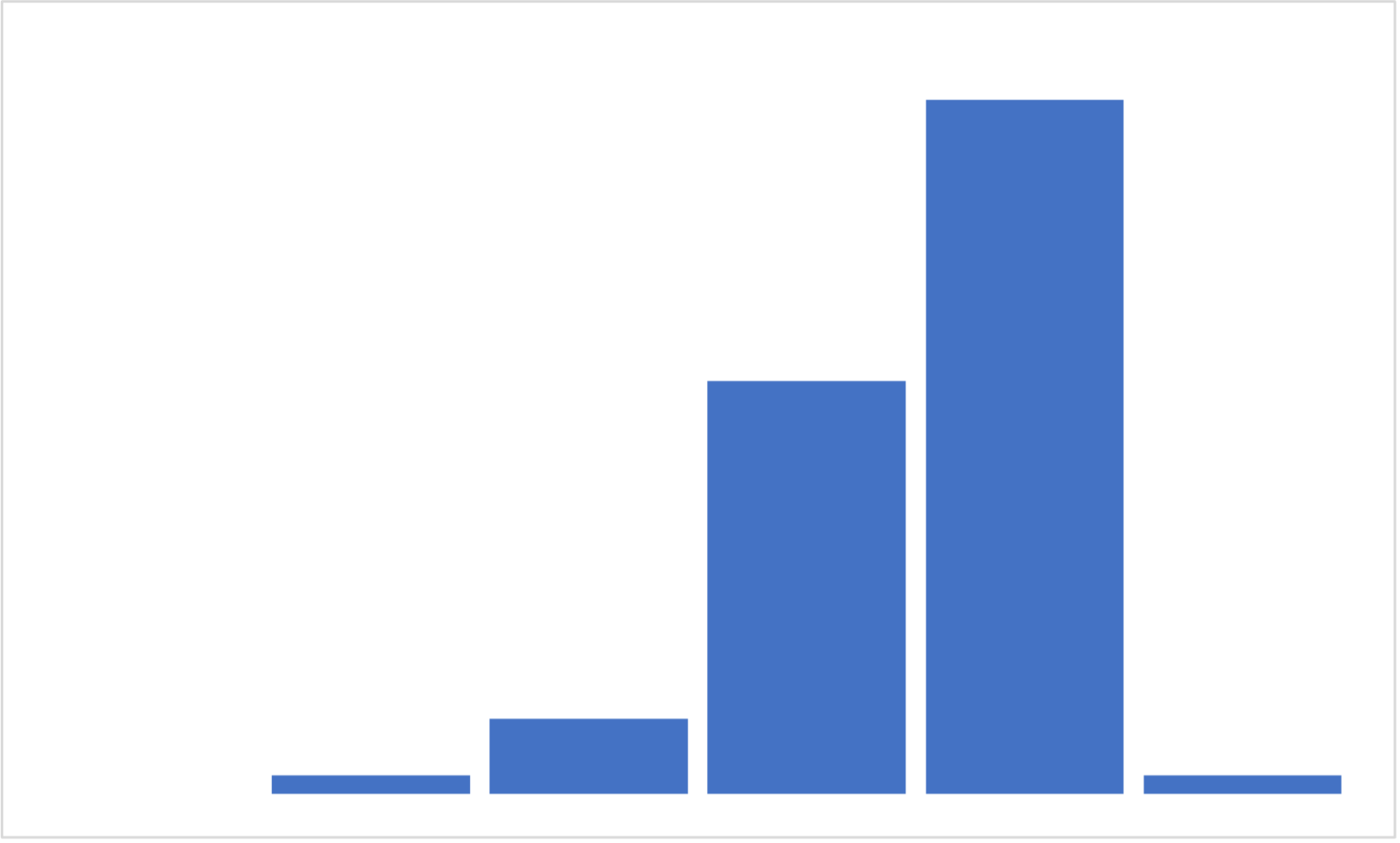} &
  4.42 \\
\multicolumn{1}{l|}{S2. SBOM data form the foundation of a potential SBOM-centric ecosystem.} &
  \multicolumn{1}{l|}{} &
  \includegraphics[width = 0.85cm, height = 0.35 cm]{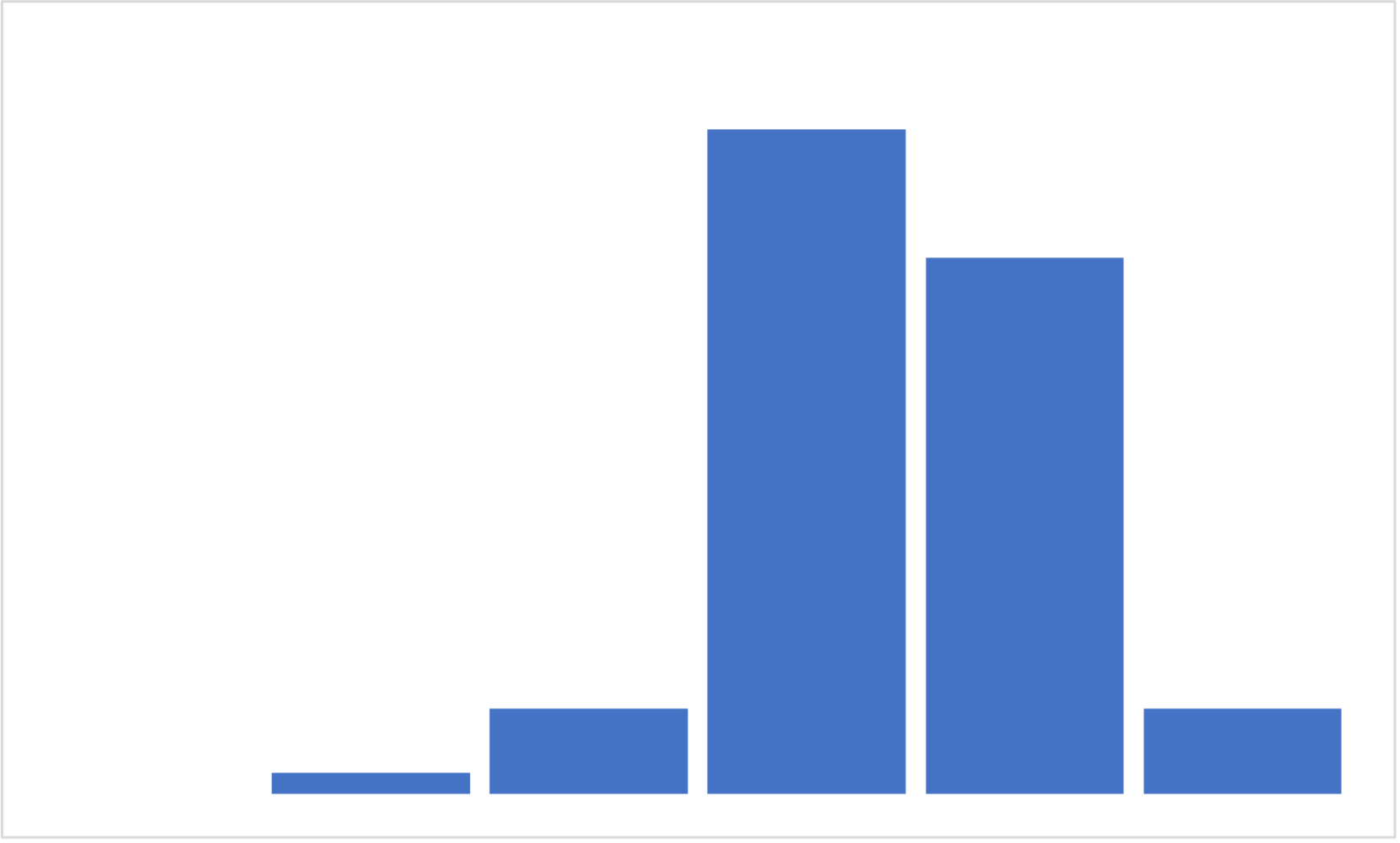} &
  4.05 \\
\multicolumn{1}{l|}{S3. The benefits brought by SBOMs outweigh the costs of SBOMs (e.g., extra learning and management of SBOMs \& tools).} &
  \multicolumn{1}{l|}{\multirow{-3}{*}{\begin{tabular}[c]{@{}l@{}}Section \ref{benefit}\\ SBOM benefits\end{tabular}}} &
  \includegraphics[width = 0.85cm, height = 0.35 cm]{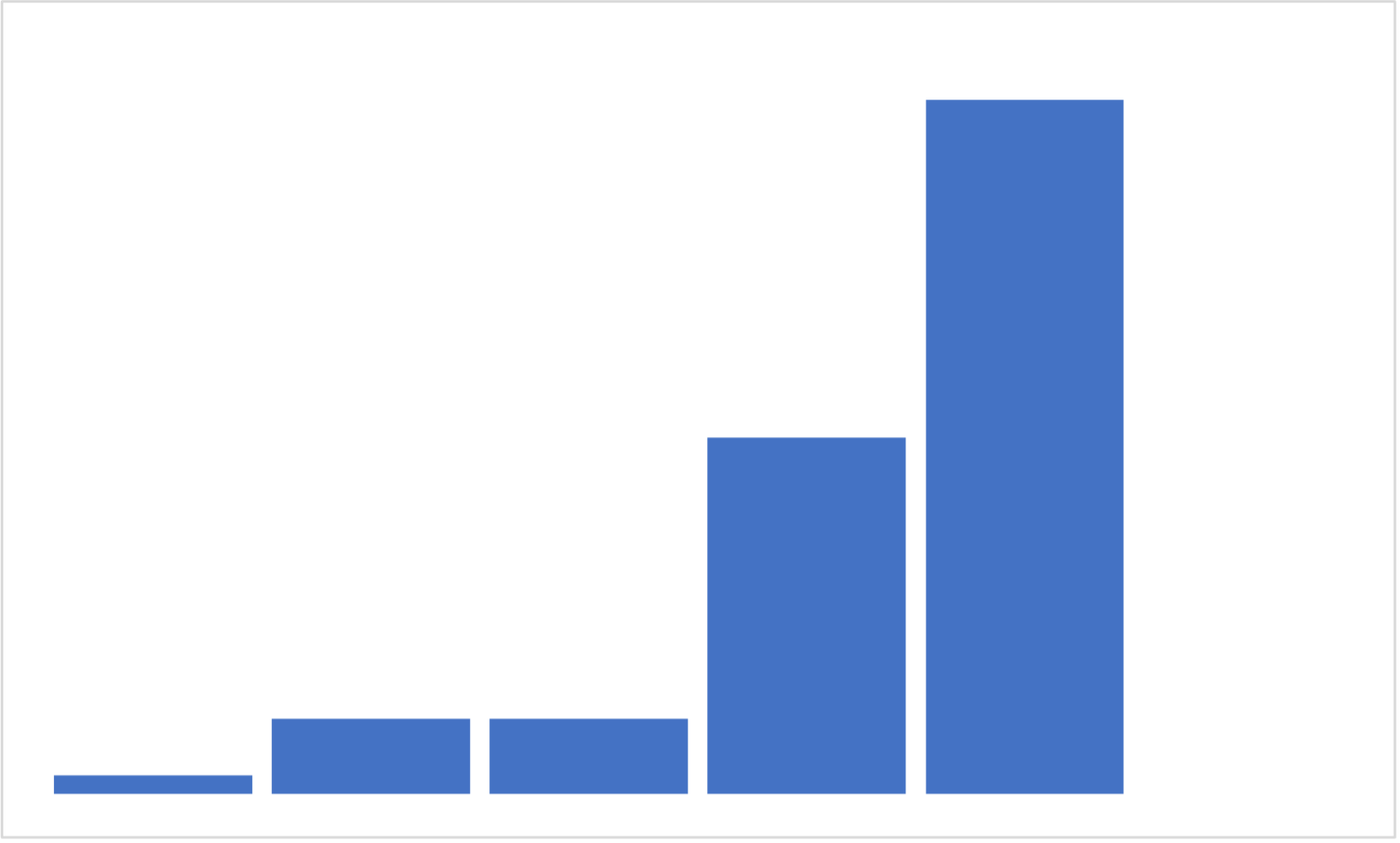} &
  4.34 \\ \hline
\multicolumn{1}{l|}{S4. Currently third-party (open source or proprietary) components are not equipped with SBOMs.} &
  \multicolumn{1}{l|}{} &
  \includegraphics[width = 0.85cm, height = 0.35 cm]{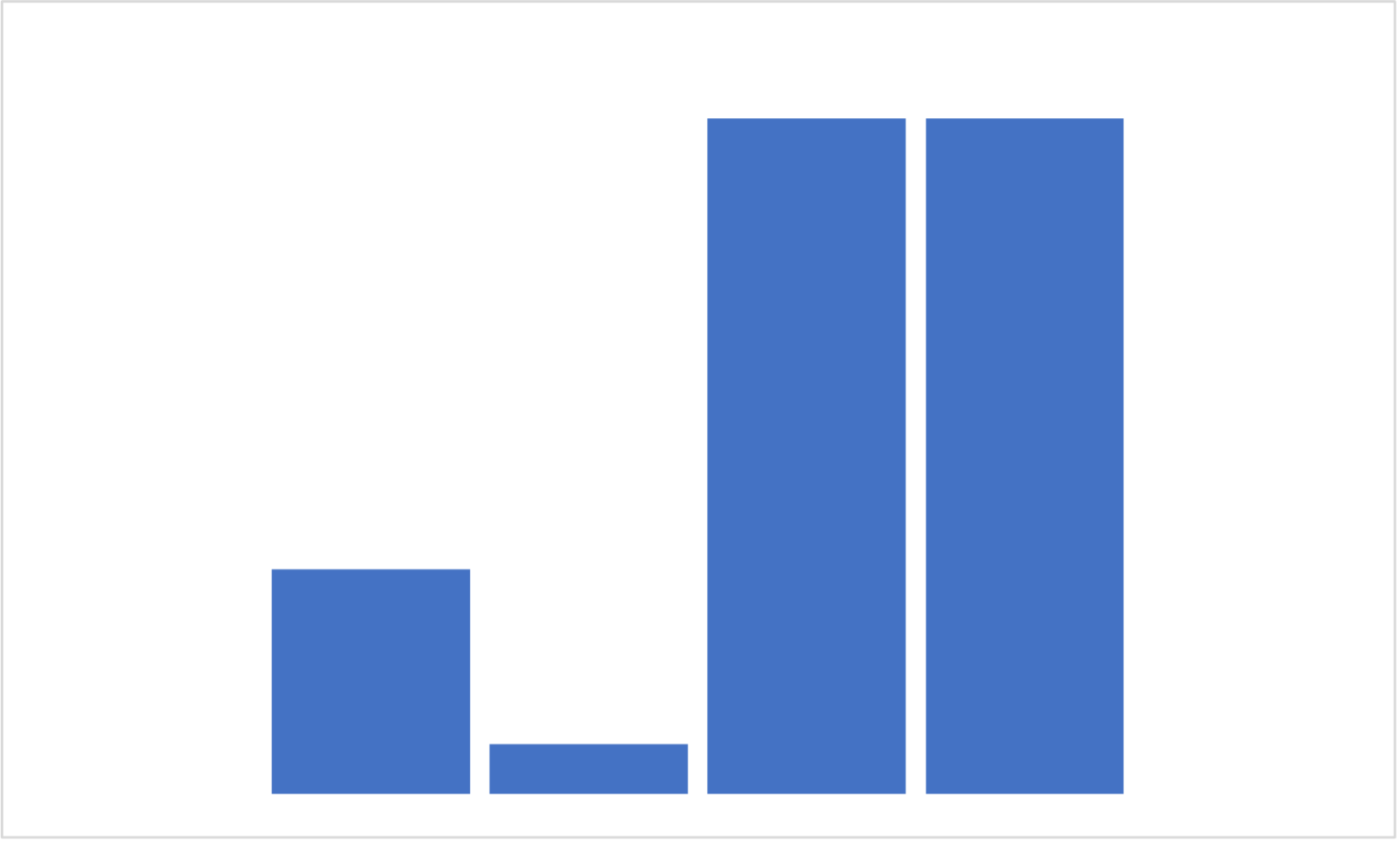} &
  4.11 \\
\multicolumn{1}{l|}{S5. SBOMs are not generated for all software products (produced/used) within an organization.} &
  \multicolumn{1}{l|}{\multirow{-2}{*}{\begin{tabular}[c]{@{}l@{}}Section \ref{adoption}\\ SBOM adoption\end{tabular}}} &
  \includegraphics[width = 0.85cm, height = 0.35 cm]{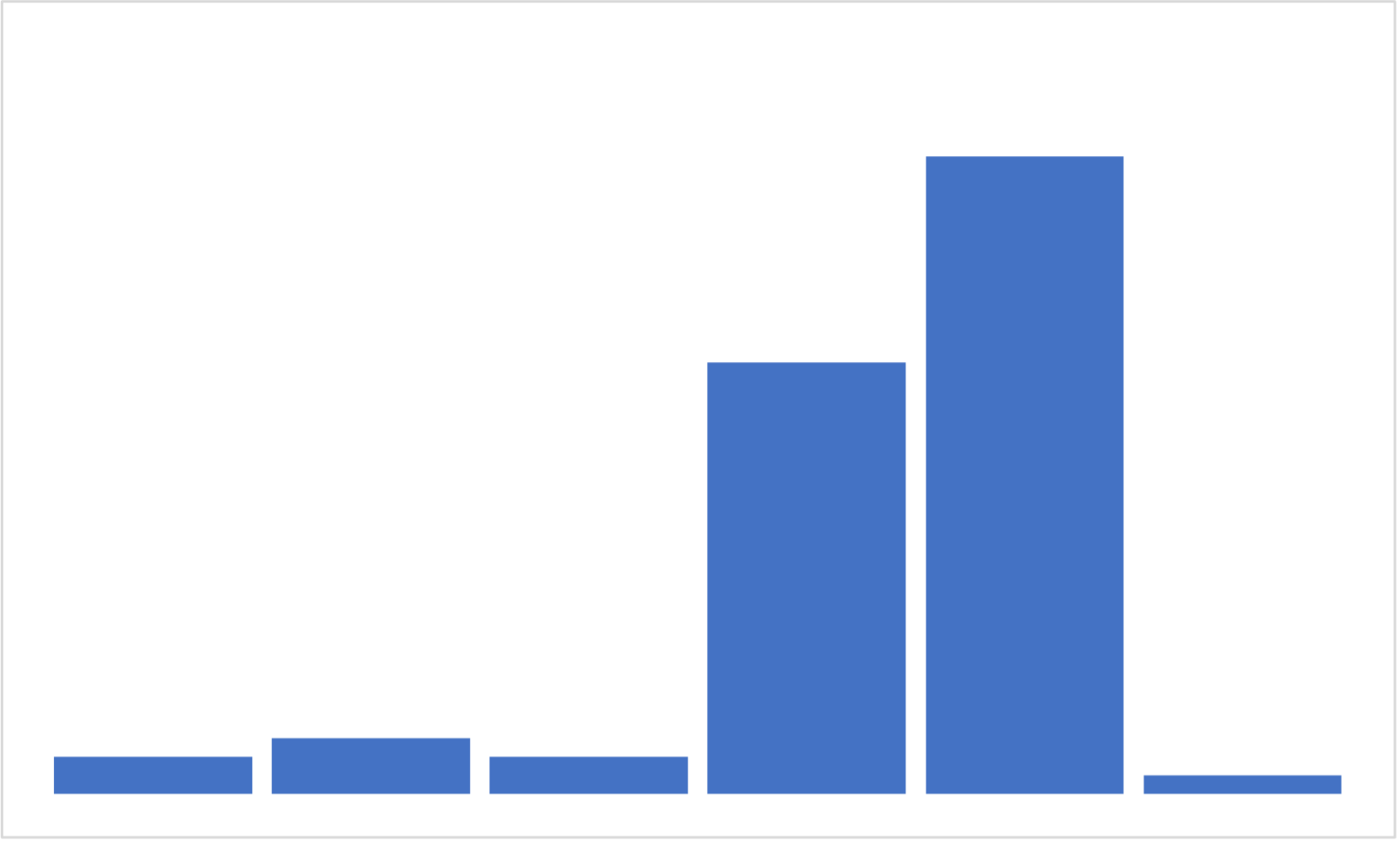} &
  4.25 \\ \hline
\multicolumn{1}{l|}{S6. SBOMs can be generated at different stages of the software development lifecycle.} &
  \multicolumn{1}{l|}{} &
  \includegraphics[width = 0.85cm, height = 0.35 cm]{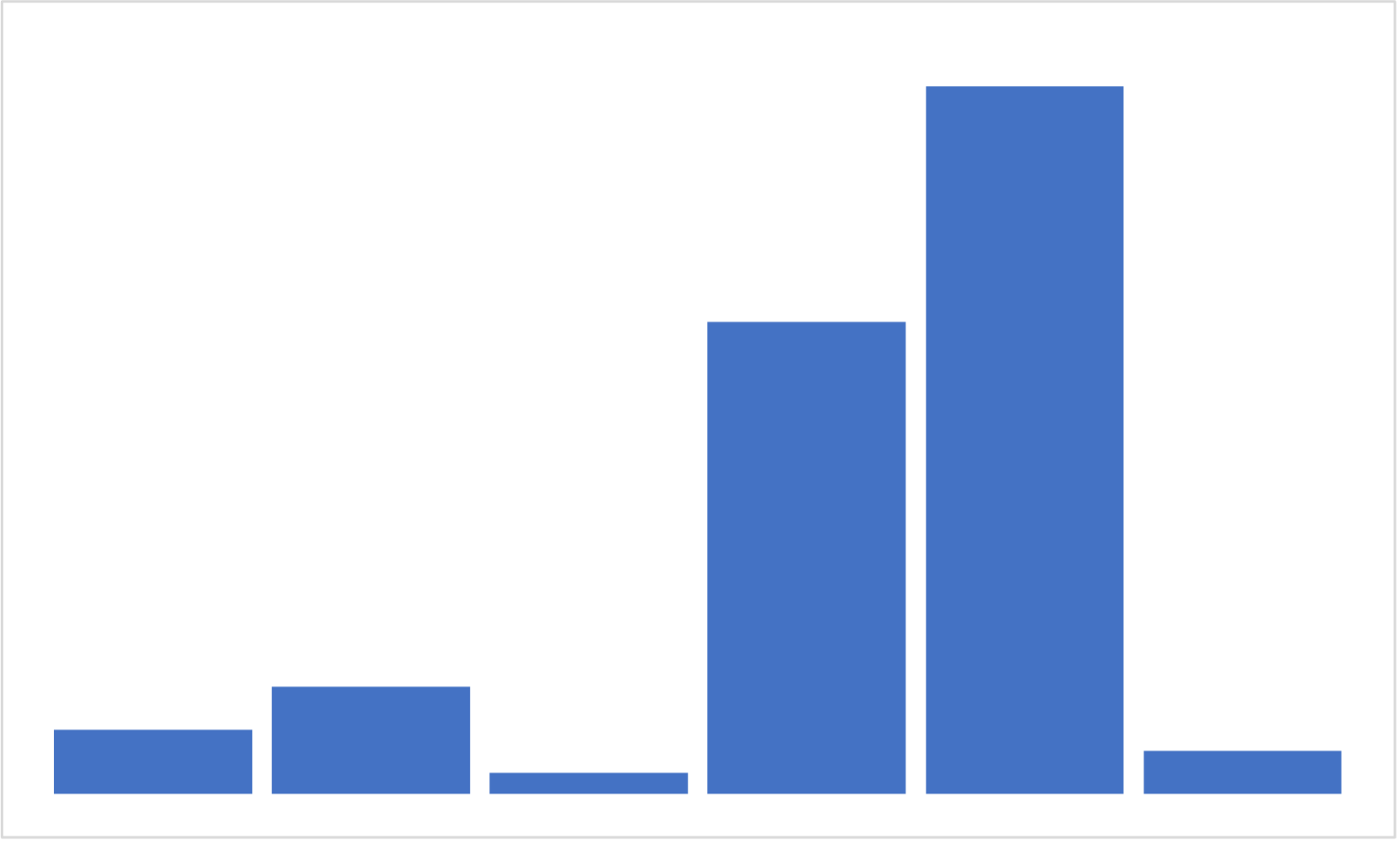} &
  4.14 \\
\multicolumn{1}{l|}{S7. Currently, a new SBOM is not always re-generated when there's any change to software artifacts.} &
  \multicolumn{1}{l|}{\multirow{-2}{*}{\begin{tabular}[c]{@{}l@{}}Section \ref{sbomgepoint}\\ SBOM generation points\end{tabular}}} &
  \includegraphics[width = 0.85cm, height = 0.35 cm]{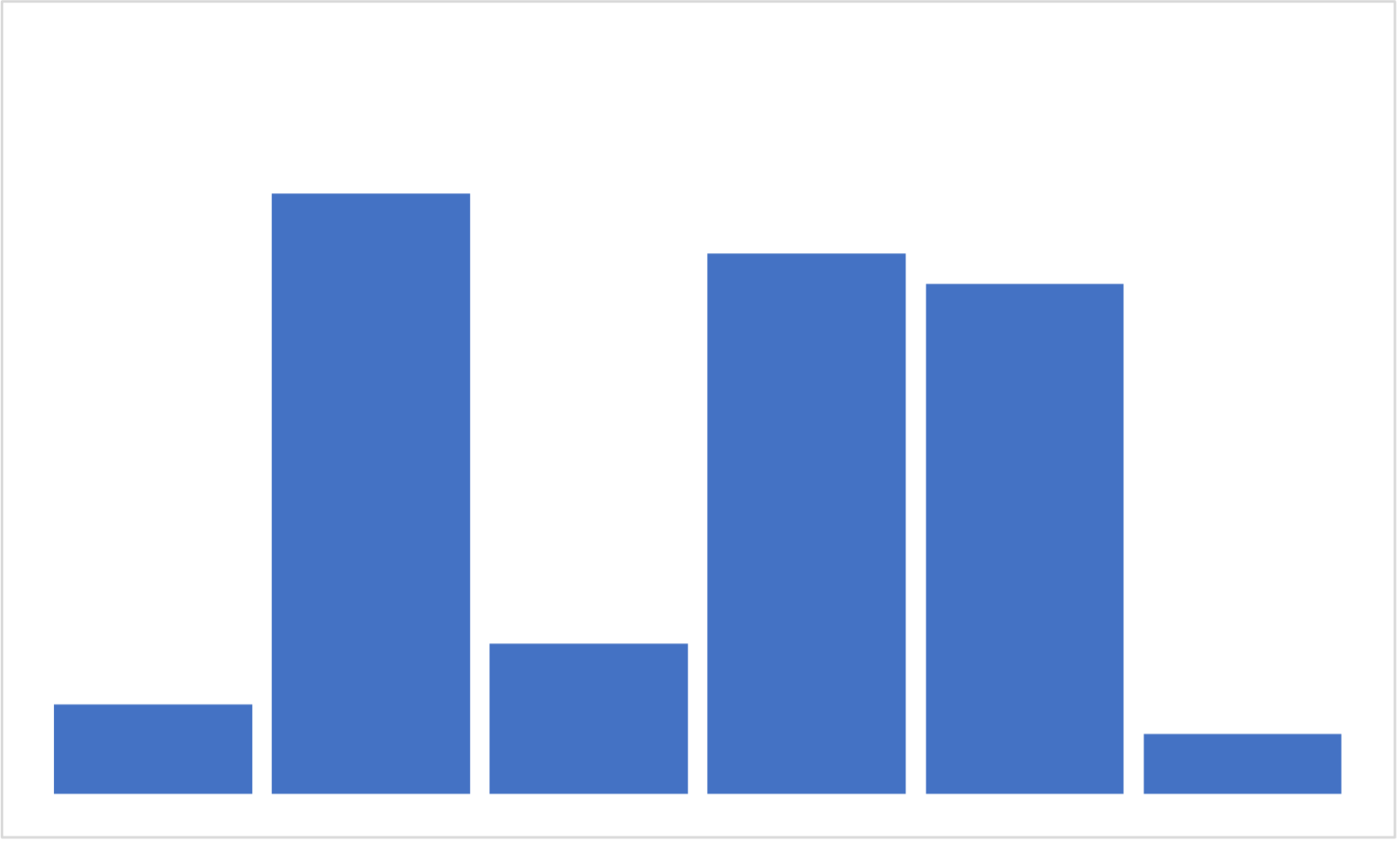} &
  3.31 \\ \hline
\multicolumn{1}{l|}{\cellcolor[HTML]{C0C0C0}S8. SBOMs are currently generated in a non-standardized format (e.g., not SPDX nor CycloneDX nor SWID).} &
  \multicolumn{1}{l|}{} &
  \includegraphics[width = 0.85cm, height = 0.35 cm]{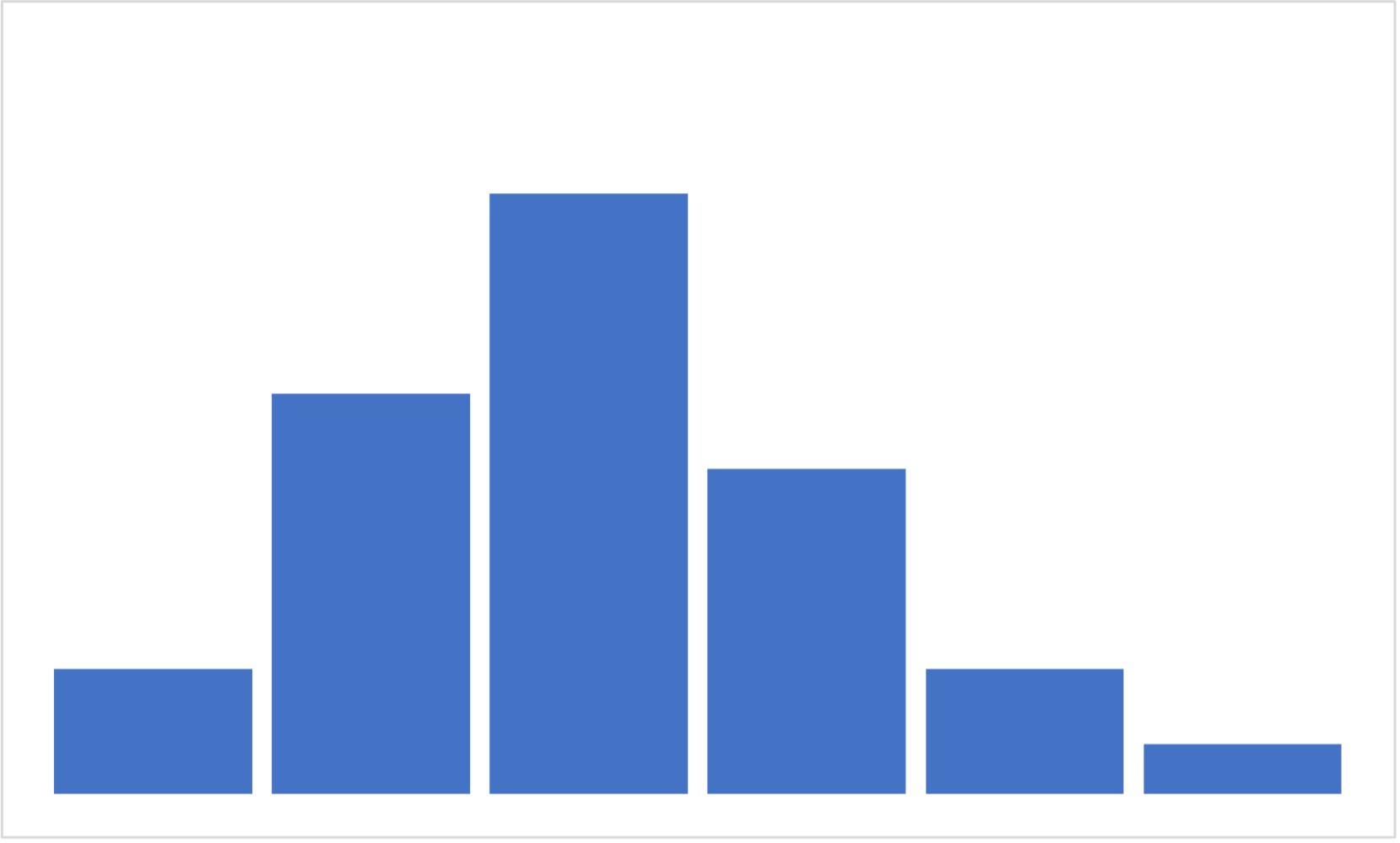} &
  \cellcolor[HTML]{C0C0C0}{\color[HTML]{333333} 2.86} \\
\multicolumn{1}{l|}{S9. Despite the 7 minimum data fields recommended by NTIA, the minimum fields are not necessarily all included in SBOMs.} &
  \multicolumn{1}{l|}{} &
  \includegraphics[width = 0.85cm, height = 0.35 cm]{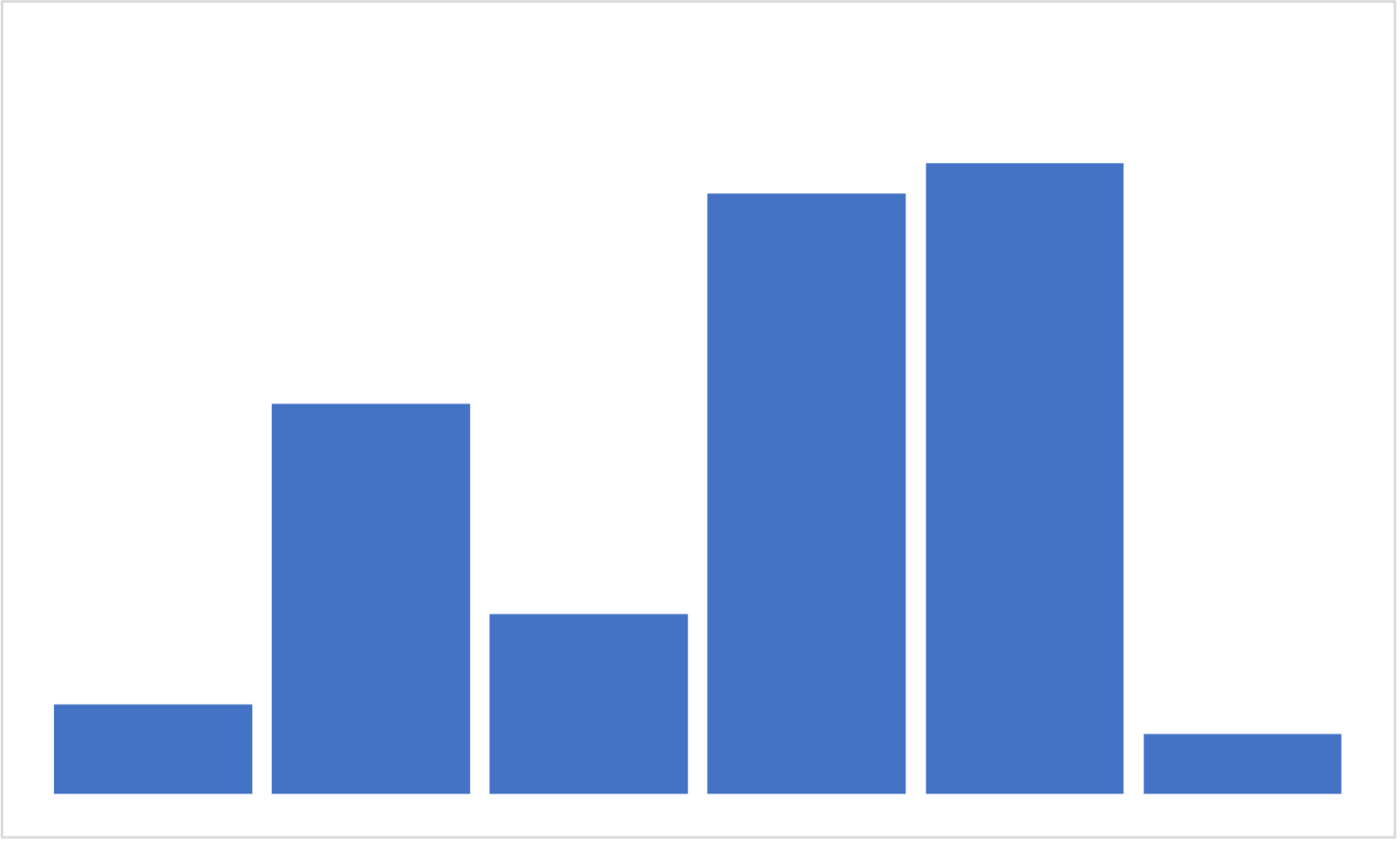} &
  3.57 \\
\multicolumn{1}{l|}{S10. In practice, SBOMs are extended with more useful data fields (other than the 7 minimum data fields) whenever possible.} &
  \multicolumn{1}{l|}{\multirow{-3}{*}{\begin{tabular}[c]{@{}l@{}}Section \ref{sbom_data}\\ SBOM data fields and\\ standarization\end{tabular}}} &
  \includegraphics[width = 0.85cm, height = 0.35 cm]{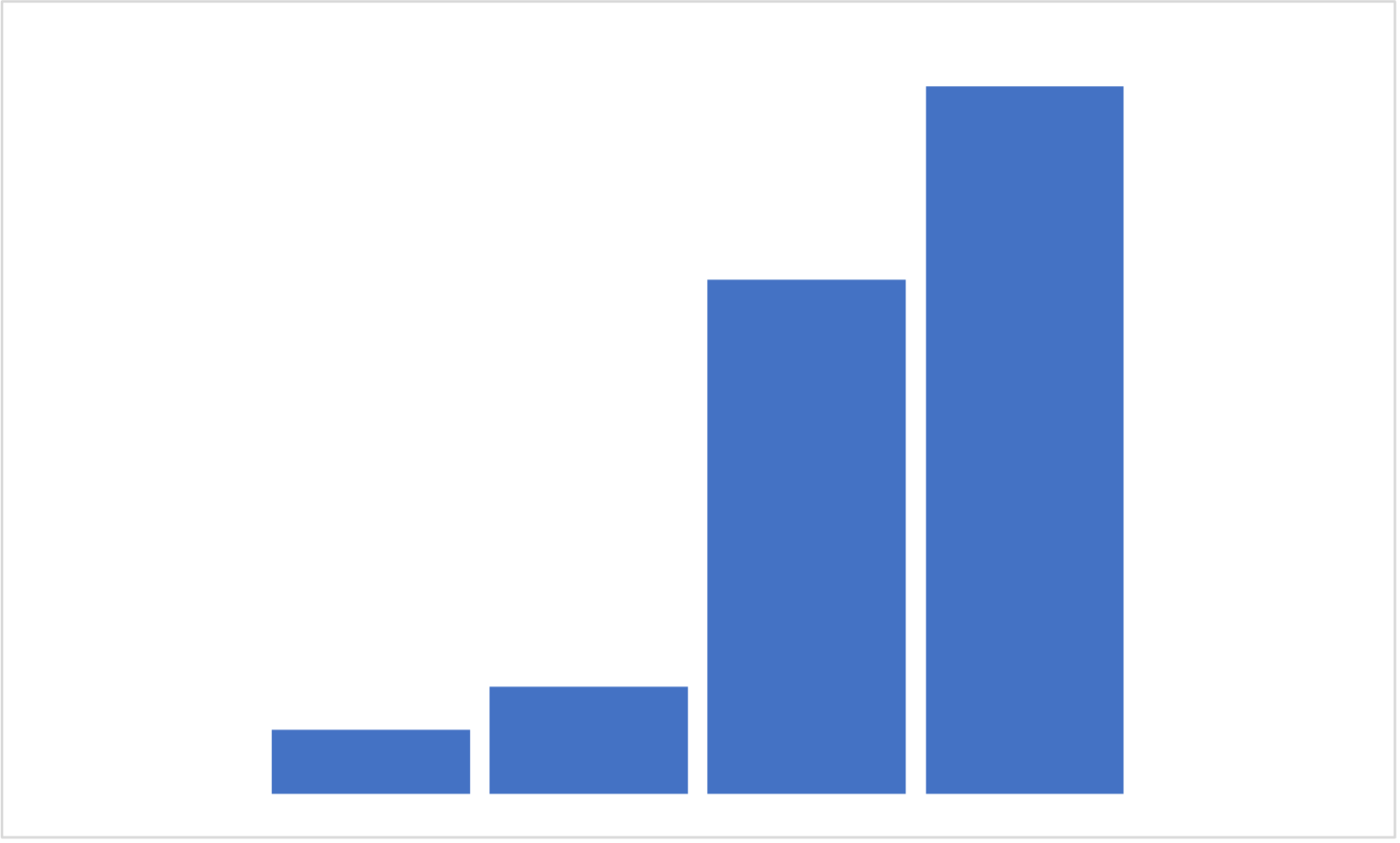} &
  4.34 \\ \hline
\multicolumn{1}{l|}{\cellcolor[HTML]{C0C0C0}S11. SBOMs are currently only generated for internal consumption.} &
  \multicolumn{1}{l|}{} &
  \includegraphics[width = 0.85cm, height = 0.35 cm]{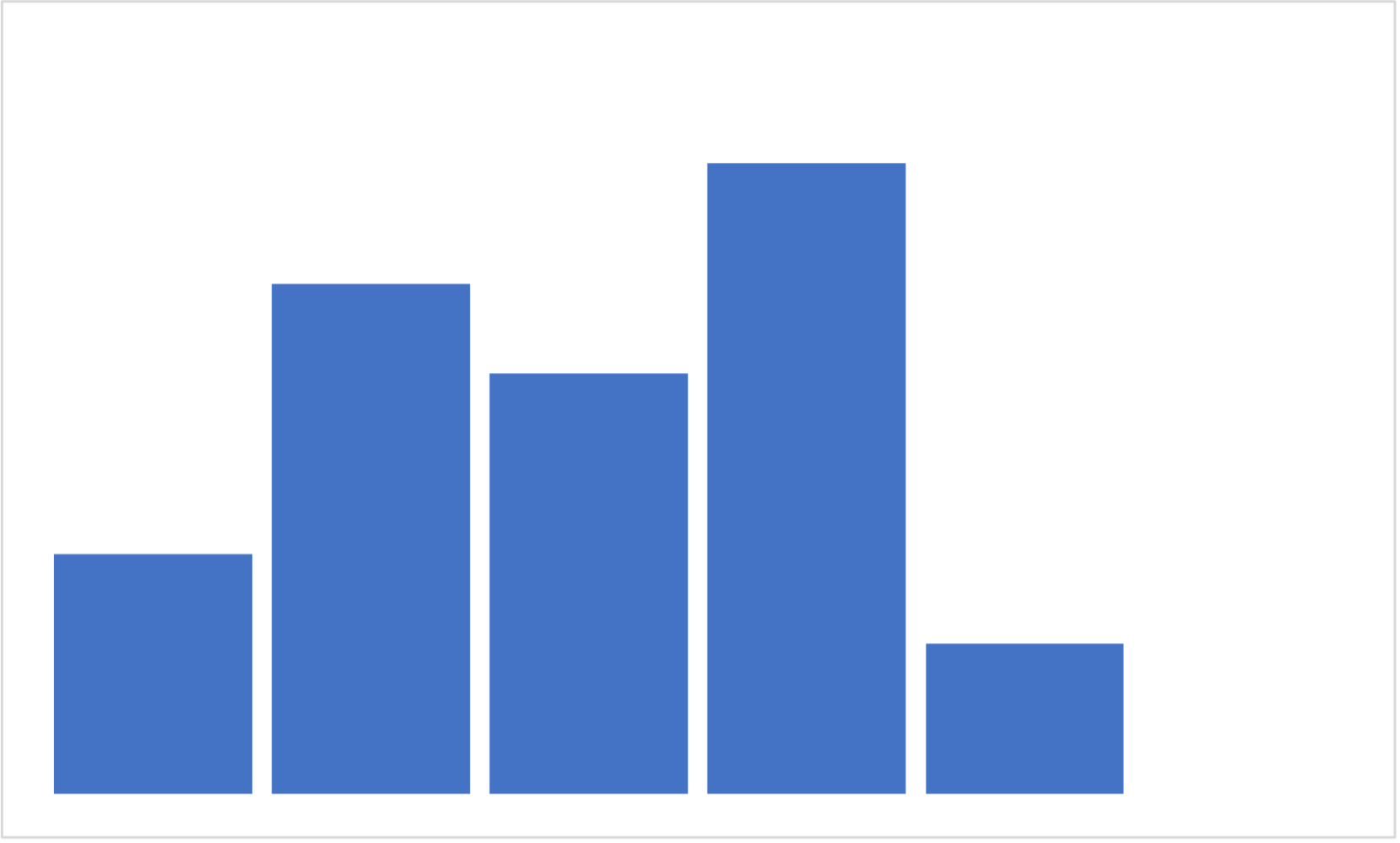} &
  \cellcolor[HTML]{C0C0C0}2.97 \\
\multicolumn{1}{l|}{S12. Access control should be required for the distribution of SBOMs for proprietary software/components.} &
  \multicolumn{1}{l|}{} &
  \includegraphics[width = 0.85cm, height = 0.35 cm]{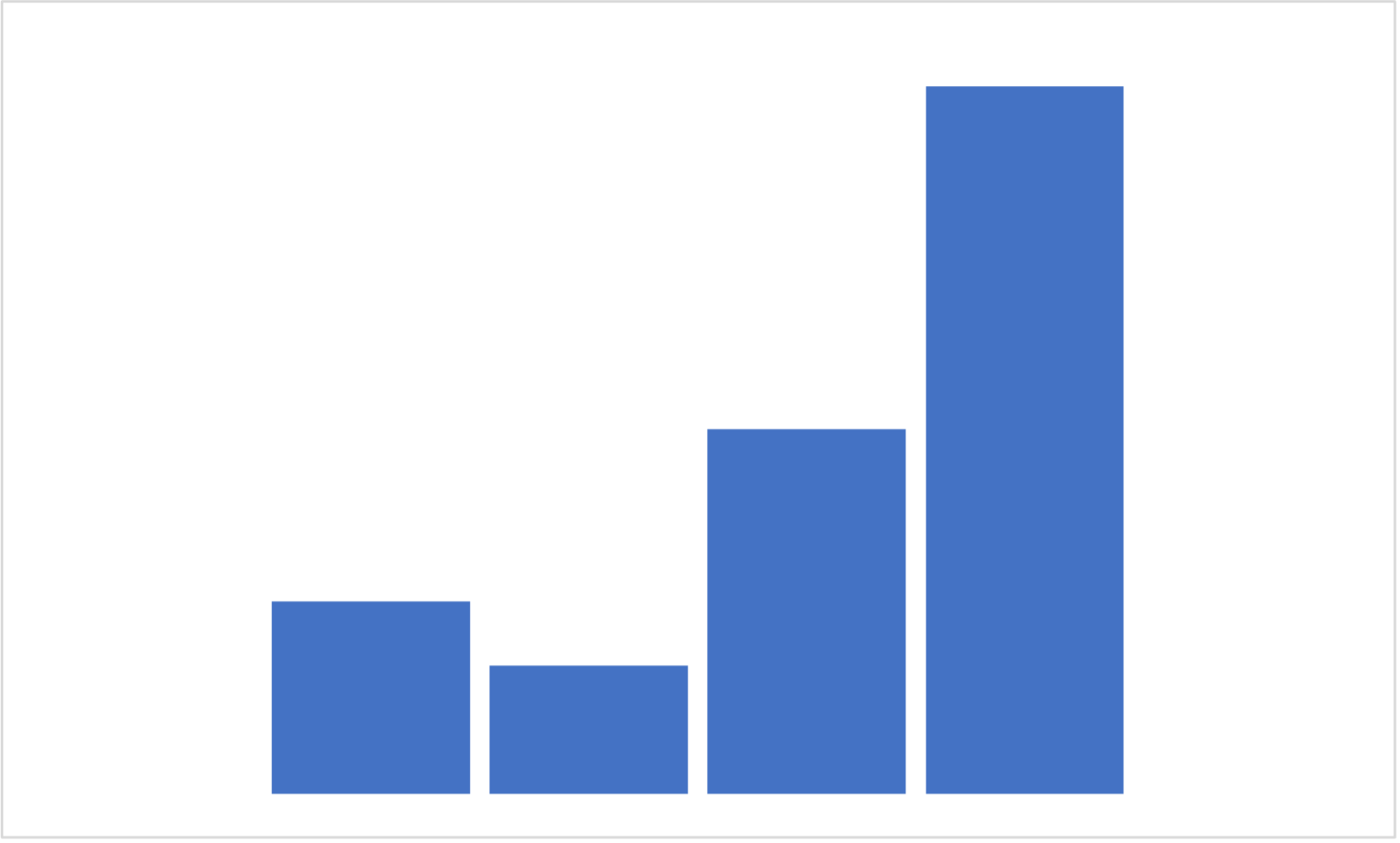} &
  4.14 \\
\multicolumn{1}{l|}{S13. Content tailoring (sharing partial SBOMs) should be required for SBOM distribution of proprietary software/components.} &
  \multicolumn{1}{l|}{\multirow{-3}{*}{\begin{tabular}[c]{@{}l@{}}Section \ref{distribution}\\ SBOM distribution\end{tabular}}} &
  \includegraphics[width = 0.85cm, height = 0.35 cm]{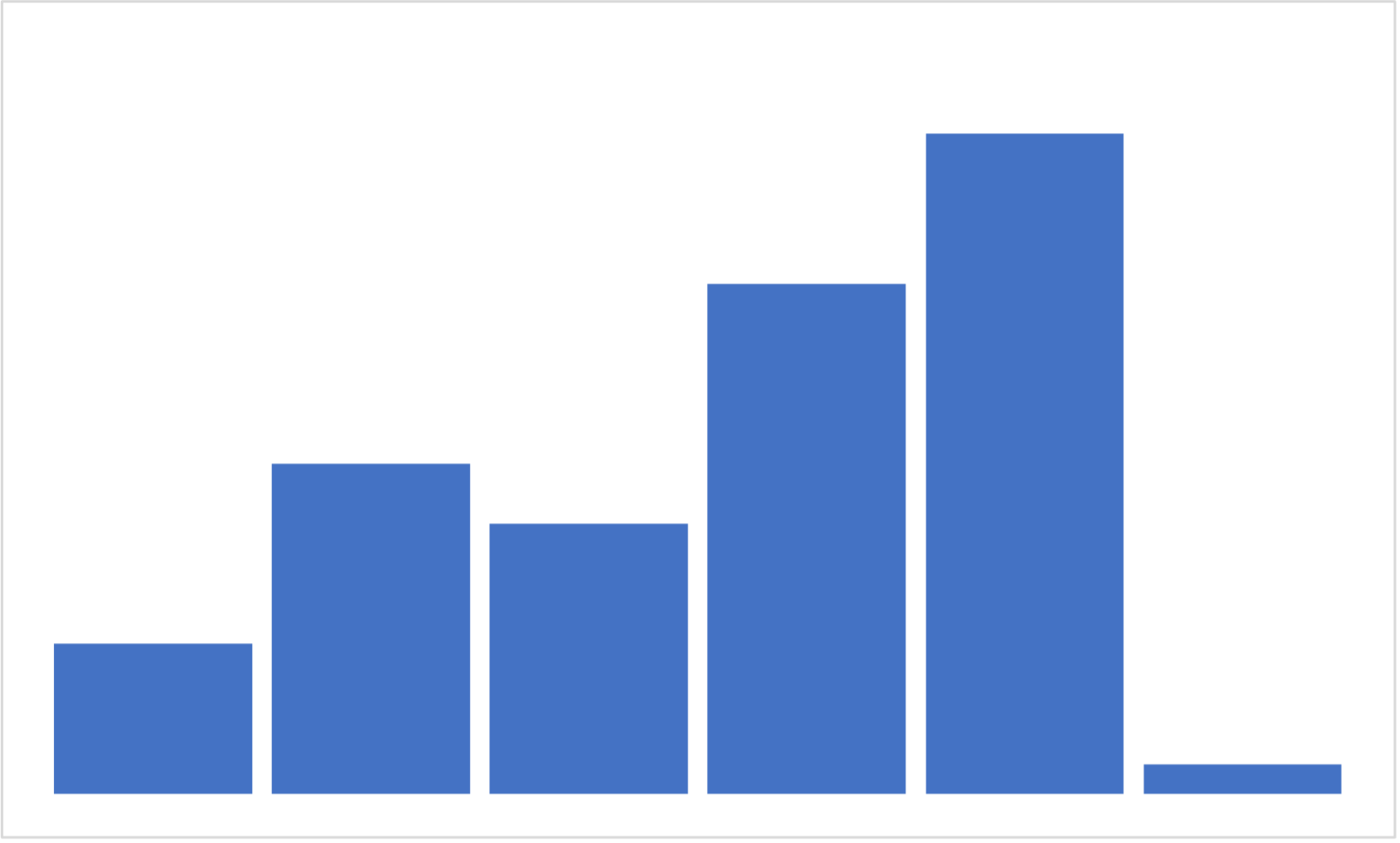} &
  3.57 \\ \hline
\multicolumn{1}{l|}{S14. SBOM producer's (i.e., software vendor) reputation is important for assessing SBOM integrity (e.g., completeness).} &
  \multicolumn{1}{l|}{} &
  \includegraphics[width = 0.85cm, height = 0.35 cm]{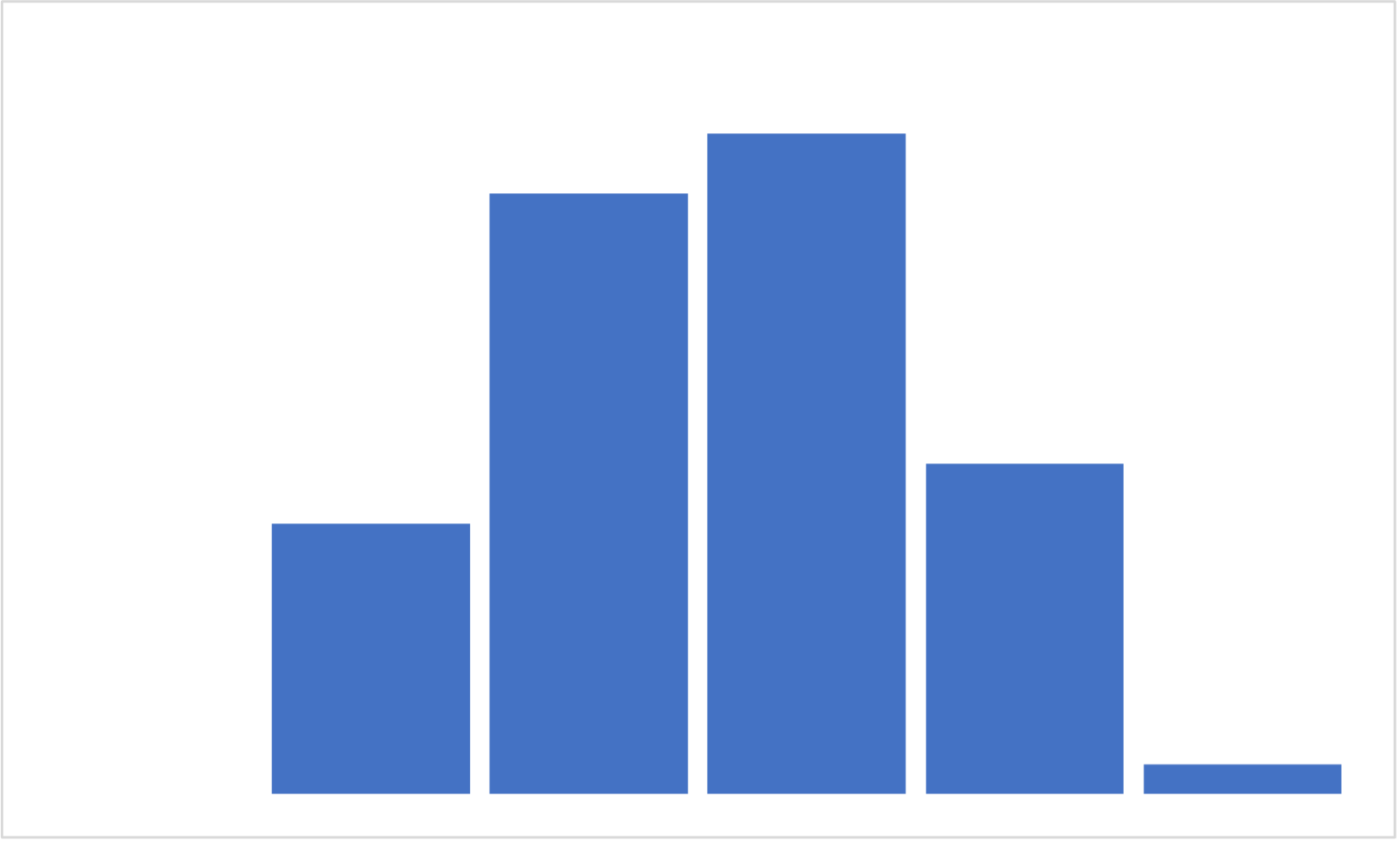} &
  3.4 \\
\multicolumn{1}{l|}{S15. Currently there are no validation mechanisms to ensure SBOM integrity (accuracy, completeness etc).} &
  \multicolumn{1}{l|}{\multirow{-2}{*}{\begin{tabular}[c]{@{}l@{}}Section \ref{sbom_validation}\\ SBOM validation\end{tabular}}} &
  \includegraphics[width = 0.85cm, height = 0.35 cm]{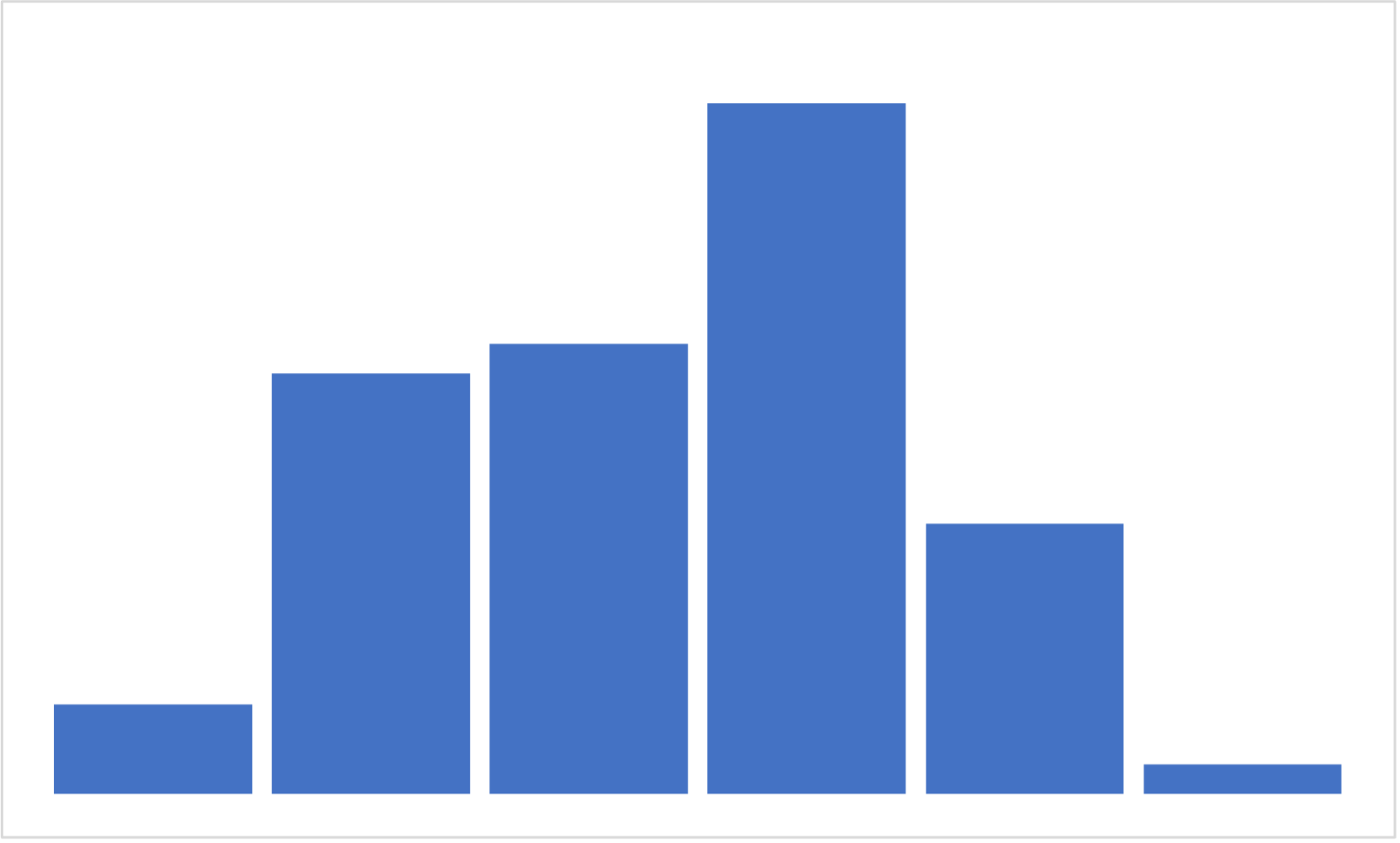} &
  3.28 \\ \hline
\multicolumn{1}{l|}{S16. Current vulnerability management with SBOMs doesn't focus on the actual exploitability of the vulnerability.} &
  \multicolumn{1}{l|}{\begin{tabular}[c]{@{}l@{}}Section \ref{vex}\\ Vul. \& exploitability\end{tabular}} &
  \includegraphics[width = 0.85cm, height = 0.35 cm]{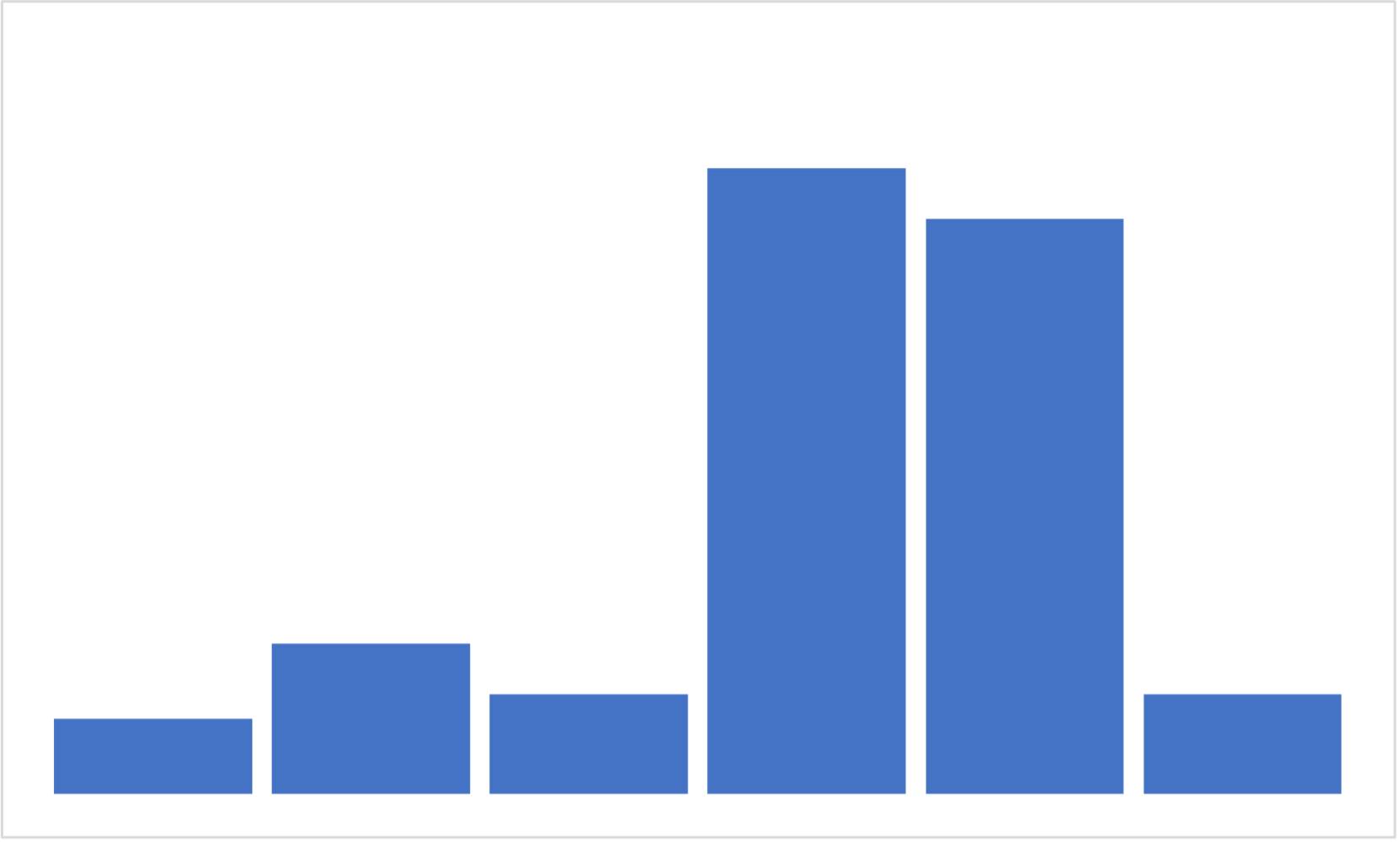} &
  3.72 \\ \hline
\multicolumn{1}{l|}{S17. SBOMs for AI software are different from SBOMs for traditional software.} &
  \multicolumn{1}{l|}{\begin{tabular}[c]{@{}l@{}}Section \ref{aibom}\\ AIBOM\end{tabular}} &
  \includegraphics[width = 0.85cm, height = 0.35 cm]{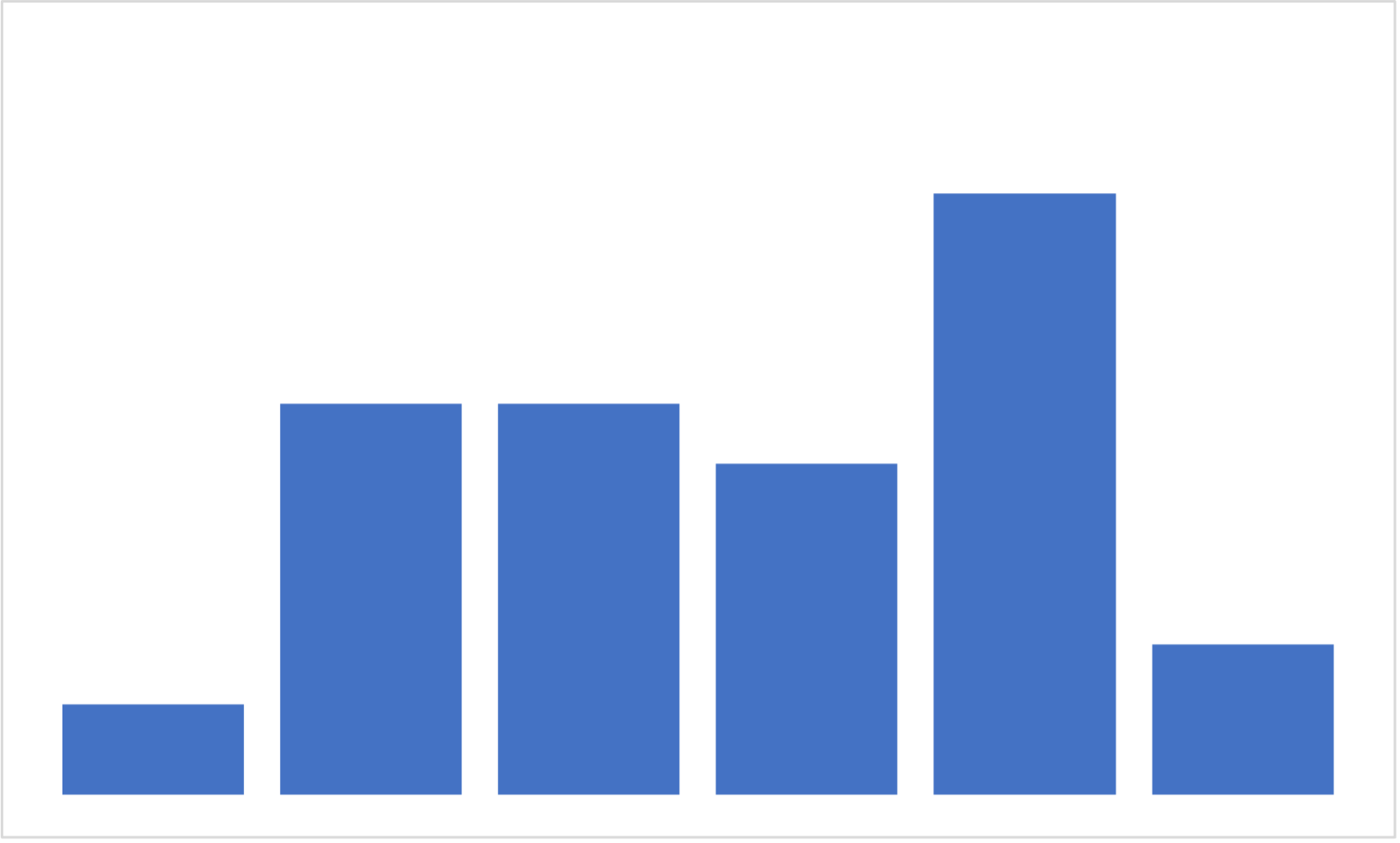} &
  3.26 \\ \hline
\textbf{T2. State of SBOM tooling support} &
   &
   &
   \\ \hline
\multicolumn{1}{l|}{\begin{tabular}[c]{@{}l@{}}S18. Although existing sources (e.g., package manager, POM.xml) are already there, it is still necessary to parse and feed\\ metadata from these sources into a standard format via SBOM tools.\end{tabular}} &
  \multicolumn{1}{l|}{\begin{tabular}[c]{@{}l@{}}Section \ref{necessity}\\ Necessity of SBOM tools\end{tabular}} &
  \includegraphics[width = 0.85cm, height = 0.35 cm]{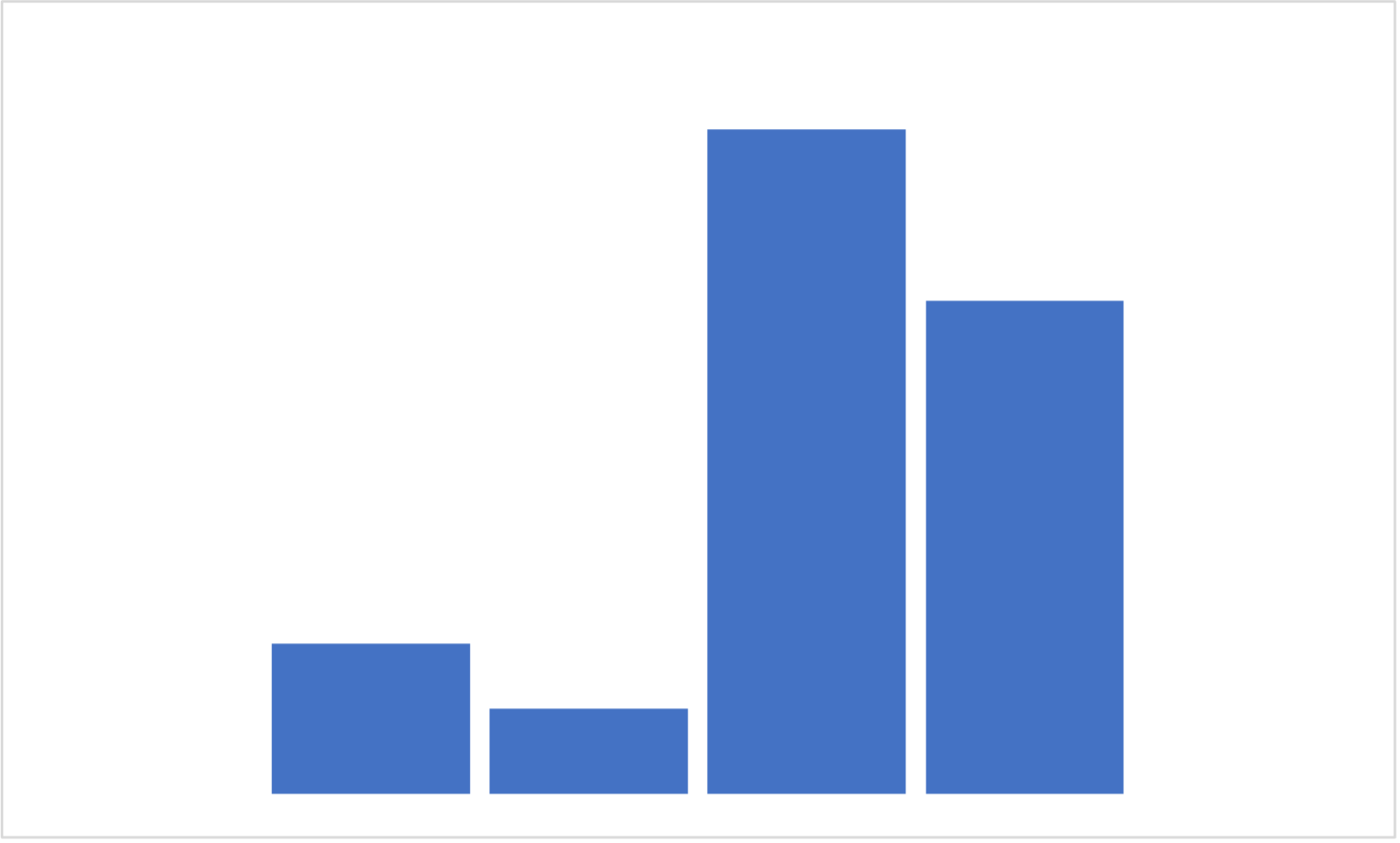} &
  4.08 \\ \cline{1-2}
\multicolumn{1}{l|}{S19. There are significantly limited tools for SBOM consumption.} &
  \multicolumn{1}{l|}{} &
  \includegraphics[width = 0.85cm, height = 0.35 cm]{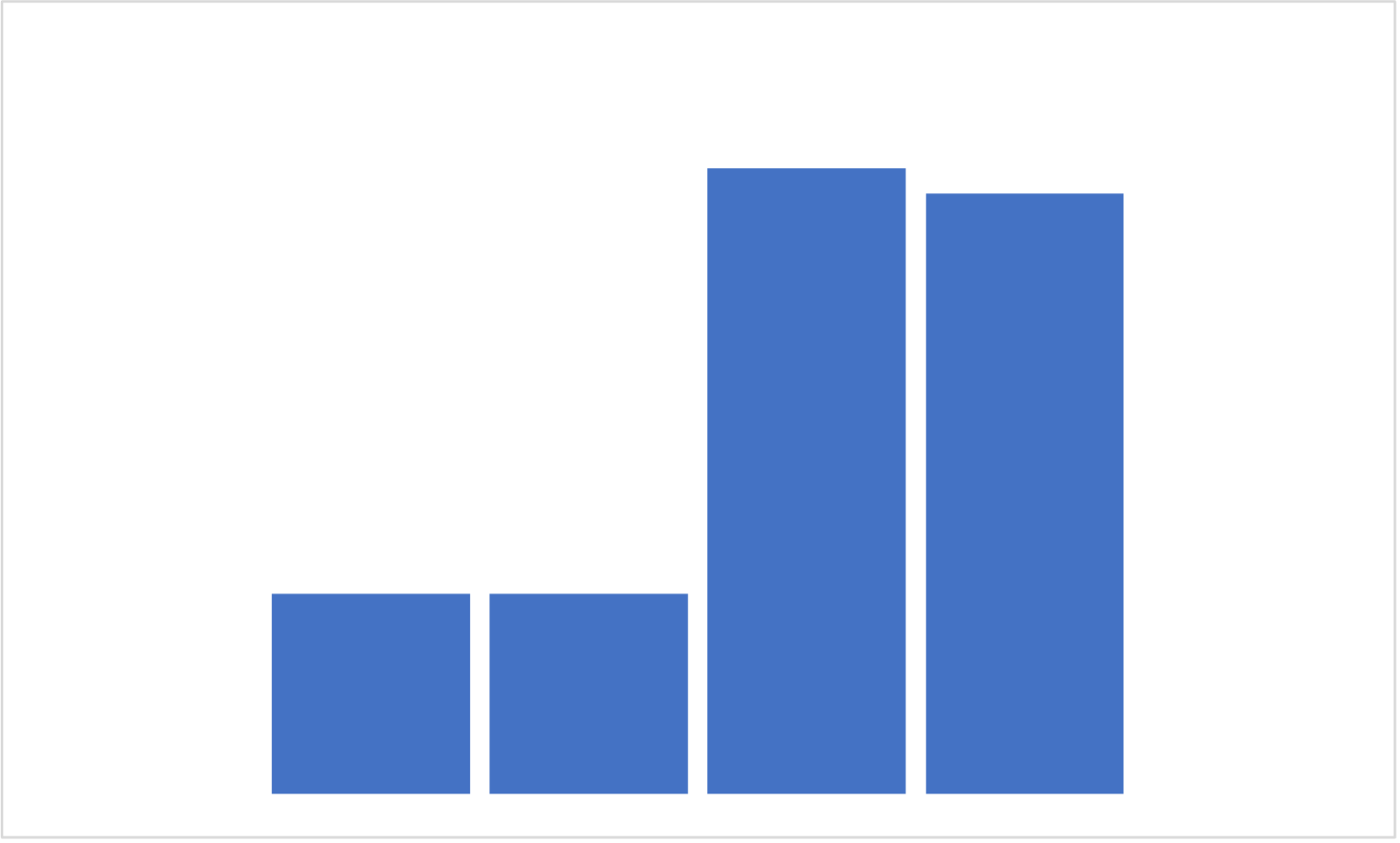} &
  4 \\
\multicolumn{1}{l|}{S20. SBOM consumption should be integrated with existing tools (e.g., vulnerability/configuration management tools).} &
  \multicolumn{1}{l|}{\multirow{-2}{*}{\begin{tabular}[c]{@{}l@{}}Section \ref{availability}\\ Availability of SBOM tools\end{tabular}}} &
  \includegraphics[width = 0.85cm, height = 0.35 cm]{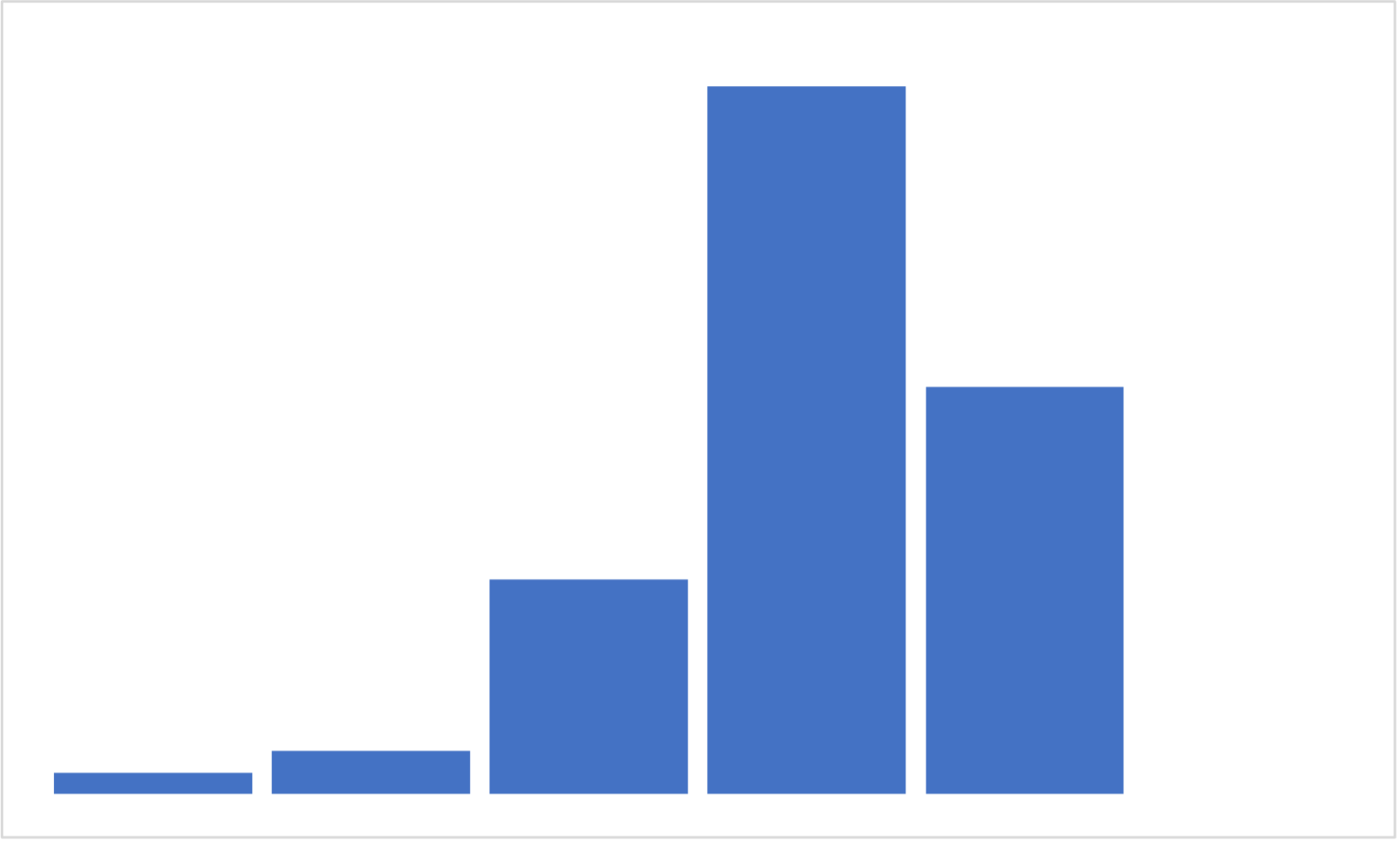} &
  4.03 \\ \cline{1-2}
\multicolumn{1}{l|}{S21. Existing SBOM tools can be hard to use (e.g., lack of usability, complexity).} &
  \multicolumn{1}{l|}{} &
  \includegraphics[width = 0.85cm, height = 0.35 cm]{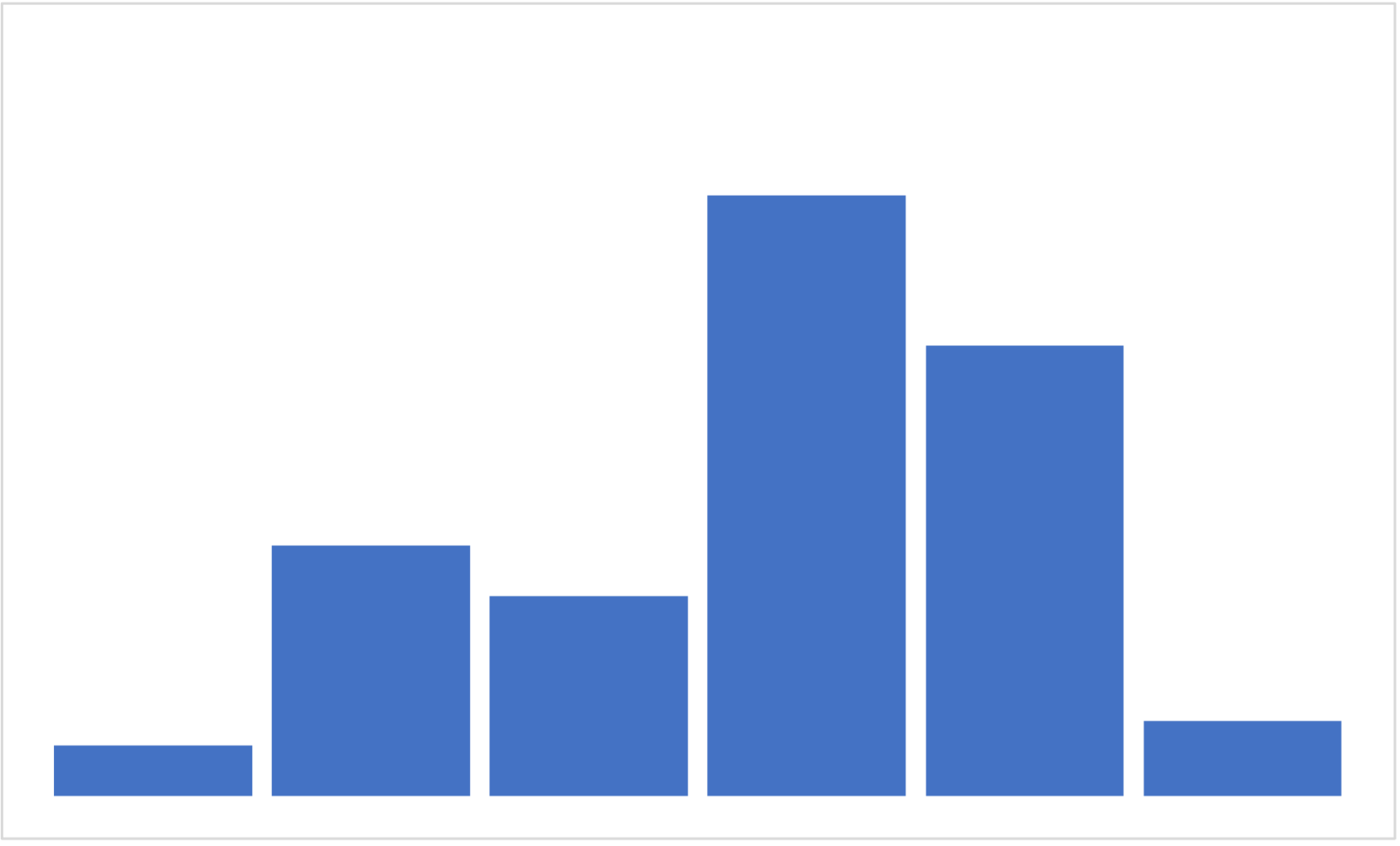} &
  3.57 \\
\multicolumn{1}{l|}{\begin{tabular}[c]{@{}l@{}}S22. SBOM tools lack of interoperability and standardization (e.g., the hash of one component generated by different tools \\ can be different).\end{tabular}} &
  \multicolumn{1}{l|}{\multirow{-2}{*}{\begin{tabular}[c]{@{}l@{}}Section \ref{usability}\\ Usability of SBOM tools\end{tabular}}} &
  \includegraphics[width = 0.85cm, height = 0.35 cm]{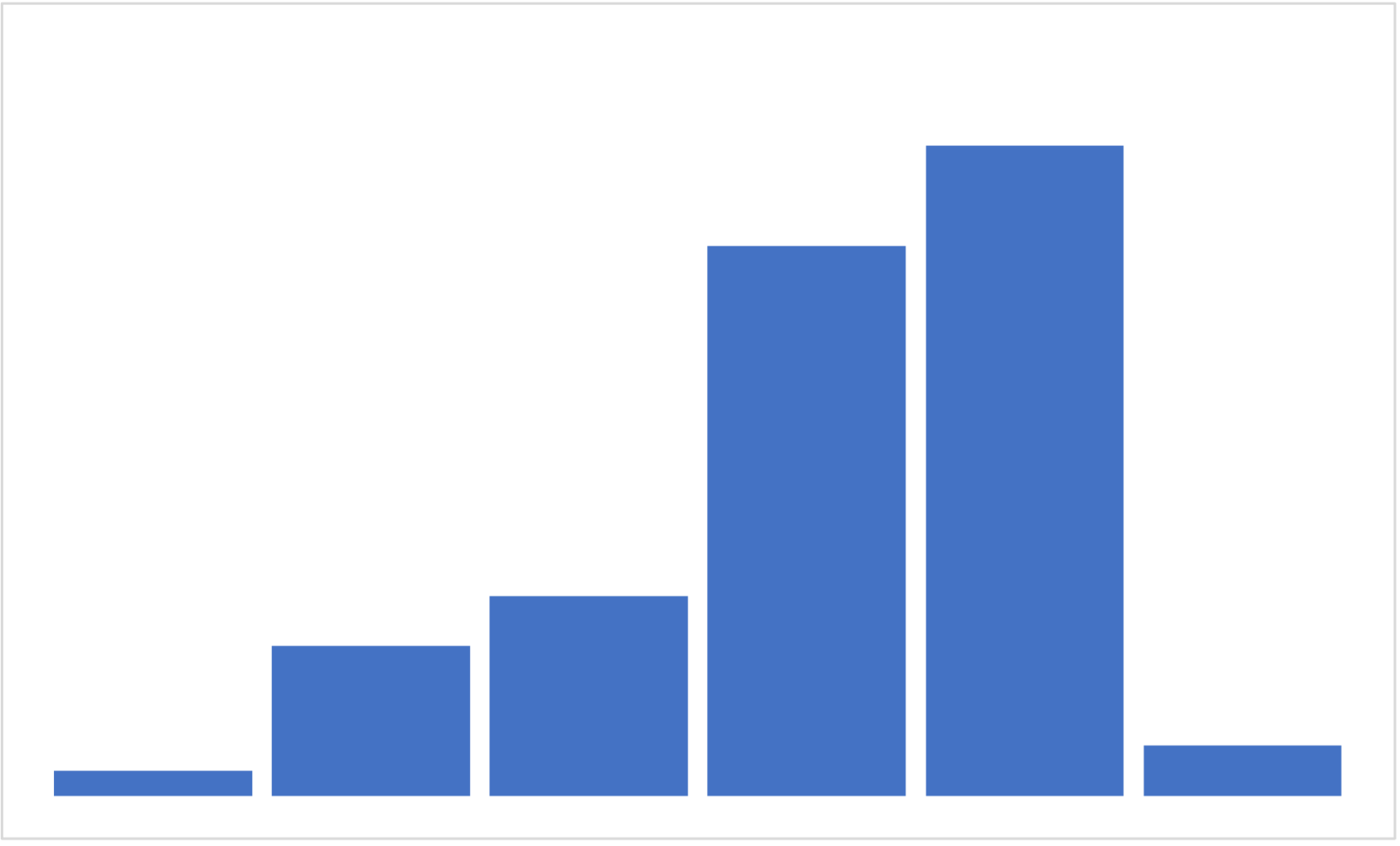} &
  3.92 \\ \cline{1-2}
\multicolumn{1}{l|}{S23. End users can't validate the integrity (e.g., accuracy and completeness) of the generated SBOMs by existing tools.} &
  \multicolumn{1}{l|}{\begin{tabular}[c]{@{}l@{}}Section \ref{tool_validation}\\ Validation of SBOM tools\end{tabular}} &
  \includegraphics[width = 0.85cm, height = 0.35 cm]{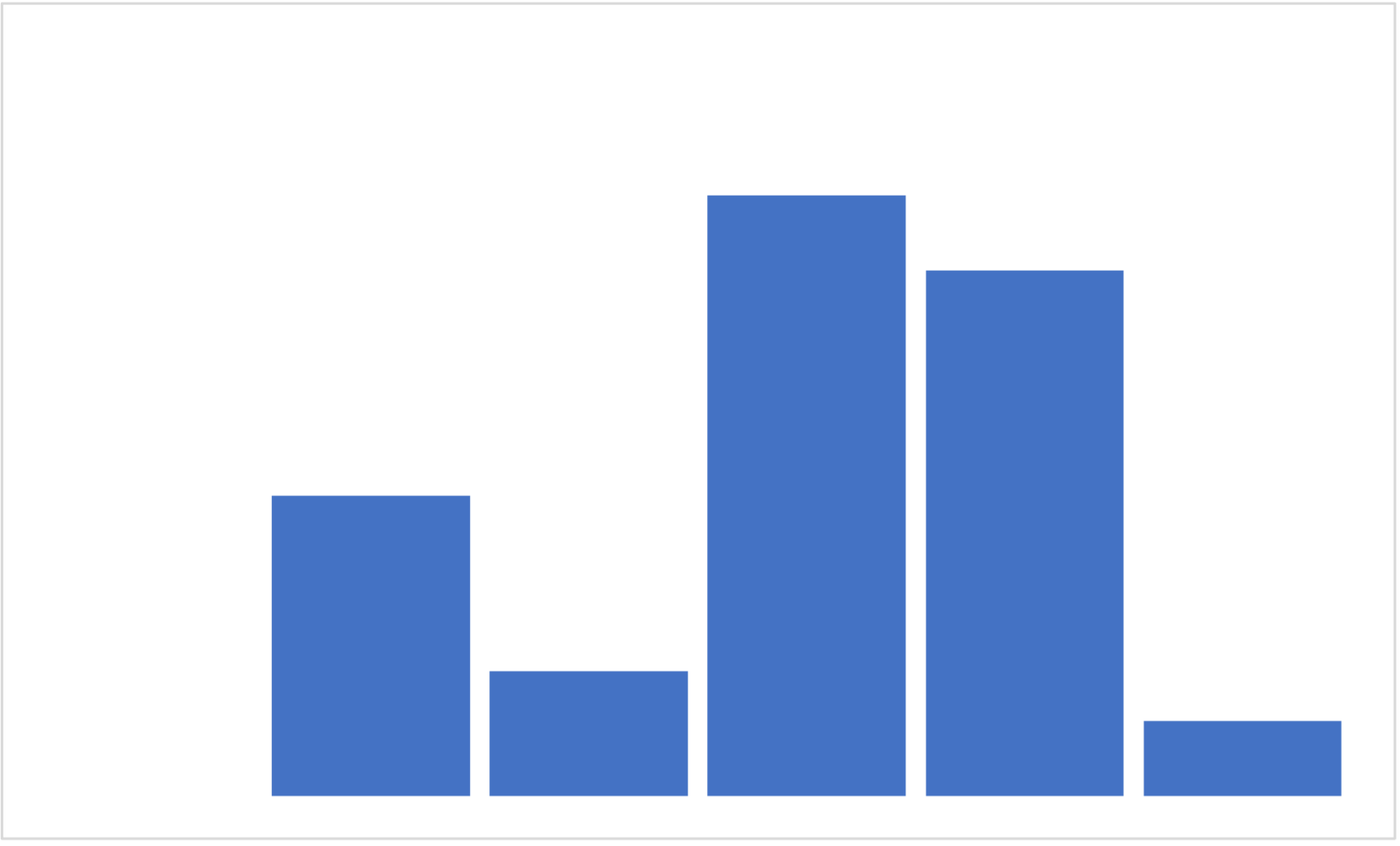} &
  3.69 \\ \hline
\multicolumn{4}{l}{\textbf{T3. SBOM issues \& concerns}} \\ \hline
\multicolumn{1}{l|}{\cellcolor[HTML]{C0C0C0}S24. Existing SBOM standards don't meet current market demands (e.g., the standards support only limited fields).} &
  \multicolumn{1}{l|}{} &
  \includegraphics[width = 0.85cm, height = 0.35 cm]{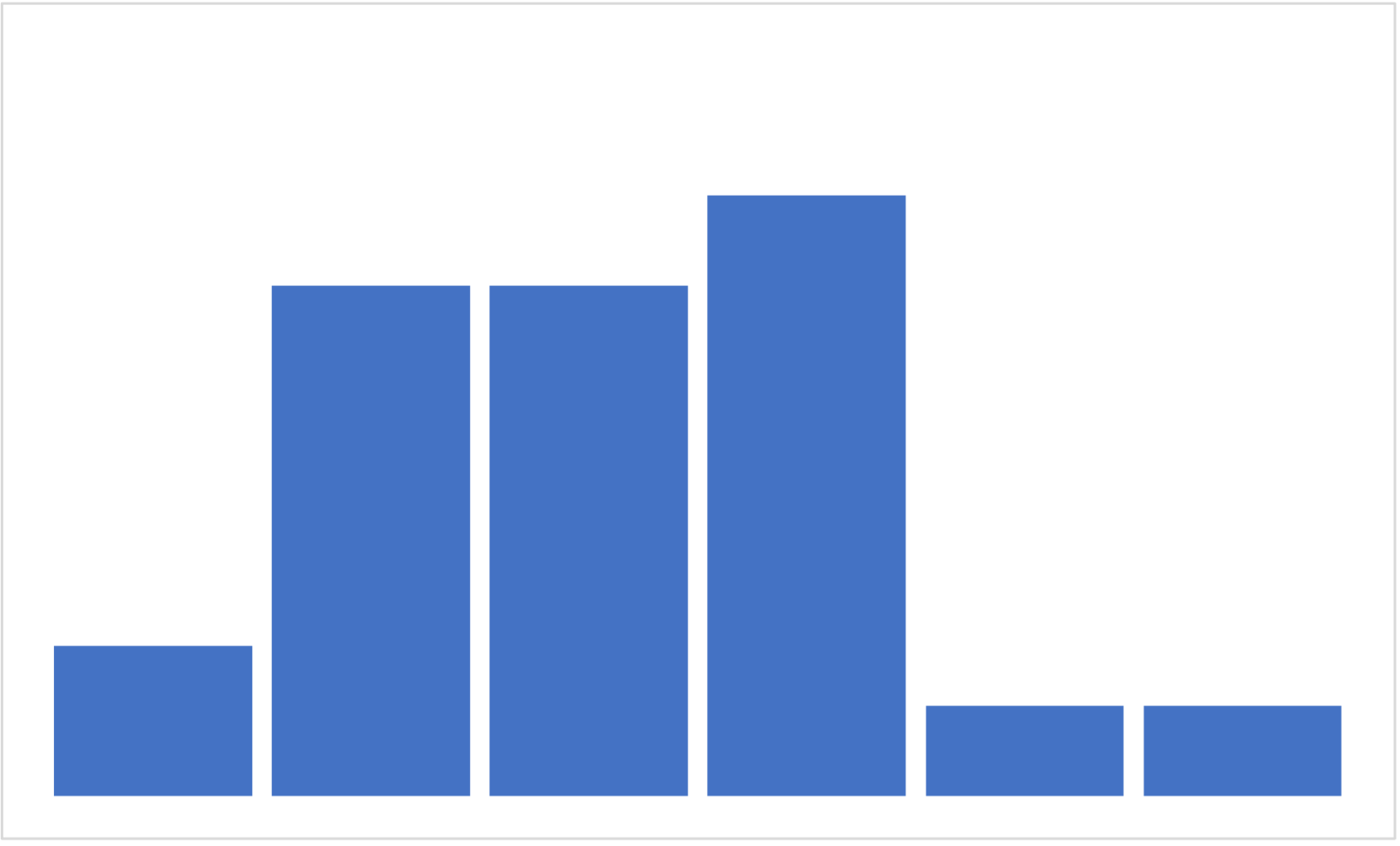} &
  \cellcolor[HTML]{C0C0C0}2.85 \\
\multicolumn{1}{l|}{S25. Attackers can take advantage of the information contained in SBOMs.} &
  \multicolumn{1}{l|}{} &
  \includegraphics[width = 0.85cm, height = 0.35 cm]{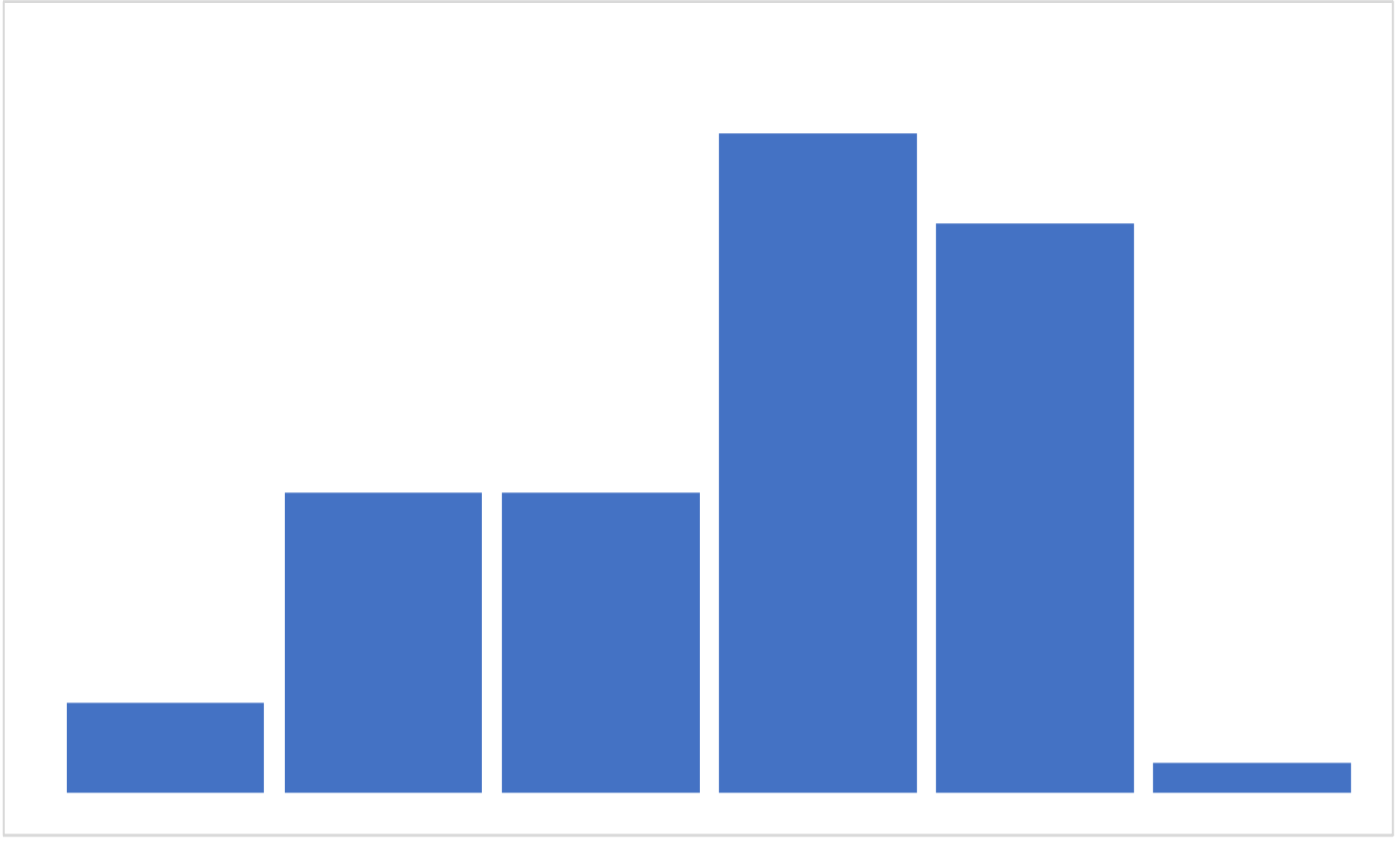} &
  3.63 \\
\multicolumn{1}{l|}{S26. There is hesitation in adopting SBOMs due to various concerns (e.g., lack of basic IT asset management).} &
  \multicolumn{1}{l|}{\multirow{-3}{*}{\begin{tabular}[c]{@{}l@{}}Section \ref{concern}\\ SBOM concerns\end{tabular}}} &
  \includegraphics[width = 0.85cm, height = 0.35 cm]{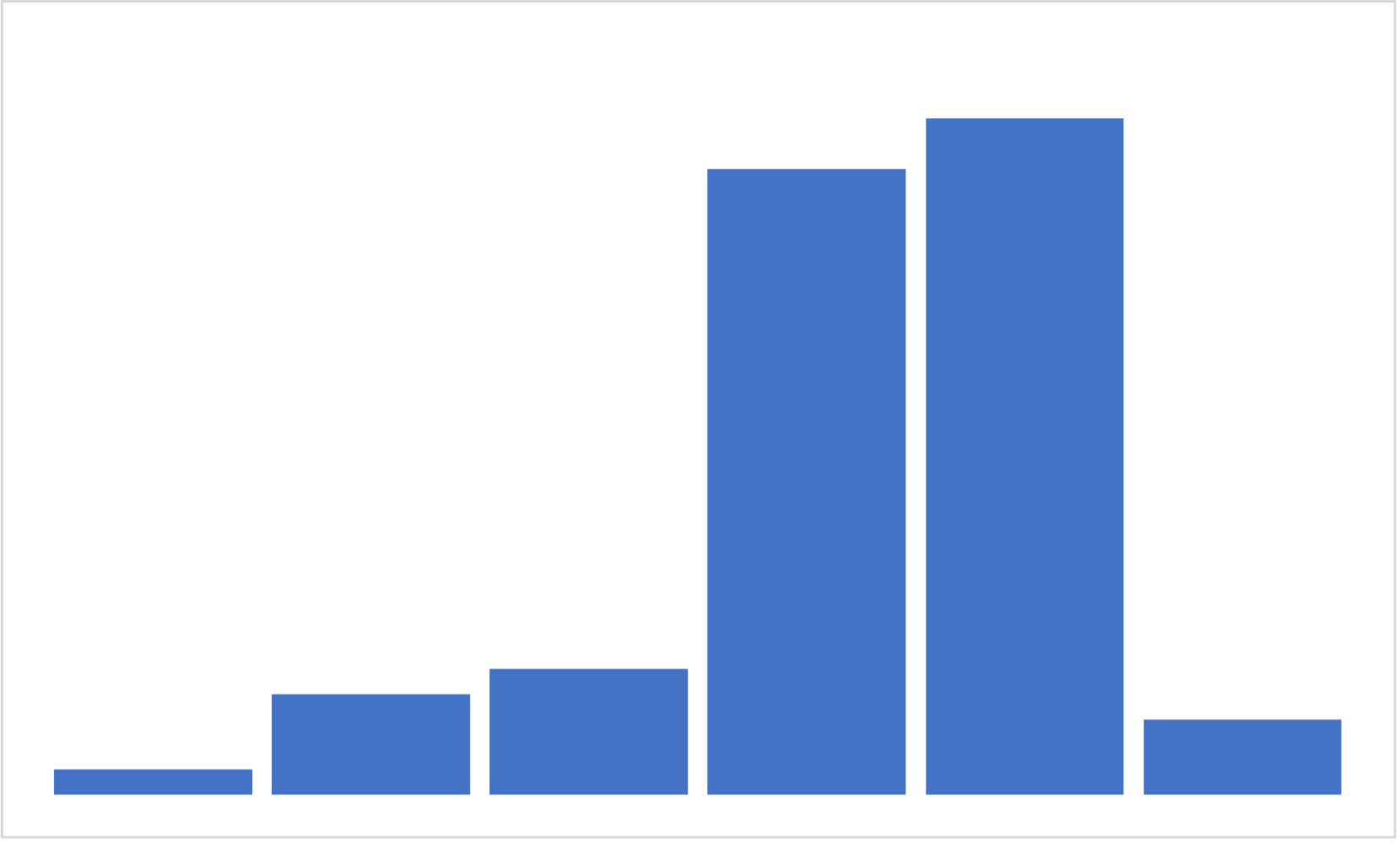} &
  3.98 \\ \hline
\end{tabular}%
}
\end{table*}

\section{Study results}
\label{4results}
This section reports the study results. In Table \ref{statements}, we drew the Linkert distribution graphs for 26 statements that were re-organized into four topics and calculated the overall scores based on: Strongly Disagree (1), Disagree (2), Neutral (3), Agree (4), Strongly Agree (5), and Not Sure (0).
We calculated the percentages of ``agrees" (strongly agree and agree) and ``disagrees" (strongly disagree and disagree) of each statement.

Following the SBOM background in Section \ref{background}, this section starts by introducing the current SBOM practices
% including the benefits of the SBOM, how adopted the SBOM is, and how SBOMs are generated, distributed, and validated 
(Section \ref{sbompractice}).
Then Section \ref{tooling} investigates the current tooling status of SBOMs.
Finally, Section \ref{concern} presents practitioners' primary concerns for SBOMs.

% \note{For each finding in this section, we first give a summary based on the empirical data, then added our insights/interpretations towards a goal/solution (in bold). The goals are then summarized into a goal model in Section \ref{goalmodel}}

% \note{We classify the findings into 3 categories (i.e.,  Aim, benefit, Concern) and list the involved stakeholders. We also attach a score to the finding based on the survey statistics. If a statement is based on several findings, we calculate the average and assign the average as the final score.} Maybe just add the percentage of agree and disagree to the statements in the text.

% \zc{Can the results discussion be somehow organized in a more structured way like concern severity, short-medium-long term objectives, agreed-on versus controversial? Seems that the statements are survey questions. Maybe can do this based on survey results?}

% \zc{I personally do not like reading long paragraphs of this type of empirical results. Can we organize the content with a concise heading of relevant statement highlighted with key aspects like concern severity, short-medium-long term objectives, agreed-on versus controversial. It would be great if the readers can obtain all key take-home messages by just looking at these headings without much needs to read the lengthy discussion.}

\subsection{RQ1: What is the current state of SBOM practice?}
\label{sbompractice}
To answer RQ1, we discussed 16 statements in this section based on T1 (see Table \ref{statements}).
Our results suggest that SBOMs are not widely adopted.
SBOM generation and distribution require further standardization and maturer mechanisms.
SBOM data validation is generally neglected.
% (agreed by 49.2\% of survey respondents)
% which causes potential risks as SBOMs can be easy to tamper with.
For the typical SBOM use case of vulnerability management, the exploitability status classification should be more than binary.

\subsubsection{SBOM benefits}
\label{benefit}

% {\color{blue} TB: some writing issues, should we use past tense? }

We summarized 3 statements (i.e., S1-S3) on SBOM benefits based on the interviews.

15 of 17 interviewees mentioned that the enhanced transparency of the software supply chain is one of the exceptional advantages of SBOMs [\textbf{S1}, 90.8\% agree, 1.5\% disagree]. Transparency brings a lot of favorable consequences, such as end-of-life software management, vulnerability tracking, and license compliance checking \cite{lf_2022}.
\textit{``The biggest benefit is knowing exactly what is being bundled in your software, right? So it is to assure our customers that, if there's a vulnerability reported, you immediately know [whether] you are impacted or not, \textbf{if the SBOM is accurate}." (I13-Dev.\&Sec.\&Res.)}

Another benefit originates from the SBOM data.
% With SBOM getting more adopted and standardized, and SBOM tools becoming more developed, the value of SBOM data will manifest.
The unification of software composition details provided by SBOMs is beneficial as the standardized SBOM data has the potential to be further built upon.
\textit{``The SBOM itself isn't the valuable part. The valuable part is, how do we turn that data into intelligence, into action." (I17-Cnslt.\&Adv.)}
Based on the SBOM data, it is promising that there will be SBOM-centric ecosystems emerging [\textbf{S2},  86.2\% agree, 1.5\% disagree]. However, due to the lack of SBOM adoption (see Section \ref{adoption}), such ecosystems are long-term goals not to be achieved soon.

Thirdly, although adopting SBOMs requires extra efforts (e.g., additional tools and processes, education to related personnel), the benefits of SBOMs outweigh the costs [\textbf{S3}, 86.2\% agree, 7.7\% disagree].
\faThumbsODown \,Although some organizations are \textit{``worried whether SBOMs would increase the cost of a software product" (I1-Dev.)},
\faThumbsOUp \, the majority favor the benefits brought by SBOMs as the potential loss without SBOMs can be devastating.
\textit{``What I think about is, what is the cost when a vulnerability is exploited? [...] if you look at SolarWinds event, that cost was \$800 million." (I6-Dev.)}

\begin{center}
\begin{tcolorbox}[breakable, colback=gray!10,
colframe=black, 
width=\columnwidth,
arc = 1mm,
boxrule=0.5 pt,
]
\textbf{Finding 1}: The transparency brought by SBOMs can enable accountability, traceability and security, but there is a lack of \textbf{systematic consumption-scenario-driven design of SBOM features}.
%\textbf{Supporter}: SBOMs are essential for software supply chain security.\\
%\textbf{Opposer}: Introducing SBOMs leads to extra efforts.
\end{tcolorbox}
\end{center}

% \begin{center}
% \begin{tcolorbox}[breakable, colback=gray!10,
% colframe=black, 
% width=\columnwidth,
% arc = 1mm,
% boxrule=0.5 pt,
% ]
% \textbf{Finding 1}: The transparency brought by SBOMs can enable accountability, traceability and security.
%     \begin{tcolorbox}[boxrule=0.5 pt]
%     \textbf{Authors interpretation}: But there is a lack of \textbf{systematic consumption-scenario-driven design of SBOM features}.
%     \end{tcolorbox}
% %\textbf{Supporter}: SBOMs are essential for software supply chain security.\\
% %\textbf{Opposer}: Introducing SBOMs leads to extra efforts.
% \end{tcolorbox}
% \end{center}

\subsubsection{SBOM adoption}
\label{adoption}

We summarized 2 statements (i.e., S4-S5) on SBOM adoption status (see Table \ref{statements}).

Despite the benefits of SBOMs and the SBOM readiness report's\cite{lf_2022} optimistic results that 90\% of the 412 sampled organizations have started or are planning their SBOM journey, \textbf{the adoption of SBOM is not as optimistic} according to our interviews and survey. For example, most existing third-party software or components, either open source or proprietary, are not equipped with SBOMs [\textbf{S4}, 83.1\% agree, 13.8\% disagree].
As I2 stated,
\textit{``when introducing third-party components to our organization, we need to try to generate SBOMs for them because not all of them have SBOMs.'' (I2-Dev.)}
However, considering the prevalence of OSS, the unavailability of SBOMs for (open-source) software/components also holds software vendors back from SBOM adoption as they may wonder whether SBOM adoption is an industrial consensus.
% As I4
% \textit{``people are banging the doors dying to get SBOMs for open source projects." (I4-Dev.\&Sec.)} 

In addition, SBOMs are not generated for all software even inside a software vendor organization [\textbf{S5}, 87.7\% agree, 7.7\% disagree].
% (i.e., software vendors only generate SBOMs for a small portion of their software products). 
% Especially, if procurers are not asking for SBOMs for certain software products, software vendors are less likely to generate SBOMs for them. 
As stated by I10, \textit{``software vendors may be producing SBOMs for some customer products. 
But I bet most of them don't generate SBOMs for the financial software they are using." (I10-Cnslt.\& Adv.)} 
\begin{center}
\begin{tcolorbox}[breakable, colback=gray!10,
colframe=black, 
width=\columnwidth,
arc = 1mm,
boxrule=0.5 pt
]
\textbf{Finding 2}: A large portion of widely used software, especially OSS, does not have SBOMs. \textbf{The incentives for generating SBOMs for OSS and proprietary software need to be propagated}.
\end{tcolorbox}
\end{center}

\subsubsection{SBOM generation point}
\label{sbomgepoint}
We summarized 2 statements (i.e., S6-S7 in Table \ref{statements}) on SBOM (re-)generation.

SBOMs can be generated at a set of different stages in the software development life cycle (e.g., build time, run time, before delivery, after third-party components introduction, etc.) [\textbf{S6}, 84.6\% agree, 12.3\% disagree], which differs from case to case as \textit{``the challenge here sort of builds on the sheer diversity of software" (I17-Cnslt.\&Adv.)} (e.g., modern/legacy software, container images, cloud-based software). For example, \textit{``for legacy software that's already out there in the wild, maybe you don't have reproducible builds. You only have that built artifacts [...] it's like you can't do it (generate SBOMs) at build time anymore [...] it's just too difficult [...] But you could probably do a run time [SBOM] generator." (I8-Dev.\&Cnslt)}

As stated by I8, \textit{``I don’t really think there is a perfect time [for SBOM generation]". (I8-Dev.\&Cnslt)}
Since software development goes through a life cycle, \textbf{ideally an SBOM should be generated at the early stages and then gradually enriched with more information from the latter stages}, which was supported by I12 and I14.
\textit{``I think the way to actually include the most full SBOM is to have visibility towards the whole cycles, from the report to the build to the factory." (I12-Cnslt.\&Adv.)}
\textit{``I do not think we should produce an SBOM as a one-shot process, but rather we should be carrying evidence and partial SBOMs and enriching them in every single operation." (I14-Dev.\&Sec.\&Res.)}

As for SBOM re-generation, it is evident that whenever any change happens to any software artifact, the corresponding SBOMs should be timely re-generated to reflect this change, which is hardly the current practice followed by every SBOM producer [\textbf{S7}, 53.8\% agree, 35.4\% disagree].
As stated by I10,
\textit{``that is more of an aspirational goal - it's not something that will be realized right away. Because right now it would be just great if they put out a new SBOM whenever they did a new major version, and that is better than nothing."(I10-Cnslt.\&Adv.)}
However, since some organizations are re-generating SBOMs upon each change, and if SBOM re-generation is to be a standard practice, solutions like SBOM version control are needed for managing the SBOMs.
As stated by I8, with different versions of SBOMs, \textit{``you need to version control your SBOMs, and figure out how to distribute that information to your customers." (I8-Dev.\&Cnslt)}

\begin{center}
\begin{tcolorbox}[breakable, colback=gray!10,
colframe=black, 
width=\columnwidth,
arc = 1mm,
boxrule=0.5 pt,
]
\textbf{Finding 3}: 
% SBOM generation can take place in different stages of the software development lifecycle, depending on software type (e.g., container image, legacy software) and SBOM producers' choice. 
% Although SBOMs can be generated at different stages of the software development lifecycle, ideally, an SBOM should be generated at an early stage and gradually enriched along the development life cycle.\\
% SBOM re-generation should happen whenever any software artifact changes, which requires version control to track this evolution.
SBOM generation is belated and not dynamic, while ideally SBOMs are expected to be \textbf{generated during early software development stages and continuously enriched/updated}.
\end{tcolorbox}
\end{center}

\subsubsection{SBOM data fields standardization}
\label{sbom_data}
In this subsection, we investigated what data fields generated SBOMs contain, based on S8-S10 in Table \ref{statements}, since not knowing what to include in an SBOM is the second biggest concern for producing SBOMs according to the SBOM readiness report\cite{lf_2022}.

First, despite the existence and relative prevalence of two major SBOM standards (i.e., SPDX and CycloneDX),
some organizations generate SBOMs based on their customized non-standard formats [\textbf{S8}, 27.7\% agree, 32.3\% disagree].
Based on the survey results, although more respondents agree with standardized SBOM formats, over one quarter agree with customized SBOM generation.
As I4 stated,
\textit{``the people I spoke to [in some organizations] might have been doing this [generating SBOMs], but they might not have been doing a standard-format SBOM. They would just be keeping an inventory of all their software components." (I4-Dev.\&Sec.)}

Second, although there are 7 minimum SBOM data fields recommended by NTIA\cite{sbom_mini}, in practice, generated SBOMs don't always meet the minimum bar [\textbf{S9}, 63.1\% agree, 24.6\% disagree] for two main reasons (i.e., software vendor customization, data availability).
Some software vendors choose only to include a subset of the minimum data fields, or customize their minimum requirements to meet their respective needs. According to I3, his/her organization
\textit{``has its own minimum requirements [different from NTIA's]" (I3, Dev.)}.
Software vendors sometimes do not include all the minimum data fields as the relevant data is not always attainable.
\textit{``The truth is, I tried to put in as much as what you said (7 minimum data fields). I don't have access all the time to all the details." (I11-Dev.\&Sec.(SBOM tool))}
As a result, when relevant information of certain data fields is unavailable, the generated SBOMs can be \textit{``full of non-assertion elements [...] In practice, this means that, yes, the standards themselves can support the NTIA recommendation; No, the tools are omitting those fields or leaving them blank." (I14-Dev.\&Sec.\&Res.)}

Third, for some organizations producing SBOMs or building SBOM tools to produce SBOMs, they want to include/support as much ``useful" information in SBOMs [\textbf{S10}, 87.7\% agree, 4.6\% disagree].
For the former, since an SBOM can effectively help with internal software supply chain management, they prefer to generate more comprehensive SBOMs with additional information such as vulnerability.
As I13 mentioned,
\textit{``we want to produce the best information [...] then the developers can look at it and then do their work" (I13-Dev.\&Sec.\&Res.)}
For the latter, the more comprehensive information their SBOM tools support, the more competitive they are in the market. As stated by I11, \textit{``I also have a big scope of metadata depending on the target other than the base." (I11-Dev.\&Sec.(SBOM tool))}
Furthermore, as I14 pointed out, \textit{``something relatively worse happens [...] to create business value [...] they are trying to extend it [an SBOM] with things that may or may not be relevant to the problem." (I14-Dev.\&Sec.\&Res.)}
% The above discussion on SBOM generation leads to the finding that \textbf{the generation of SBOMs needs to be more standardized}.
% \note{For example, for different types of software, there should be official recommendations for generating SBOMs during different stages based on different situations, so that the software producers can have some examples to refer to when generating SBOMs. The minimum SBOM data fields recommendations should also be a standard requirement for SBOMs instead of just recommendations, while non-assertions are allowed in cases certain information cannot be acquired.}
\begin{center}
\begin{tcolorbox}[breakable, colback=gray!10,
colframe=black, 
width=\columnwidth,
arc = 1mm,
boxrule=0.5 pt,
]
\textbf{Finding 4}: Despite official recommendations on minimum SBOM data fields, there is still a lack of consensus on what to include in SBOMs.
% \textbf{Further standardization on SBOM-included data fields is needed}.

% Considering the lack of consensus on the very baseline of SBOM data fields (i.e., minimum data fields), further standardization is needed. The baseline standardization could be set based on different industrial sectors or software types, and non-assertions should be allowed, given the possible unavailability of specific data.
% \note{add LF support}. Apart from these ``musts", more data fields can be added when needed.
\end{tcolorbox}
\end{center}

\subsubsection{SBOM distribution}
\label{distribution}
In this subsection we discussed 3 statements (i.e., S11-S13 in Table \ref{statements}).

% Although 38.5\% of respondents disagree that the generated SBOMs are meant for internal consumption instead of sharing with downstream procurers [\textbf{S11}], 40\% agree that 
% % generated SBOMs are used internally.As mentioned by I11, some
A considerable portion of the respondents agree that
\textit{``organizations generate SBOMs for internal consumption, rather than giving them to customers" (I11-Dev.\&Sec.(SBOM tool)} [\textbf{S11}, 40\% agree, 38.5\% disagree].
The authors notice a lack of consensus on this statement from the survey respondents. We believe this is consistent with the general ``lack of consensus" status of the SBOM as in the SBOM readiness report\cite{lf_2022}, resulting from the relative recentness and lack of adoption.

However, since SBOMs are now being distributed in practice, proper distribution mechanisms are needed.
According to the SBOM readiness report\cite{lf_2022}, one of the leading concerns for SBOM production is that some information inside an SBOM is too sensitive and risky to be public. 
Since the source code is already publicly available for OSS, their SBOMs should be public.
For proprietary software/components, although some of the SBOMs can be public, \textit{``authenticated [access control] is going to be the norm" (I4-Dev.\&Sec)} [\textbf{S12}, 76.9\% agree, 13.8\% disagree], depending on the software vendors' policies.
\textit{``At least for some segments of the market, I think access management is part of it. You need to be able to share your SBOMs with whomever you want to share, and not have all of the world get access to it." (I12-Cnslt.\&Adv.)}

Apart from access control, content tailoring (selective sharing) is also helpful for mitigating the above concern.
There can be a negotiated compromise between the software vendor and its downstream procurers on what to include in the distributed SBOMs, instead of sharing the complete SBOMs [\textbf{S13}, 60\% agree, 24.6\% disagree].
\textit{``There's got to be a mechanism... like a router in the middle, that takes the SBOMs or VEXs produced by the suppliers and routes them down to each end user, exactly what they need" (I0-Cnslt.\&Adv.)}, so that \textit{``only the right people can see the right information." (I13-Dev.\&Sec.\&Res.)}
% Based on our survey, xx\% and xx\% respondents (strongly) agree with the need for access control and content tailoring (i.e., sharing only partial SBOM), respectively.

% In practice, neither access control (agreed by xx\% respondents) nor content tailoring (agreed by xx\% respondents) is implemented during SBOM distribution [\textbf{S14, S15}].
% For access control, as stated by I4, a solution for the moment is a user portal which is already in use in various scenarios. 
% \textit{``I suspect probably 90\% of them are going to start off by a customer portal, the same way you share other stuff that you don't want everyone to get their hands on." (I4)}
% In terms of content tailoring, it is one step further since not only user access is constraint, but also the specific content each user is entitled to access.
% \textit{``I think software vendors should generate APIs for everyone to consume different parts of the SBOM. If you get to that point then people will be much more comfortable sharing SBOMs." (I5)}

\begin{center}
\label{f5}
\begin{tcolorbox}[breakable, colback=gray!10,
colframe=black, 
width=\columnwidth,
arc = 1mm,
boxrule=0.5 pt,
]
\textbf{Finding 5}:
% Proprietary software vendors tend not to share (complete) SBOMs with procurers considering security risks or different customer needs. As a solution, access control and content tailoring (customization) are needed for SBOM distribution so that only permitted people can have access to specific SBOM content.
Proprietary and sensitive information in SBOMs introduces barriers to SBOM distribution. \textbf{Selective sharing (content tailoring) and access control
% , and confidential computing (e.g., zero-knowledge proofs, secure multiparty computation for sharing without access) 
mechanisms need to be considered}.
% With these mechanisms, SBOMs can be partially distributed to specific user groups, thus mitigating the SBOM producers' concern that certain information is too sensitive and risky to share with everyone.\\
\end{tcolorbox}
\end{center}

\subsubsection{SBOM validation}
\label{sbom_validation}
This subsection is based on S14 and S15 (see Table \ref{statements}).

The lack of SBOM integrity validation is a shared problem mentioned by 13 out of 17 interviewees. Without reliable validation measures [\textbf{S15}, 49.3\% agree, 26.2\% disagree], software procurers may only roughly assess the quality of an SBOM by referring to the SBOM producer's reputation [\textbf{S14}, 50.8\% agree, 13.8\% disagree].
SBOM integrity is two-fold:
a) \textbf{SBOM data integrity} (whether the SBOM has been tampered with), and b) \textbf{SBOM tooling integrity} (\textit{tooling capability} as to the competence to generate complete and accurate SBOMs; and \textit{tooling security} as to whether the SBOM generation tools are hacked). We discuss SBOM tooling integrity in Section \ref{tool_validation}.

SBOM data tampering can come from outside and inside an organization (i.e., \textbf{external/internal tampering}).
External tampering is more straightforward as an SBOM can be \textit{``easy to tamper [with] and easy to fake" (I14-Dev.\&Sec.\&Res.)} without reliable validation methods.
Thus, proper validation mechanisms are needed (e.g., signing using \href{https://docs.sigstore.dev/cosign/overview}{sigstore's Cosign}).
Inside tampering means a software vendor may change the SBOM data considering customer acceptance and security issues. For example,
I14 and I15 mentioned instances of internal tampering based on their experiences:\\
a)
\textit{``You could have a release engineer at the last minute, realizing that they wanted to change the SBOM just because otherwise, the customer wouldn't take it. It's not that the whole organization lied. But it does mean that they got to tamper with the SBOM that doesn't again faithfully represent the product that they weren't given." (I14-Dev.\&Sec.\&Res.)}\\
b)
\textit{``Whenever we are going to use an open source project, there has to be a security check [...] if some kind of [vulnerable] code is there, we just need to remove it [...]  we are not actually passing those kinds of changes to the public." (I15-Dev.)}

\begin{center}
\begin{tcolorbox}[breakable, colback=gray!10,
colframe=black, 
width=\columnwidth,
arc = 1mm,
boxrule=0.5 pt,
]
\textbf{Finding 6}:
% As an SBOM could be externally maliciously, or internally intentionally tampered with, there should be mechanisms for validating SBOM data to ensure SBOM integrity.
Trust in SBOM data needs to be assured considering tampering threats. \textbf{SBOM data validation/verification mechanisms and integrity services are needed}.
\end{tcolorbox}
\end{center}

\subsubsection{Vulnerability and exploitability}
\label{vex}
In this section we discuss SBOMs for vulnerability management and the exploitability of vulnerabilities (i.e., S16 in Table \ref{statements}).

Although vulnerability management is a representative SBOM use case\cite{9174365}, vulnerability management currently barely considers the actual exploitability [\textbf{S16}, 73.8\% agree, 13.8\% disagree].
Nevertheless, the exploitability of a vulnerability should be taken seriously, as a vulnerability may not necessarily be exploitable \cite{yin_apply_2020}.

\textit{``I think it's a very, very interesting point, and it's a very legit issue. Vulnerability and exploitability are totally different. You can't simply send a long report to the developers and ask them to update each and every vulnerable dependency that has been flagged. I know the development team might end up ignoring your report, or come back at you saying, `are you able to exploit this vulnerability? No? Then why should I go and update it if you are unable to exploit it?'" (I9-Sec.)}
% This can lead to unnecessary work and lower efficiency. As \textit{``19 out of 20 vulnerabilities that you identify for components are not actually exploitable. You’re going to waste 95\% of your time literally if you start chasing after all of those every 20 vulnerabilities, and then only 1 of those is actually exploitable." (I10)}

As mitigation, vulnerabilities are often selectively fixed based on the criticality (e.g., The Common Vulnerability Scoring System (CVSS) score). As stated by I9, \textit{``if it's a critical or a high vulnerability [...] then make sure it is updated. But when it comes to [a] medium or low [criticality vulnerability], then ignore it." (I9-Sec.)}
Although there are efforts towards exploitability, such as \href{https://www.cisa.gov/known-exploited-vulnerabilities-catalog}{CISA's Known Exploited Vulnerabilities Catalog} that serves as a ``must patch list", there are only limited records (around 800 as of August 2022) in this catalog.

Vulnerability Exploitability eXchange (VEX) has emerged as a tailored method to cope with such problems. \faThumbsOUp\,A VEX is a security advisory produced by a software vendor that allows the assertions of the vulnerability status of a software product \cite{vex}. 
As companion artifacts to SBOMs \cite{sbom_sharing}, VEXs provide SBOM operators with clearer understanding of the vulnerabilities and suggested remediation.
% Although initially developed for the SBOM-related use cases, the application of VEXs is beyond SBOM.
\faThumbsODown \,However, current exploitability evaluation is manual and subjective to the domain knowledge of the security experts\cite{6754581,slava_end_nodate}.
Also, \textit{``the ability to differentiate between whether it’s exploitable or not is a hard thing to do itself" (I11-Dev.\&Sec.(SBOM tool))},
\textit{``especially if you want to automate it". (I13--Dev.\&Sec.\&Res.)}
What is more, \textit{``there is no way to confirm this [VEX], and it's actually very, very hard to prove a negative. So if you see a VEX entry that says this [vulnerability] doesn't hit me, the only way to prove it wrong is to make an exploit yourself. Again, if you are able to do that, then you're almost making things worse, right? (I14-Dev.\&Sec.\& Res.)"}

\faThumbsODown \,To further complicate this problem, \textbf{there is hardly guaranteed unexploitability}.
As I11, an SBOM tool developer with security (hacker) experience, stated,
\textit{``I agree the more valuable these exploitable vulnerabilities are, but I don't agree that the ones defined less exploitable are not valuable. I used to be on the attacker’s side [...] Hackers can take their time, and they can find a way to put together a lot of things that look very not exploitable, and at the end of the day, find themselves with very easy and exploitable access." (I11-Dev.\&Sec.(SBOM tool))}
A possible solution is there should be \textit{``potential exploitability" (I14-Dev.\&Sec.\& Res.)}.

\begin{center}
\begin{tcolorbox}[breakable, colback=gray!10,
colframe=black, 
width=\columnwidth,
arc = 1mm,
boxrule=0.5 pt,
]
\textbf{Finding 7}: It is unclear what to do with vulnerabilities with limited exploitability exposed by SBOMs/VEXs.
% \textbf{There need to be risk-based flexible policies to communicate unfixed vulnerabilities via VEXs (or SBOMs)}.
% Although VEX is promising in that differentiating between vulnerability and exploitability is essential for efficient vulnerability management, exploitability evaluation is subject to evaluators' knowledge. Considering hackers may still be able to exploit a seemingly unexploitable vulnerability, ``potential exploitability" should exist.
% \textbf{Supporter}: VEX is promising, and \textit{``there's a fever of development (around VEX) going on at the moment" (I10-Cnslt.\&Adv.)}.\\
% \textbf{Opposer}: Determining exploitability is hard, and exploiting an ``unexploitable" vulnerability results in degraded trust.
\end{tcolorbox}
\end{center}

\subsubsection{AIBOM}
\label{aibom}
This subsection is based on S16 (see Table \ref{statements}).

AI software is software with AI components. Compared with SBOMs for traditional software, SBOMs for AI software (i.e., AIBOMs) are different [\textbf{S16}, 47.7\% agree, 24.6\% disagree].
Although some interviewees thought an AIBOM \textit{``contains only additional AI package information" (I1-Dev.)}, the AI artifacts (e.g., data, code, model, configuration) also need provenance and co-versioning\cite{lu_towards_2022, barclay_providing_nodate}.
An AIBOM (see Fig. \ref{aissc_fig}) records not only the software composition information as a traditional SBOM, but also contains information about the data/model/code/configuration co-versioning registries, allowing transparency and accountability into the AI artifacts for AI model training and evaluation.
Considering the AI software deployment is continuous progress (e.g., continuous training in case of data/concept drift), these AI artifacts' co-versioning registries are more dynamic and subject to change, while the component inventory information is relatively static.
To reduce frequent re-generation of the AIBOM, the co-versioning registries can be independent of the AIBOMs, instead of being embedded in the AIBOMs.

\subsection{RQ2: What is the current state of SBOM tooling support?}
\label{tooling}
This section discusses 6 statements (see T2 in Table \ref{statements}).
Although some practitioners argue that SBOM tools were not necessary, the importance and necessity of SBOM tools are recognized by most participants.
However, the existing tools still lack maturity in general and require further development.
% most existing tools are open source and focus on SBOM generation, while SBOM consumption on the procurers' side is significantly deficient.
% Moreover, existing SBOM tools can be hard to use for various reasons (e.g., complexity), and there is no comprehensive validation of SBOM tools.
% Another issue for SBOM tooling is how to validate SBOM tools, such as their ability to generate accurate and complete SBOMs and their security against malicious manipulation.

\subsubsection{Necessity of SBOM tools}
\label{necessity}
This subsection is based on S18 in Table \ref{statements}.

The most interesting argument about SBOM tooling is the necessity of using SBOM tools to generate SBOMs. 
% (\zc{is this just the symptom? I think the more root cause factor is how SBOM differs from something already existing. If so, should we make this info difference as part of RQ1?})
Since currently few organizations (e.g., the US government agencies) are actually requiring SBOMs to be provided upon software delivery, \faThumbsODown \,some interviewees think they do not have to generate SBOMs, especially when most SBOM tools serve like a ``proxy": they merely feed the existing metadata (e.g., package manager) into a standard format, but the data is already there with or without SBOMs.
\textit{``We do not use SBOMs internally [...] because we have tools, which is where the SBOM information is coming anyway: package manager. Those tools generally just take a look at the files in the system and the configuration of the system, whereas SBOM tools just essentially parse those and then try to use that in a format. So, if we don't have a lot of end users... why would we introduce this (SBOM)? It's like a middleman that really doesn't produce much." (I14-Dev.\&Sec.\&Res.)}

\faThumbsOUp \,That being said, most survey respondents think \textbf{generating SBOMs using SBOM tools is necessary} [\textbf{S18}, 83.1\% agree, 10.8\% disagree], which is consistent with the benefit of standardization and unification of the software composition data enabled by SBOMs discussed in Section \ref{benefit}.
Generating SBOMs is more than simply putting metadata from different sources into a standard format as SBOMs are usually enriched with information such as licenses.

\subsubsection{Availability of SBOM tools}
\label{availability}
In this subsection, we discussed statements S19-S20 in Table \ref{statements}.

Tooling is an integral part of SBOM, as SBOMs are not manually generated nor intended for direct human consumption. The generation and consumption of SBOMs rely on SBOM tools.
``Shift left" originates from DevSecOps\cite{rajapakse_challenges_2022}, which means shifting the security work to earlier stages of the software development life cycle so that security issues can be identified and fixed earlier.
Considering there is a \textit{``lack of more developer-oriented (SBOM) tools that are more familiar by developers" (I7-Sec.)}, and SBOMs are tightly coupled with security tasks, SBOM tools should also consider ``shift left".

Despite there is a lot of existing SBOM tools as mentioned in Section \ref{background}, contrary to the finding in the SBOM readiness report\cite{lf_2022} that ``SBOM consumption mirrors SBOM production", our finding shows that currently,
the \textit{``SBOM generation is ahead of SBOM consumption" (I17-Cnslt.\&Adv.)}, and there are significantly limited tools for SBOM consumption [\textbf{S19}, 75.4\% agree, 12.3\% disagree].
As stated by I17,
\textit{``the large bucket of what we don't have today in 2022 is SBOM consumption." (I17-Cnslt.\&Adv.)}
Without SBOM consumption tools, even if an SBOM was provided to a software procurer, the procurer would wonder, \textit{``what do I do with the SBOMs? How do I process them? How do I analyze them?" (I12-Cnslt.\&Adv.)}
Besides dedicated SBOM consumption tools, a possible solution is to feed SBOMs into existing IT asset management tools [\textbf{S20}, 41.5\% agree, 12.3\% disagree], which requires functional extensions.
% \textit{``SBOM should just be a natural part of the data that feeds into those tools (Transcript – I17, Pos. 11)}

\subsubsection{Usability of SBOM tools}
\label{usability}
This section discusses S21-S22 in Table \ref{statements}.

% Being easy-to-use contributes to a tool's wide adoption.
Although the SBOM tools market is proliferating with the expectation to ``explode" in 2022 and 2023\cite{lf_2022}, the usability of existing tools remains an issue.
SBOM tools can be hard to use due to various reasons. (e.g., complexity, aggressivity, lack of generalization) [\textbf{S21}, 64.6\% agree, 18.5\% disagree].
For example,
\textit{``to use the CycloneDX Maven plugin, it’s required to import this plugin in the POM.xml file. For open source software, it is all right. But for proprietary software, this introduces invasion, which can be a problem". (I3-Dev.)}

Although four interviewees (i.e., I9-I12) mentioned that there were user-friendly tools such as \href{https://dependencytrack.org/}{Dependency-track}, the interviewees also acknowledged that most SBOM tools were open source and not enterprise-ready.
A problem with open source tools is, 
\textit{``an organization needs to have the capability of knowing open source projects, running them, fine-tuning them towards its needs, maintaining them" (I12-Cnslt.\&Adv.)}, which can be a considerable problem for smaller-scale organizations and start-ups.

% \note{delete or not} Even though the SBOM tooling market is growing rapidly, and there are a growing number of tools (\zc{explain a bit tools are produced by who? This will help people understand why these tools are scattered in different places of the lifecycle}), xx\% respondents agree these tools lack detailed user manuals [\textbf{S24}]. (\zc{Is this really about user manuals? should be "when and where to use which tool?" Feel that this can be correlated to SBOM generation finding?})
% That is,
% \textit{``there are different collections of SBOM tools used in different places of the lifecycle. And there's not a good understanding of which part of the lifecycle they belong to.} (I14)

Tooling interoperability and standardization also hinder the usability of SBOM tools [\textbf{S22}, 73.8\% agree, 10.8\% disagree].
As mentioned by I17, SBOM tooling is also \textit{``an area where we need further harmonization and standardization." (I17-Cnslt.\& Adv.)}
For instance, the SBOM data (e.g., component hash) of the same software/components generated by different tools can be different\cite{sbom_mini}, while
\textit{``the whole point of a hash is that it should be the same [for the same component], so that [...] downstream users can validate it". (I17-Cnslt.\& Adv.)}

% Apart from the factors above, tooling generalization and integration with the build process

\subsubsection{Integrity of SBOM tools}
\label{tool_validation}
In this subsection, we discuss SBOM tooling integrity based on S23 in Table \ref{statements}.

As mentioned in Section \ref{sbom_validation}, SBOM integrity consists of SBOM data integrity and SBOM tooling integrity. SBOM tooling integrity also includes two aspects: a) \textit{tooling competence}: the completeness and accuracy of the accuracy caused by SBOM tooling capability; and b) \textit{tooling security}: whether the SBOM generation toolchain has been maliciously altered.

Most respondents agree that the integrity of SBOMs generated by existing tools cannot be validated [\textbf{S23}, 69.2\% agree, 18.5\% disagree]. The accuracy and completeness of the generated SBOMs caused by tooling competence is a common concern for generating SBOMs. To the best of our knowledge, there is no comprehensive measure or validation against such unintentional mistakes from the end users’ point of view.
However, the intentional tampering resulting from compromised toolchains is another story. One possible solution is to evaluate the SBOM tools' assurance based on Automated Rapid Certification Of Software (ARCOS)\cite{martin_automated_nodate} though it is still a work in progress.

\begin{center}
\begin{tcolorbox}[breakable, colback=gray!10,
colframe=black, 
width=\columnwidth,
arc = 1mm,
boxrule=0.5 pt,
]
\textbf{Finding 8}: There is a lack of maturity in SBOM tooling. More reliable, user-friendly, standard-conformable, and interoperable enterprise-level SBOM tools, especially SBOM consumption tools, are needed.
% (Comments: Some aspects in this point seem to overlap with some of the above)
% The pre-maturity of existing SBOM tools results from several aspects: lack of enterprise-level tools, lack of consumption tools for SBOM procurers, lack of output consistency (standardization) from different tools, and lack of tooling integrity validation, which also results in the lack of SBOM adoption.
% \textbf{Supporter}: SBOM tools are essential for automated SBOM generation and consumption.\\
% \textbf{Opposer}: Most existing tools are not enterprise-ready. Even the need for SBOM tools remains a question to some people.
\end{tcolorbox}
\end{center}

\subsection{RQ3: What are the main concerns for SBOM?}
\label{concern}

This section investigates practitioners' main concerns for SBOMs based on T3 in Table \ref{statements}.
During the interviews, SBOM tool developers’ common concern lay with the SBOM standard formats.
Most respondents remained concerned about SBOMs being ``roadmaps for attackers"\cite{NTIA_myth}.
The most fundamental issue is the lack of SBOM adoption and education.
% The prospect of the SBOM is bright. But before we can safely get there, there are several major concerns to be addressed.
% \note{refine the titles of the subsections}

\subsubsection{SBOM formats' lack of extensibility}
In this subsection we discussed S24 in Table \ref{statements}.

There are mainly two competing SBOM standard formats (i.e., SPDX and CycloneDX), and neither can fully meet current market needs [\textbf{S24}, 35.4\% agree, 33.8\% disagree]. Notably, this statement was mentioned by both interviewees (i.e., I11, I16) working on SBOM tool development. 
Although the respondents show a discrepancy and lack of consensus on this statement, it is consistent with the interview results.
% where there was a lack of consensus on whether the existing formats are well developed to meet various needs.

Interestingly, I11 considered the SBOM format standardization to be one of the most significant benefits, while agreed the existing standards remain to be developed.
\faThumbsOUp\,On the one hand,
\textit{``the big advantage [of SBOMs] is standardization. The formats allow a lot of people to understand the same language" (I11-Dev.\&Sec.(SBOM tool))}. It offers
\textit{``a unified framework to communicate software composition information" (I14-Dev.\&Sec.\& Res.)}.
\faThumbsODown\,On the other hand, some think the formats are not extensible enough. For example, \textit{``current formats only support one dependency relationship, DependsOn" (I11-Dev.\&Sec.(SBOM tool))}.
We summarized possible format extension points in Section \ref{formatextension} based on the interviews.\\
a)
\textit{``My biggest concern is the dynamicity and the ability to use the standard formats of SBOM [...] to define the things I want to do with these SBOMs." (I11-Dev.\&Sec.(SBOM tool))}\\
b)
\textit{``My biggest concern [...] is that the standardization is really kind of not good [...] competing standards [...] different properties, and different kinds of purposes. But consolidating down until reasonable sets of things are the same between the competing formats would be great." (I16-Dev.(SBOM tool))}

\begin{center}
\begin{tcolorbox}[breakable, colback=gray!10,
colframe=black, 
width=\columnwidth,
arc = 1mm,
boxrule=0.5 pt,
]
\textbf{Finding 9}: Although there is a set of standard formats, \textbf{they require further consensus, standardization as well as additional extension points}.
% Although standardization provides a unified foundation for SBOM development, current SBOM standard formats lack enough extensibility to support more diverse needs, thus needing further development.\\
% % SBOM formats and standardization cause concerns to some stakeholders. One of the reasons is the lack of .\\
% \textbf{Supporter}: Although the source of SBOM information may already be available through packager managers, SBOM formats are valuable for providing a unified mechanism to communicate the software composition information.\\
% % \note{relate to necessity of SBOM tools} \\
% \textbf{Opposer}: SBOM formats sometimes cannot meet the market's needs (especially for SBOM tool vendors). Some organizations are developing and maintaining their own standards.
\end{tcolorbox}
\end{center}

\subsubsection{SBOM information sensitivity}
In this section, we discussed S25 in Table \ref{statements}.

% Some SBOM practitioners worry that  
% This statement is consistent with the need for access control and content tailoring mechanisms in Section \ref{distribution}.
There are two types of opinions among the participants about the SBOM information sensitivity issue:
\faThumbsODown\,Some think certain information inside the SBOM is too risky and sensitive to be public, and the information inside an SBOM may serve the attackers as a "roadmap" of the software and supply chain [\textbf{S25}, 63.1\% agree, 20\% disagree].
\faThumbsOUp\,Meanwhile, others believe that there is no need to worry as the attackers do not need SBOMs because they already have tools to easily get the software composition information. SBOMs are actually roadmaps for defenders to help level the playing field\cite{NTIA_myth}.
Just as I4 stated,
\textit{``on the surface, it seems like a valid concern. But [...] if you're highly sophisticated cooperation, you probably don't necessarily need this information. As a start-up, you don't get to be a target in an attack [...] I think the benefit of that visibility is certainly on the defender's side." (I4-Dev.\&Sec.)} 

In addition, access control and content tailoring (see Section \ref{distribution}) can also play a role in mitigating this concern.

\subsubsection{SBOM adoption and education deficiency}
\label{adoptionconcern}
In this section, we discussed S26 in Table \ref{statements}.

The most fundamental and imminent concern is limited SBOM adoption [\textbf{S26}, 80\% agree, 7.7\% disagree].
% \faThumbsODown 
Organizations may have various reasons for their hesitation to SBOM adoption. For example, they may worry about the industrial consensus or SBOMs’ value to customers, or they may lack even the most basic IT asset management.\\
% As I11 stated,
a)
\textit{``I'm actually a bit worried about the people producing and consuming SBOMs because I think the market is not really ready. They need to [be] educate(d), and many people don't know what SBOMs are." (I11-Dev.\&Sec.(SBOM tool))}\\
% I6 also mentioned, 
b)
\textit{``For SBOMs to become valuable to consume, many people need to produce [SBOMs] [...] So one of the concerns I have about SBOMs is, is everybody going to follow this pattern [to produce SBOMs]? Because there’re obviously people [saying] it's not accurate so they don't want to produce it." (I6-Dev.)}

% \faThumbsOUp 
However, despite being concerned about SBOM adoption, I6 also agreed that
\textit{``even the less accurate ones are better than adding no visibility" (I6-Dev.)}. Because
\textit{``there's always going to be a maturity problem. The idea that because some people can't use SBOM so others shouldn't [...] That's not how we do security." (I17-Cnslt.\&Adv.)}

SBOM education is needed not only to educate the public for increased SBOM adoption, but the SBOM practitioners also need to be educated and realize SBOMs have unaddressed issues before they rush into generating SBOMs.
\textit{``The problem right now is that we are almost putting the cart before the horse -- we're expecting the SBOMs to fix the problems rather than fixing the problems with an SBOM." (I14-Dev.\&Sec.)}

\begin{center}
\begin{tcolorbox}[breakable, colback=gray!10,
colframe=black, 
width=\columnwidth,
arc = 1mm,
boxrule=0.5 pt,
]
\textbf{Finding 10}: There is a lack of market awareness and good value propositions for SBOM adoption. SBOM advocators need to: \textbf{a) leverage relevant regulation and use cases such as procurement evaluation and supply chain risk management to improve SBOM awareness; and b)
promote more SBOM consumption tools with clear benefits}.
% SBOM education is needed to mitigate the lack of SBOM adoption, one of the practitioners' most significant concerns. As more people realize the importance of SBOMs through education, SBOMs will be more adopted. With the increased adoption, the SBOM field will be further developed, and many current issues will be resolved.\\
% \textbf{Supporter}: Education will help improve SBOM adoption, and increased adoption will subsequently help SBOM development.\\
% \textbf{Opposer}: SBOM adoption status is worrying. 
% There are various reasons for the hesitation of SBOM adoption (e.g., lack of basic IT asset management, not sure about the value of SBOM). What's more, many people don't even know what an SBOM is.
\end{tcolorbox}
\end{center}

% \subsection{RQ3: SBOM perception (benefits and concerns)}

\section{Discussion and implications}
\label{Dis}
% \note{add figure: similar to goal model, link all the findings, stakeholders etc}

% We summarize the complete structure of this paper in Fig. \ref{summary_fig}.
% To investigate practitioners' perception of SBOMs, we focus on 3 research questions, from which we derived 15 subtopics corresponding to 26 statements and 10 findings.
% The most fundamental finding is the lack of SBOM adoption which is mitigated or hindered by the other 9 findings.
% We acknowledge there is a lack of consensus on some statements (e.g., S8, S11, and S23 in Table \ref{statements}). However, we believe this aligns with the lack of consensus on SBOM practices in the Linux Foundation's SBOM readiness report since SBOM is a relatively fresh concept.
% With this study, we hope to present an unbiased reflection of the SBOM status quo, describing what is in the SBOM field and what needs to be there in the future.
% \subsection{Discussion}

\subsection{Implications}

This section discusses below key implications for future SBOM research and development.

\begin{table*}[]
\renewcommand\arraystretch{1.15}
\centering
\caption{SBOM readiness report v.s. this paper}
\label{compare}
\resizebox{\textwidth}{!}{%
\begin{tabular}{l|l|l}
\hline
\textbf{Topics} &
  \textbf{SBOM readiness report} &
  \textbf{This paper} \\ \hline
\textbf{Benefits} &
  \begin{tabular}[c]{@{}l@{}}16 specific benefits of SBOMs (10 for producing and 6 for\\ consuming SBOMs), all enabled by transparency.\end{tabular} &
  \begin{tabular}[c]{@{}l@{}}Transparency, and subsequently enabled \textbf{accountability, traceability, and}\\ \textbf{security}. (Finding 1)\end{tabular} \\ \hline
\textbf{Adoption} &
  \begin{tabular}[c]{@{}l@{}}90\% surveyed organizations have started SBOM journey;\\ 47\% are using (i.e., producing/consuming) SBOMs.\end{tabular} &
  \begin{tabular}[c]{@{}l@{}}SBOM adoption is \textbf{worrying}: limited generation \& more limited consumption. \\(Findings 1, 2, 10)\end{tabular} \\ \hline
\textbf{Generation} &
  \begin{tabular}[c]{@{}l@{}}a) SBOMs can be generated at different SDLC stages.\\ b) More organizations favor including more than baseline\\ SBOM information.\end{tabular} &
  \begin{tabular}[c]{@{}l@{}}a) SBOMs can be generated at different SDLC stages but practitioners expect\\ \textbf{``dynamic" SBOM generation} throughout SDLC (Finding 3).\\ b) SBOM-included data fields need \textbf{further standardization} (Finding 4).\end{tabular} \\ \hline
\textbf{Distribution} &
  N/A &
  \textbf{Secure yet flexible} SBOM distribution mechanisms are needed. (Finding 5) \\ \hline
\textbf{Integrity} &
  N/A &
  SBOM \textbf{integrity assurances} are needed against tampering threats. (Finding 6) \\ \hline
\textbf{Vulnerability} &
  SBOMs should reflect vulnerability information. &
  \begin{tabular}[c]{@{}l@{}}a) Organizations \textbf{may not want to share sensitive (vulnerability) data}\\ (Finding 5).\\ b) Mechanisms are needed to \textbf{communicate vulnerabilities with limited/}\\\textbf{undetermined exploitability} (Finding 7).\end{tabular} \\ \hline
\textbf{Tooling} &
  Limited availability of SBOM tooling &
  \begin{tabular}[c]{@{}l@{}}Affirmed \textbf{necessity} but \textbf{limited availability, usability, and integrity} of SBOM\\ tooling. (Finding 8)\end{tabular} \\ \hline
\textbf{Concerns} &
  \begin{tabular}[c]{@{}l@{}}4 shared concerns for production \& consumption: industry\\ commitment, data fields consensus, value of SBOMs,\\ tooling availability. 2 additional concerns for production:\\ information privacy, correctness.\end{tabular} &
  \begin{tabular}[c]{@{}l@{}}Explicitly identifies 3 major concerns but covers more throughout\\a) \textbf{Standard formats}' lack of extensibility (Finding 9).\\ b) SBOM information sensitivity \& privacy (Finding 5).\\ c) \textbf{Adoption and education} deficiency (Finding 10).\end{tabular} \\ \hline
\textbf{AIBOM} &
  N/A &
  AIBOM should also include AI/ML-specific data. (Section \ref{aibom})\\ \hline
\end{tabular}%
}
\end{table*}
\subsubsection{Goal model}
\label{goalmodel}
\begin{figure}[htb]
\includegraphics[width=\columnwidth]{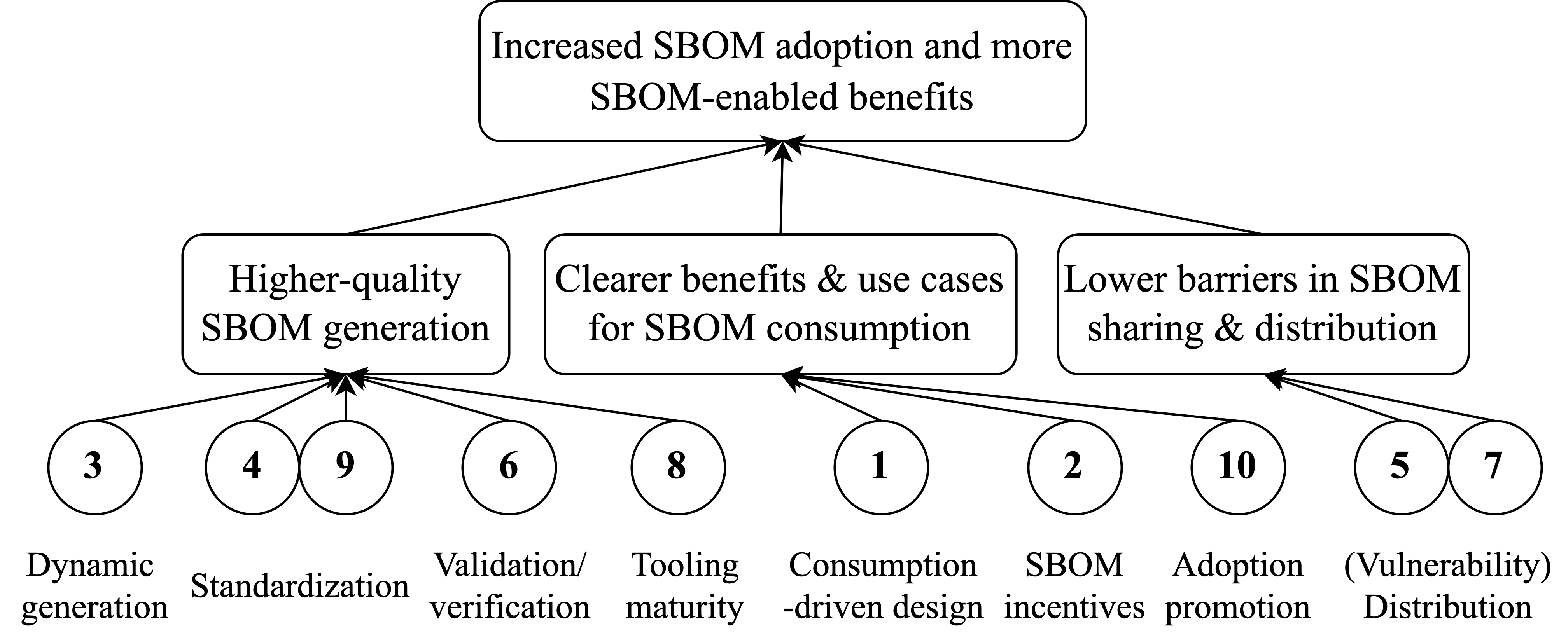}
\caption{SBOM goal model.}
\label{goalmodel_fig}
\end{figure}

Based on the study results, we present a goal model for future SBOM endeavors (see Fig. \ref{goalmodel_fig}).
As mentioned in Sections \ref{adoption} and \ref{adoptionconcern}, the lack of SBOM adoption causes a substantial obstacle to SBOM progress.
To achieve
\textbf{increased SBOM adoption and more SBOM-enabled benefits}, there are three goals to be satisfied:

\textbf{a) Higher-quality SBOM generation} (findings 3, 4, 6, 8, and 9): maturer tooling support for the generation of more standardized tamper-proof ``dynamic" SBOMs \cite{rezilion}. For instance, further standardization on SBOM-included data fields is needed. SBOM industry should strictly conform to an agreed minimum data fields, while considering different industry and business sectors when adding optional data fields.

\textbf{b) Clearer benefits and use cases for SBOM consumption} (findings 1, 2, 10): SBOM education (e.g., on SBOM-enabled benefits) results in increased SBOM adoption (including consumption); Increased SBOM adoption, in turn, leads to more developed SBOM-centric ecosystems with favorable use cases.

\textbf{c) Lower barriers in SBOM sharing and distribution} (findings 5 and 7):
The distribution and sharing of SBOMs and the vulnerability status (e.g., VEXs) need to be more flexible with proper mechanisms that meet both the software vendors' and procurers' needs. Technologies such as Blockchain, confidential computing (e.g., zero-knowledge proofs, secure multiparty computation for sharing without access) can potentially be leveraged to communicate SBOM data. During the distribution of SBOM data, there also need to be risk-based flexible policies to communicate unfixed vulnerabilities.

\subsubsection{Format extension points}
\label{formatextension}
As discussed in Section \ref{concern}, the interviewees mentioned several potential extensions to the existing SBOM formats. Other than supporting more dependency relationship types, another possible extension point is component types. \textit{``For example, the components of a Git is the commits, and CycloneDX does not have such a component [type] defined." (I11-Dev.\&Sec.(SBOM tool))}
The third extension point is file location.
\textit{``My organization has its own SBOM format [...] which has [...] things like where does it find something in a file system [...] However, formats like SPDX might not have a place for that." (I16-Dev.(SBOM tool))}
Also, Microsoft include build provenance information for tamper-evident SBOMs.
Apart from the points mentioned above, another essential point is verifiable credentials\cite{lu_towards_2022} embedded in or linked to an SBOM.
For traditional software, such credentials can prove the validity of software/components.
For AI software, responsible AI-related information such as conformance to certain AI ethics principles can be included.

\subsection{Positioning with respect to SBOM readiness report}
% Emphasising security and organizational ``readiness", SBOM readiness report investigates whether and to what extent organizations are prepared for generating and consuming SBOMs, while our paper supplements the SE angles.
This section compares the key differences between our findings and the SBOM readiness report's results.

Overall, the SBOM readiness report adopts an organizational perspective (e.g., industry-wide challenges and opportunities) while our study focuses on SBOM practitioners' perspectives and offers a more fine-grained investigation.
Specifically, this study broadens and deepens the SBOM readiness report mainly from 9 SBOM-related aspects: benefits, adoption, generation, distribution, integrity, vulnerability management, tooling, concerns, and AIBOM (See Table \ref{compare}).

\subsection{Threats to validity}
The number of survey participants may pose a threat to validity. However, considering the relative novelty of SBOM and the lack of SBOM adoption, we made our best effort and included participants from different backgrounds.
To minimize the impact of limited SBOM understanding, we provided the option for respondents to skip the survey if they are not familiar with SBOMs. Despite these measures, it's possible that the collected responses may not accurately reflect the participants' beliefs. This is a common and acceptable threat to validity in similar studies, which assume that the majority of responses are truthful.

\section{Related work}
\label{6related}
% \note{add software supply chain security papers ~5}
% \subsection{software supply chain security}
% \subsection{Software bill of materials}
% \subsection{SBOM-related work}
On the one hand, there has been work on SSC security.
For instance,
Ohm et al.\cite{ohm2020backstabber} summarized 174 OSS packages used for real-world malicious SSC attacks.
% do Amaral et al. \cite{do2021integrating} proposed to incorporate Zero Trust into SSC risk management.
Blockchain has been applied to SSC security (e.g., \cite{8473517,9653024,8890486}. Especially, Marjanović et al. \cite{marjanovic2021improving} used blockchain-based techniques to record the software composition details.
On the other hand, there are limited scholarly papers on SBOMs.
Martin et al. \cite{9174365} introduced the concept of the SBOM and listed nine possible use scenarios.
In 2021, Carmody et al. \cite{carmody_building_2021} presented a high-level overview of how SBOMs help build resilient medical SSCs. They illustrated the benefits of SBOMs for software producers, consumers and regulators, as well as relevant progress on SBOMs.
Back in 2019, Barclay et al. \cite{barclay2019towards} introduced their ideas on applying BOMs to data ecosystems for transparency and traceability, where they detailed a conceptual model to combine a BOM (static) and a bill of lots (dynamic) to jointly record the static data components and the dynamic data of a specific experiment. 
% Although their work was only conceptual, the idea can be borrowed and apply to data-driven AI software.
Based on their previous work, Barclay et al. \cite{barclay2022providing} recently introduced their work using a BOM as a verifiable credential for transparency into the AI SSCs, which is a step towards AIBOM.

In recent years, there have been numerous studies related to SBOMs, such as software composition analysis (SCA), release engineering, and reproducible builds. For example, Imtiaz et al. \cite{10.1145/3475716.3475769} compared various SCA tools for vulnerability reporting, and Ombredanne et al. \cite{9206429} reviewed SCA tools for license compliance. The application of SCA in the automotive industry was discussed by Kengo Oka et al. \cite{9821841}, and Mackey et al. \cite{mackey2018building} explored integrating automated SCA into DevOps processes. In the area of release engineering, Openja et al. \cite{9240667} studied modern release engineering topics and challenges on StackOverflow, while Laukkanen et al. \cite{laukkanen2018comparison} compared release engineering practices between a mature company and a startup. Bi et al. \cite{bi2020empirical} investigated the production and use of release notes. Lastly, the production of verifiable builds was explored in works such as \cite{9465650, 10.1145/3510457.3513050, 10.1145/3510003.3510102}.

\section{conclusion and future work}
\label{7conclusion}
SBOMs are essential to SSC security considering the transparency enabled by SBOMs and the subsequently enhanced accountability, traceability and security. In this study, we interviewed 17 and surveyed 65 SBOM practitioners on their perception of SBOMs. Despite the promising SSC transparency and security enabled by SBOMs, %potential and promising as SBOMs are, 
there are still open challenges to be addressed.
%questions to answer and problems.
% Currently, the main task is to educate the public about the value of SBOMs and convince them that the government and industry are committed to addressing SBOMs.
%For SBOMs to be more widely adopted and accepted,
To accelerate the adoption of SBOMs, higher-quality SBOM generation, clearer benefits and use cases in SBOM consumption, and lower barriers in SBOM sharing are prerequisites which need to be further studied, mitigated and addressed. In addition, SBOMs for AI software (i.e., AIBOM) are an inevitable trend given the popularity of AI applications, and AIBOMs need to consider the co-evolution of data/model/code/configuration.

\section*{Acknowledgment}
The authors would like to sincerely thank all the interview and survey participants for their help and support. This work could never have been completed without them.

% *\section*{References}
\Urlmuskip=0mu plus 1mu\relax
\bibliographystyle{IEEEtran}
\bibliography{bibli}

\begin{thebibliography}{10}
\providecommand{\url}[1]{#1}
\csname url@samestyle\endcsname
\providecommand{\newblock}{\relax}
\providecommand{\bibinfo}[2]{#2}
\providecommand{\BIBentrySTDinterwordspacing}{\spaceskip=0pt\relax}
\providecommand{\BIBentryALTinterwordstretchfactor}{4}
\providecommand{\BIBentryALTinterwordspacing}{\spaceskip=\fontdimen2\font plus
\BIBentryALTinterwordstretchfactor\fontdimen3\font minus
  \fontdimen4\font\relax}
\providecommand{\BIBforeignlanguage}[2]{{%
\expandafter\ifx\csname l@#1\endcsname\relax
\typeout{** WARNING: IEEEtran.bst: No hyphenation pattern has been}%
\typeout{** loaded for the language `#1'. Using the pattern for}%
\typeout{** the default language instead.}%
\else
\language=\csname l@#1\endcsname
\fi
#2}}
\providecommand{\BIBdecl}{\relax}
\BIBdecl

\bibitem{framing_ntia}
\BIBentryALTinterwordspacing
NTIA, ``{Framing Software Component Transparency: Establishing a Common
  Software Bill of Material (SBOM)}.'' [Online]. Available:
  \url{https://ntia.gov/files/ntia/publications/framingsbom_20191112.pdf}
\BIBentrySTDinterwordspacing

\bibitem{enwiki:1104117684}
{Wikipedia contributors}, ``Solarwinds --- {Wikipedia}{,} the free
  encyclopedia,''
  \url{https://en.wikipedia.org/w/index.php?title=SolarWinds&oldid=1104117684},
  2022, [Online; accessed 25-August-2022].

\bibitem{sonatype_report}
\BIBentryALTinterwordspacing
Sonatype, ``The 2021 {State} of the {Software} {Supply} {Chain} {Report}.''
  [Online]. Available:
  \url{https://www.sonatype.com/resources/state-of-the-software-supply-chain-2021}
\BIBentrySTDinterwordspacing

\bibitem{ohm2020backstabber}
M.~Ohm, H.~Plate, A.~Sykosch, and M.~Meier, ``Backstabber’s knife collection:
  A review of open source software supply chain attacks,'' in
  \emph{International Conference on Detection of Intrusions and Malware, and
  Vulnerability Assessment}.\hskip 1em plus 0.5em minus 0.4em\relax Springer,
  2020, pp. 23--43.

\bibitem{lf_2022}
\BIBentryALTinterwordspacing
``\BIBforeignlanguage{en-US}{The {State} of {Software} {Bill} of {Materials}
  ({SBOM}) and {Cybersecurity} {Readiness}}.'' [Online]. Available:
  \url{https://www.linuxfoundation.org/tools/the-state-of-software-bill-of-materials-sbom-and-cybersecurity-readiness/}
\BIBentrySTDinterwordspacing

\bibitem{singh2015open}
A.~Singh, R.~Bansal, and N.~Jha, ``Open source software vs proprietary
  software,'' \emph{International Journal of Computer Applications}, vol. 114,
  no.~18, 2015.

\bibitem{sbom_mini}
\BIBentryALTinterwordspacing
NTIA, ``The {Minimum} {Elements} {For} a {Software} {Bill} of {Materials}
  {(SBOM)}.'' [Online]. Available:
  \url{https://www.ntia.doc.gov/files/ntia/publications/sbom_minimum_elements_report.pdf}
\BIBentrySTDinterwordspacing

\bibitem{gartner_sbom}
\BIBentryALTinterwordspacing
{Gartner Research}, ``{Innovation Insight for SBOMs}.'' [Online]. Available:
  \url{https://www.gartner.com/en/documents/4011501}
\BIBentrySTDinterwordspacing

\bibitem{jiao2000generic}
J.~Jiao, M.~M. Tseng, Q.~Ma, and Y.~Zou, ``Generic
  bill-of-materials-and-operations for high-variety production management,''
  \emph{Concurrent Engineering}, vol.~8, no.~4, pp. 297--321, 2000.

\bibitem{google_white_paper}
\BIBentryALTinterwordspacing
{Google Cloud}, ``Delivering software securely.'' [Online]. Available:
  \url{https://cloud.google.com/resources/delivering-software-securely-whitepaper}
\BIBentrySTDinterwordspacing

\bibitem{sbom_his}
\BIBentryALTinterwordspacing
B.~Bensing, ``History of the {Software} {Bill} of {Materials} {(SBOM)}.''
  [Online]. Available:
  \url{https://billbensing.com/software-supply-chain/history-software-bill-of-material-sbom/}
\BIBentrySTDinterwordspacing

\bibitem{house_executive_2021}
\BIBentryALTinterwordspacing
``\BIBforeignlanguage{en-US}{Executive {Order} on {Improving} the {Nation}'s
  {Cybersecurity}},'' May 2021. [Online]. Available:
  \url{https://www.whitehouse.gov/briefing-room/presidential-actions/2021/05/12/executive-order-on-improving-the-nations-cybersecurity/}
\BIBentrySTDinterwordspacing

\bibitem{sca}
{E-SPIN}, ``{The Evolution of Software Composition Analysis (SCA)},''
  \url{https://www.e-spincorp.com/the-evolution-of-software-composition-analysissca/},
  2018, [Online; accessed 25-August-2022].

\bibitem{easterbrook2008selecting}
S.~Easterbrook, J.~Singer, M.-A. Storey, and D.~Damian, ``Selecting empirical
  methods for software engineering research,'' in \emph{Guide to advanced
  empirical software engineering}.\hskip 1em plus 0.5em minus 0.4em\relax
  Springer, 2008, pp. 285--311.

\bibitem{smith2015qualitative}
J.~A. Smith, ``Qualitative psychology: A practical guide to research methods,''
  \emph{Qualitative psychology}, pp. 1--312, 2015.

\bibitem{bi2022accessibility}
T.~Bi, X.~Xia, D.~Lo, J.~Grundy, T.~Zimmermann, and D.~Ford, ``Accessibility in
  software practice: A practitioner’s perspective,'' \emph{ACM Transactions
  on Software Engineering and Methodology (TOSEM)}, vol.~31, no.~4, pp. 1--26,
  2022.

\bibitem{cohen1960coefficient}
J.~Cohen, ``A coefficient of agreement for nominal scales,'' \emph{Educational
  and psychological measurement}, vol.~20, no.~1, pp. 37--46, 1960.

\bibitem{kitchenham2008personal}
B.~A. Kitchenham and S.~L. Pfleeger, ``Personal opinion surveys,'' in
  \emph{Guide to advanced empirical software engineering}.\hskip 1em plus 0.5em
  minus 0.4em\relax Springer, 2008, pp. 63--92.

\bibitem{xiaxin}
\BIBentryALTinterwordspacing
X.~Hu, X.~Xia, D.~Lo, Z.~Wan, Q.~Chen, and T.~Zimmermann, ``Practitioners'
  expectations on automated code comment generation,'' in \emph{Proceedings of
  the 44th International Conference on Software Engineering}, ser. ICSE
  '22.\hskip 1em plus 0.5em minus 0.4em\relax New York, NY, USA: Association
  for Computing Machinery, 2022, p. 1693–1705. [Online]. Available:
  \url{https://doi.org/10.1145/3510003.3510152}
\BIBentrySTDinterwordspacing

\bibitem{9174365}
R.~A. Martin, ``Visibility \& control: Addressing supply chain challenges to
  trustworthy software-enabled things,'' in \emph{2020 IEEE Systems Security
  Symposium (SSS)}, 2020, pp. 1--4.

\bibitem{yin_apply_2020}
\BIBentryALTinterwordspacing
J.~Yin, M.~Tang, J.~Cao, and H.~Wang, ``Apply transfer learning to
  cybersecurity: {Predicting} exploitability of vulnerabilities by
  description,'' \emph{Knowledge-Based Systems}, vol. 210, p. 106529, 2020.
  [Online]. Available:
  \url{https://www.sciencedirect.com/science/article/pii/S0950705120306584}
\BIBentrySTDinterwordspacing

\bibitem{vex}
CISA, ``\BIBforeignlanguage{en}{Vulnerability {Exploitability} {eXchange}
  ({VEX}) - {Status} {Justifications}},'' p.~13, 2022.

\bibitem{sbom_sharing}
\BIBentryALTinterwordspacing
NTIA, ``{Sharing and Exchanging SBOMs}.'' [Online]. Available:
  \url{https://www.ntia.doc.gov/files/ntia/publications/ntia_sbom_framing_sharing_july9.pdf}
\BIBentrySTDinterwordspacing

\bibitem{6754581}
A.~A. Younis, Y.~K. Malaiya, and I.~Ray, ``Using attack surface entry points
  and reachability analysis to assess the risk of software vulnerability
  exploitability,'' in \emph{2014 IEEE 15th International Symposium on
  High-Assurance Systems Engineering}, 2014, pp. 1--8.

\bibitem{slava_end_nodate}
\BIBentryALTinterwordspacing
B.~Slava, ``The end goal for {SBOMs} - lessons learned from parallel
  cybersecurity markets {\textbar} {LinkedIn}.'' [Online]. Available:
  \url{https://www.linkedin.com/pulse/end-goal-sboms-lessons-learned-from-parallel-markets-slava-bronfman/?trackingId=L48%2BvRWJfddIMba1jDSHNQ%3D%3D}
\BIBentrySTDinterwordspacing

\bibitem{lu_towards_2022}
Q.~Lu, L.~Zhu, X.~Xu, J.~Whittle, and Z.~Xing, ``Towards a {Roadmap} on
  {Software} {Engineering} for {Responsible} {AI},'' in \emph{2022 {IEEE}/{ACM}
  1st {International} {Conference} on {AI} {Engineering} – {Software}
  {Engineering} for {AI} ({CAIN})}, 2022, pp. 101--112.

\bibitem{barclay_providing_nodate}
\BIBentryALTinterwordspacing
I.~Barclay, A.~Preece, I.~Taylor, S.~K. Radha, and J.~Nabrzyski,
  ``\BIBforeignlanguage{en}{Providing assurance and scrutability on shared data
  and machine learning models with verifiable credentials},''
  \emph{\BIBforeignlanguage{en}{Concurrency and Computation: Practice and
  Experience}}, vol. n/a, no. n/a, p. e6997, \_eprint:
  https://onlinelibrary.wiley.com/doi/pdf/10.1002/cpe.6997. [Online].
  Available: \url{https://onlinelibrary.wiley.com/doi/abs/10.1002/cpe.6997}
\BIBentrySTDinterwordspacing

\bibitem{rajapakse_challenges_2022}
\BIBentryALTinterwordspacing
R.~N. Rajapakse, M.~Zahedi, M.~A. Babar, and H.~Shen, ``Challenges and
  solutions when adopting {DevSecOps}: {A} systematic review,''
  \emph{Information and Software Technology}, vol. 141, p. 106700, 2022.
  [Online]. Available:
  \url{https://www.sciencedirect.com/science/article/pii/S0950584921001543}
\BIBentrySTDinterwordspacing

\bibitem{martin_automated_nodate}
\BIBentryALTinterwordspacing
W.~Martin, ``Automated {Rapid} {Certification} {Of} {Software}.'' [Online].
  Available:
  \url{https://www.darpa.mil/program/automated-rapid-certification-of-software}
\BIBentrySTDinterwordspacing

\bibitem{NTIA_myth}
\BIBentryALTinterwordspacing
``Sbom {Myths} vs. {Facts}.'' [Online]. Available:
  \url{https://www.ntia.gov/files/ntia/publications/sbom_myths_vs_facts_nov2021.pdf}
\BIBentrySTDinterwordspacing

\bibitem{rezilion}
{Rezilion}, ``{Dynamic SBOM: A Comprehensive Guide},''
  \url{https://www.rezilion.com/blog/dynamic-sbom-a-comprehensive-guide/},
  2022, [Online; accessed 31-August-2022].

\bibitem{8473517}
M.~Mylrea and S.~N.~G. Gourisetti, ``Blockchain for supply chain cybersecurity,
  optimization and compliance,'' in \emph{2018 Resilience Week (RWS)}, 2018,
  pp. 70--76.

\bibitem{9653024}
E.~Bandara, S.~Shetty, A.~Rahman, and R.~Mukkamala, ``Let'strace —
  blockchain, federated learning and tuf/in-toto enabled cyber supply chain
  provenance platform,'' in \emph{MILCOM 2021 - 2021 IEEE Military
  Communications Conference (MILCOM)}, 2021, pp. 470--476.

\bibitem{8890486}
H.~Zhang, T.~Nakamura, and K.~Sakurai, ``Security and trust issues on digital
  supply chain,'' in \emph{2019 IEEE Intl Conf on Dependable, Autonomic and
  Secure Computing, Intl Conf on Pervasive Intelligence and Computing, Intl
  Conf on Cloud and Big Data Computing, Intl Conf on Cyber Science and
  Technology Congress (DASC/PiCom/CBDCom/CyberSciTech)}, 2019, pp. 338--343.

\bibitem{marjanovic2021improving}
J.~Marjanovi{\'c}, N.~Dal{\v{c}}ekovi{\'c}, and G.~Sladi{\'c}, ``Improving
  critical infrastructure protection by enhancing software acquisition process
  through blockchain,'' in \emph{7th Conference on the Engineering of Computer
  Based Systems}, 2021, pp. 1--7.

\bibitem{carmody_building_2021}
\BIBentryALTinterwordspacing
S.~Carmody, A.~Coravos, G.~Fahs, A.~Hatch, J.~Medina, B.~Woods, and J.~Corman,
  ``\BIBforeignlanguage{en}{Building resilient medical technology supply chains
  with a software bill of materials},'' \emph{\BIBforeignlanguage{en}{npj
  Digital Medicine}}, vol.~4, no.~1, pp. 1--6, Feb. 2021, number: 1 Publisher:
  Nature Publishing Group. [Online]. Available:
  \url{https://www.nature.com/articles/s41746-021-00403-w}
\BIBentrySTDinterwordspacing

\bibitem{barclay2019towards}
I.~Barclay, A.~Preece, I.~Taylor, and D.~Verma, ``Towards traceability in data
  ecosystems using a bill of materials model,'' \emph{arXiv preprint
  arXiv:1904.04253}, 2019.

\bibitem{barclay2022providing}
I.~Barclay, A.~Preece, I.~Taylor, S.~K. Radha, and J.~Nabrzyski, ``Providing
  assurance and scrutability on shared data and machine learning models with
  verifiable credentials,'' \emph{Concurrency and Computation: Practice and
  Experience}, p. e6997, 2022.

\bibitem{10.1145/3475716.3475769}
\BIBentryALTinterwordspacing
N.~Imtiaz, S.~Thorn, and L.~Williams, ``A comparative study of vulnerability
  reporting by software composition analysis tools,'' in \emph{Proceedings of
  the 15th ACM / IEEE International Symposium on Empirical Software Engineering
  and Measurement (ESEM)}, ser. ESEM '21.\hskip 1em plus 0.5em minus
  0.4em\relax New York, NY, USA: Association for Computing Machinery, 2021.
  [Online]. Available: \url{https://doi.org/10.1145/3475716.3475769}
\BIBentrySTDinterwordspacing

\bibitem{9206429}
P.~Ombredanne, ``Free and open source software license compliance: Tools for
  software composition analysis,'' \emph{Computer}, vol.~53, no.~10, pp.
  105--109, 2020.

\bibitem{9821841}
\BIBentryALTinterwordspacing
D.~Kengo~Oka, \emph{Software Composition Analysis in the Automotive
  Industry}.\hskip 1em plus 0.5em minus 0.4em\relax Wiley, 2021, pp. 91--110.
  [Online]. Available: \url{https://ieeexplore.ieee.org/document/9821841}
\BIBentrySTDinterwordspacing

\bibitem{mackey2018building}
T.~Mackey, ``Building open source security into agile application builds,''
  \emph{Network Security}, vol. 2018, no.~4, pp. 5--8, 2018.

\bibitem{9240667}
M.~Openja, B.~Adams, and F.~Khomh, ``Analysis of modern release engineering
  topics : – a large-scale study using stackoverflow –,'' in \emph{2020
  IEEE International Conference on Software Maintenance and Evolution (ICSME)},
  Sep. 2020, pp. 104--114.

\bibitem{laukkanen2018comparison}
E.~Laukkanen, M.~Paasivaara, J.~Itkonen, and C.~Lassenius, ``Comparison of
  release engineering practices in a large mature company and a startup,''
  \emph{Empirical Software Engineering}, vol.~23, no.~6, pp. 3535--3577, 2018.

\bibitem{bi2020empirical}
T.~Bi, X.~Xia, D.~Lo, J.~Grundy, and T.~Zimmermann, ``An empirical study of
  release note production and usage in practice,'' \emph{IEEE Transactions on
  Software Engineering}, vol.~48, no.~6, pp. 1834--1852, 2020.

\bibitem{9465650}
Y.~Shi, M.~Wen, F.~R. Cogo, B.~Chen, and Z.~M. Jiang, ``An experience report on
  producing verifiable builds for large-scale commercial systems,'' \emph{IEEE
  Transactions on Software Engineering}, vol.~48, no.~9, pp. 3361--3377, Sep.
  2022.

\bibitem{10.1145/3510457.3513050}
\BIBentryALTinterwordspacing
J.~Xiong, Y.~Shi, B.~Chen, F.~R. Cogo, and Z.~M.~J. Jiang, ``Towards build
  verifiability for java-based systems,'' in \emph{Proceedings of the 44th
  International Conference on Software Engineering: Software Engineering in
  Practice}, ser. ICSE-SEIP '22.\hskip 1em plus 0.5em minus 0.4em\relax New
  York, NY, USA: Association for Computing Machinery, 2022, p. 297–306.
  [Online]. Available: \url{https://doi.org/10.1145/3510457.3513050}
\BIBentrySTDinterwordspacing

\bibitem{10.1145/3510003.3510102}
\BIBentryALTinterwordspacing
Z.~Ren, S.~Sun, J.~Xuan, X.~Li, Z.~Zhou, and H.~Jiang, ``Automated patching for
  unreproducible builds,'' in \emph{Proceedings of the 44th International
  Conference on Software Engineering}, ser. ICSE '22.\hskip 1em plus 0.5em
  minus 0.4em\relax New York, NY, USA: Association for Computing Machinery,
  2022, p. 200–211. [Online]. Available:
  \url{https://doi.org/10.1145/3510003.3510102}
\BIBentrySTDinterwordspacing

\end{thebibliography}

\end{document}